\documentclass[a4paper,11pt]{article}

\usepackage[ left=1.0in,  
top=0.6in,
right=1.0in,
bottom=1.0in,
headsep=0.25in,
headheight=20pt,
heightrounded,
a4paper]
{geometry}

\usepackage{amsmath,amssymb,amsthm}
\newcommand\inlineeqno{\stepcounter{equation}\ (\theequation)}
\usepackage{colortbl}
\usepackage{textcomp}
\usepackage{mathtools}
\usepackage{eurosym}
\usepackage{booktabs}
\usepackage{longtable}
\usepackage{rotating}
\usepackage[USenglish]{babel}

\usepackage{hyperref}

\usepackage{graphicx,subcaption}

\usepackage{pdfpages}
\usepackage{pdflscape}

\begin{document}

\title{The sun as colliding beam, betatron cosmic ray factory}
\author{Richard M. Talman\\
Laboratory for Elementary-Particle Physics\\
Cornell University, Ithaca, NY, USA\\
}

\maketitle

\tableofcontents

\clearpage

\begin{abstract}
A theory of cosmic ray production within the solar system is presented.  Contrary to existing theories of
cosmic ray generation that assume cosmic rays to be extra-galactic in origin, this paper describes cosmic
rays as originating primarily within the solar system.  The sun's time variable magnetic flux linkage makes it
a natural, all-purpose, betatron storage ring, with semi-infinite acceptance aperture,
capable of storing and accelerating counter circulating, opposite-sign, colliding beams.

The puzzle of how positrons and anti-protons can be well represented even at very high energies, is
addressed and accounted for.  It is a consequence, initially, of the low energy capture of particles
of either sign, as well as, later, the capture of anti-particles produced in QED beam-beam collisions of
sufficiently high energy.  Initially the low energy beams are captured primarily by the sun's magnetic dipole
field.  Later, as the magnet field bending becomes negligible compared to the gravitational bending, both
positive and negative beams will have survived the gradual transition from magnetic to gravitational
bending.  As an aside: this requires a ``fine-tuning'' of the sun's magnetic bending field.

Though the solar magnetic period is routinely quoted as 22 years, the actual data is better represented
by $22\pm2$\,year. Currently there is no explanation for this variability.  With little justification at
present, this paper suggests that Jupiter, with three orders of magnitude higher magnetic field reversal rate,
is a significant asynchronous source of solar cosmic rays.  In any case, the cosmic ray energy inherited from outside
the solar system is negligibly small. 

The high quality of cosmic ray data collected over recent decades, at steadily increasing energies, especially
by the  International Space Station (ISS), make the study of cosmic ray production mechanisms both timely and
essential. Highly energetic cosmic ray nuclear particles of all A-values, along with anti-protons, and electrons
of either sign are observed at present in nature.

The paper proceeds to describe how longitudinal electric fields, explained by the Parker theory of the solar
wind, can enable the sun to serve as a ``booster'' accelerator of cosmic rays, which increases the maximum
cosmic ray energies by many orders of magnitude.  High energy particle collision processes also maintain the
particle abundances at every energy.  The dynamic range of energies of the full process produces the
observed 13 orders of magnitude maximum particle energy and the energy flux needed to maintain the cosmic
ray atmosphere equilibrium.

A steady state mechanism is described, based on semi-quantitative discussion of a relativistic Hamilton-Jacobi
formalism, according to which the highest energy cosmic rays observed can have been produced by the Parker
longitudinal electric field component, during fractionally brief, but periodic, semi-circular turns
centered on the sun.
\end{abstract}

\section{Introduction}
This paper refers to my recent arXiv paper,
\href{https://github.com/jtalman/ual2/blob/main/docs/Sun-as-colliding-beam-breeder.pdf}{}
but already out-dated, due to a significant error concerning the
cosmic ray start-up mechanism.  It is available in the arXiv\,\cite{Talman-solar-cosmic-ray-1}; a paper that
emphasizes laboratory-based astrophysical experiments while the present paper discusses only solar cosmic ray
production.  To access the previous arXiv paper, click on the link:\quad \href{https://arxiv.org/abs/2508.19296v1}{"Appendix"}

The present paper is based on recent experimental cosmic ray studies, mainly aboard the International Space Station (ISS).
Some figures specific to cosmic ray production are common, often modified, to both papers. There is only one lengthy tables
in the present paper.  In this respect the arXiv paper can be regarded as an "appendix" to the present paper.
\footnote{``High energy accelerator physics units'' are employed in the present paper. 
as in $v=\beta c$. But the units are ``natural'' in another sense; namely because the value of $c$ can be
taken to be ``1.0'' in numerical evaluations, should the reader's preference be the same as the author's.
Planck's (modified) constant  ``$\hbar$'' does not appear, but if it did, its numerical value would also be ``1.0''.
These conventions cause magnetic units to be confusing, causing them to be avoided to the extent possible.  The numerical
values of energies will normally be expressed in GeV units.  ``Joules'' will (almost) never appear, and ``ergs'' never.
Any errors or actual inconsistency resulting from this convention should be flagged as typographical errors.  Units, especially
magnetic, are discussed in greater detail in Appendix~A.}

Another significant difference is that this paper emphasizes protons, Z=1, N=0,,
while the previous paper emphasizes ``alpha-particle-like'' nuclear isotopes with equal numbers of protons
and neutrons, Z=N, A=2Z; in particular, \emph{not protons}.
A reason for the Z=N ``alpha-like'' emphasis in the arXiv paper is that all such nuclei have
identical ``magnetic rigidities'', meaning they have identical orbits in magnetic fields. This is
helpful for understanding the Parker solar wind theory, but is probably unimportant in present context.
In this paper it is the solar wind (serving as source of particles for acceleration) that is emphasized.

\begin{figure}[hbt!]
  \centering
  \includegraphics[scale=0.53]{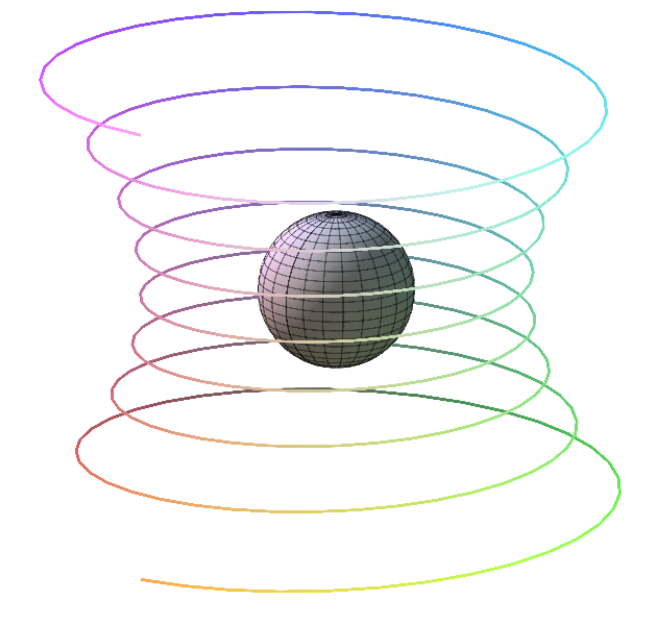}
  \includegraphics[scale=0.39]{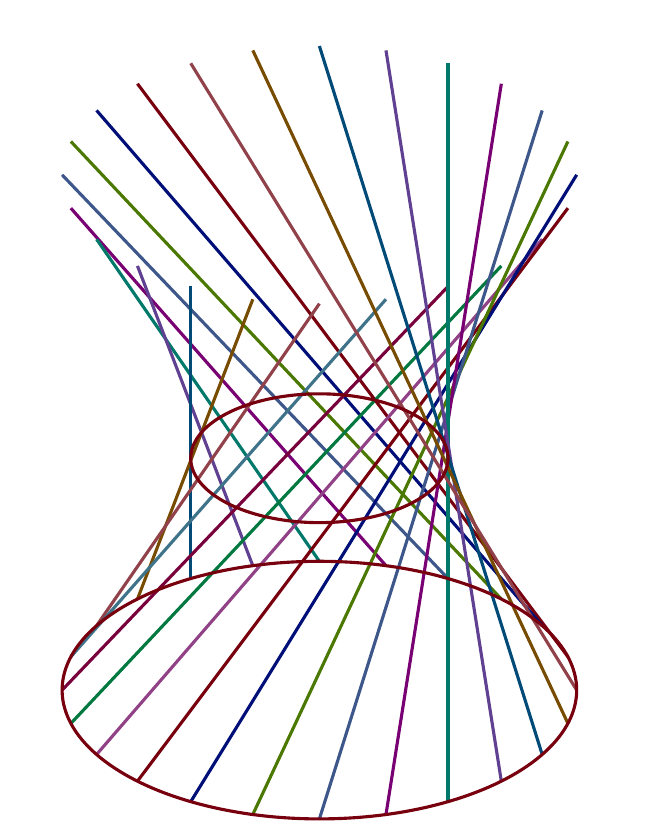}
  \includegraphics[scale=0.44]{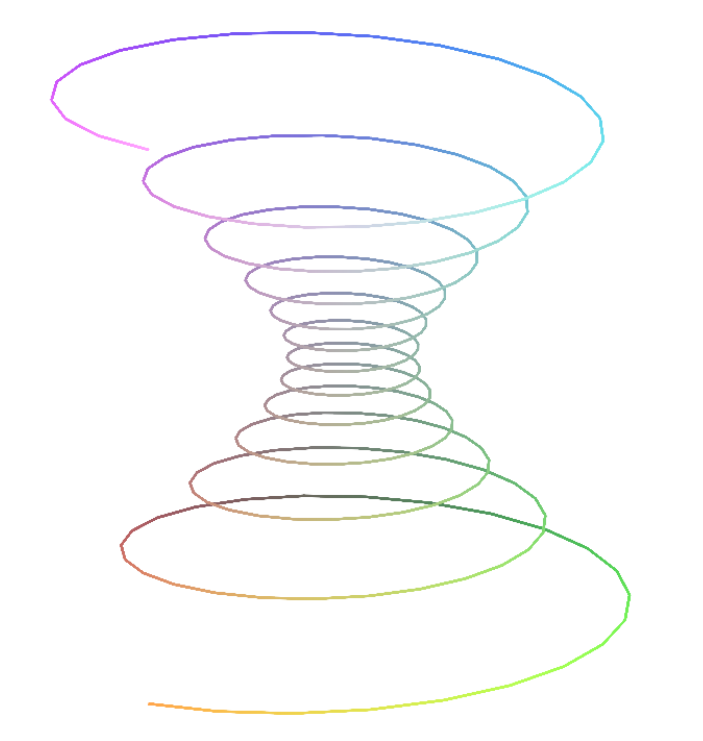}
  \caption{\label{fig:Beam-falling-in-views}Selected topologies of particles falling toward the sun:
     {\bf Left: }corkscrew-shaped orbit, ``captured briefly'' but then released; {\bf Center: } grazing the sun;
    {\bf Right: }falling into the sun.} 
\end{figure}

In this paper, once high energy proton cosmic rays are understood, so also will be all nuclear
isotope cosmic rays particles, as well as ant-protons and positrons.
Figure~(\ref{fig:Beam-falling-in-views}) shows a range of three topologically similar orbit shapes of
nuclear particles being pulled gravitationally toward the sun.    
The important case is on the left.  Only two full cork-screw rotations more or less in the equatorial plane of
the sun are shown, but there could be many more.  This is the solar cosmic ray geometry to be
described and analyzed in this paper.

\subsection{The Parker model of solar wind}
In 1926, A. A. Milne, a British writer, famously penned a puzzle in the form of a verse;\\
\noindent
``No one can tell me, Nobody knows, Where the wind comes from, Where the wind goes.``\\
Milne was referring, presumably, to the meteorological earth wind.  Though a mathematician by training,
Milne had not researched this very carefully, since, at that date, the earth's wind was fairly well
understood.  In 2025 the Milne puzzle can be applied to the solar wind.  As to ``where the wind comes from''
Milne has again become outdated; Parker has explained this.  But ``Where the solar wind goes'' remains an
active puzzle for astrophysics.

The challenge for this paper is to demonstrate that the solar wind can be treated as the injected beam,
into an accelerator consisting of the the sun itself, as betatron, producing ultrahigh energy cosmic ray
particles. 

\subsection{Solar magnetic fields}
Magnetic fields in the interior of the sun are hopelessly chaotic. Interior magnetic field structures, referred to as 
``flux tubes'' are distributed throughout the sun's interior. These are in the form of more or less cylindrical, randomly
oriented, structures that are quite well modeled in isolation. Fortunately, for present purposes, for reasons to be
explained, it will not be necessary to understand how the internal stellar matter is organized.  Just the magnetic fields
at the surface of the sun are sufficient, as boundary conditions, to establish the external magnetic fields.  It is,
of coarse, important for these surface fields to be reliably known. 

Magnetic fields external to the sun are shown in Figure~\ref{fig:sun-exterior-magnetic-field}.  Since magnetic flux
lines are continuous the average magnetic fields, even in the interior of the sun, would seem to be required to match
the magnetic fields shown in the upper two plots in Figure~\ref{fig:sun-exterior-magnetic-field}.  It is clear, in
the bottom picture, that the surface itself is represented by actual data, made available from the NASA's Goddard Space
Flight Center.  From their description it seems likely that the external magnetic field lines have been constrained to
best fit the available measurements to the theoretically well known magnetic dipole pattern outside the sun. The field
outside a uniformly magnetized sphere is the same as the field of an equivalent point dipole located at its center;
\begin{equation}
B(r,z)\Big|_{z=0} = \frac{\mu_0}{4\pi}\,\frac{(AI)_{\rm sun}}{(z^2 + r^2)^{3/2}})\Big|_{z=0}; \hbox{\ for\ } r>R_{\rm sun},
\label{eq:mag-dipole} 
\end{equation}
where $(AI)_{\rm sun}=2\pi R^2_{_{\rm sun}} I_{\rm eff.}$, and $I_{\rm eff.}$ is the effective current flowing
along the sun's equator that would produce the observed magnetic field in the free space outside the sun.
This is the (changing) magnetic field that produces the acceleration of particles during their temporary
circulation  and acceleration while near the sun, possibly for multiple turns, before their return to
precessing elliptical orbits.

These magnetic fields need to be treated carefully.  Their role in the Parker solar wind theory, that
explains the source of the  longitudinal electric field which accelerates charged particles, is essential.
At the same time it must be confirmed that any radial magnetic force, either centripetal or centrifugal,
does not compete too seriously with the centripetal gravitational pull on charged particles circulating
above the sun's equator.  This region is illustrated and labeled in
Figure~{\ref{fig:sun-exterior-magnetic-field}.  When referring to the sun as a storage ring or accelerator,
the ``aperture'' is as shown in this figure.  Of course, no vacuum chamber is required, and the ``acceptance''
is ``semi-infinite''.
\begin{figure}[hbt!]
   \centering
   \includegraphics[scale=0.33]{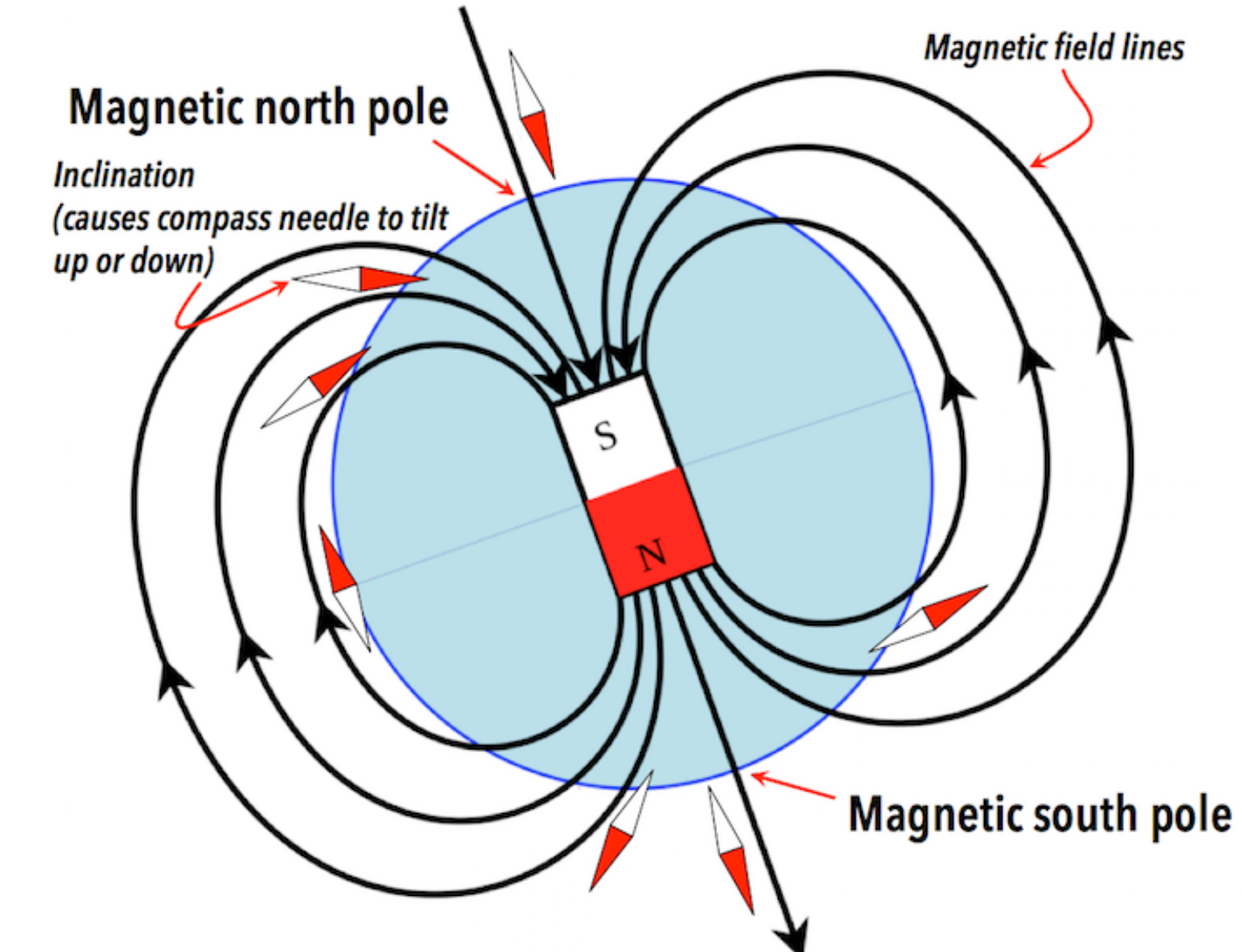}
   \includegraphics[scale=0.33]{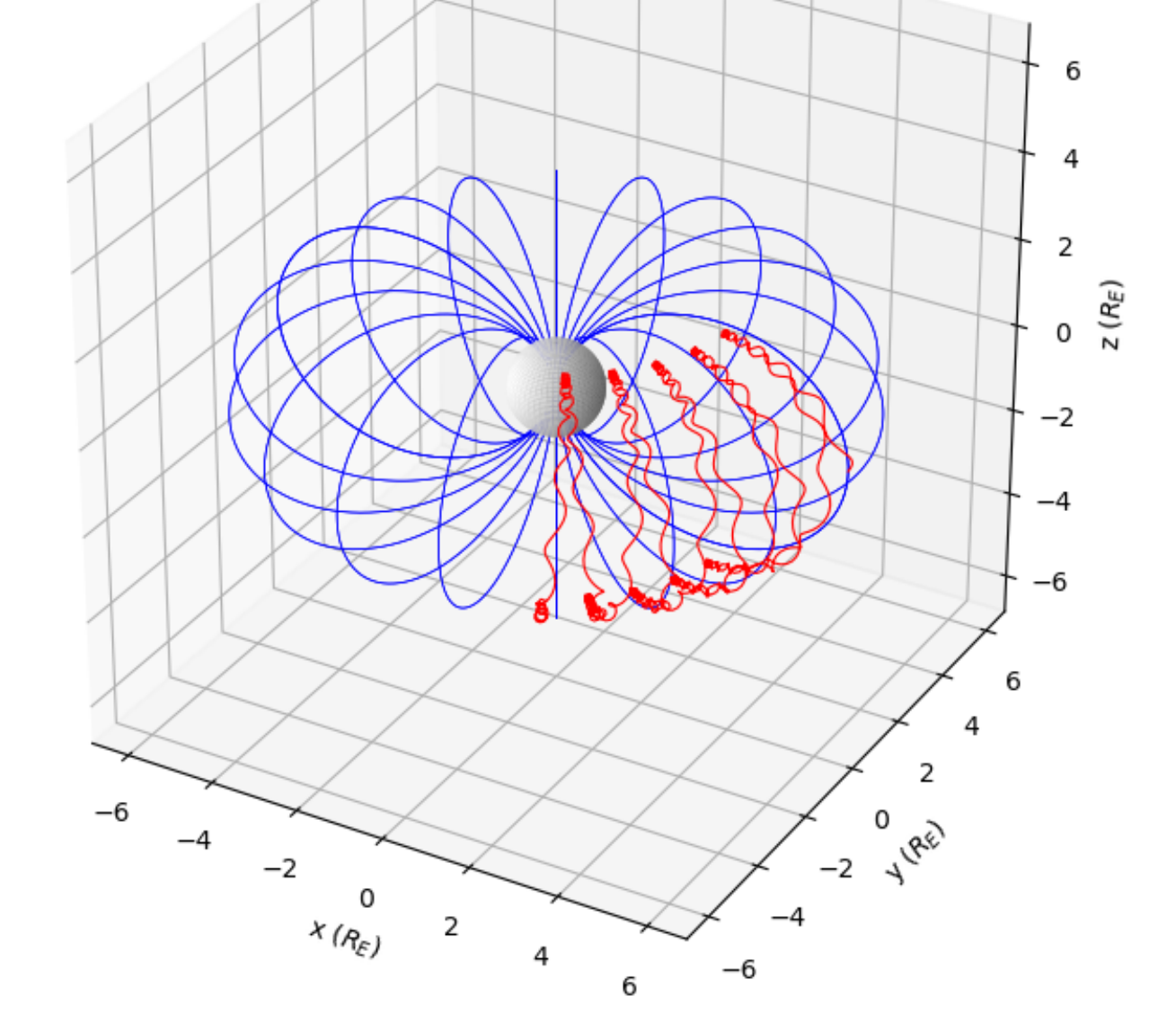}
   \includegraphics[scale=0.40]{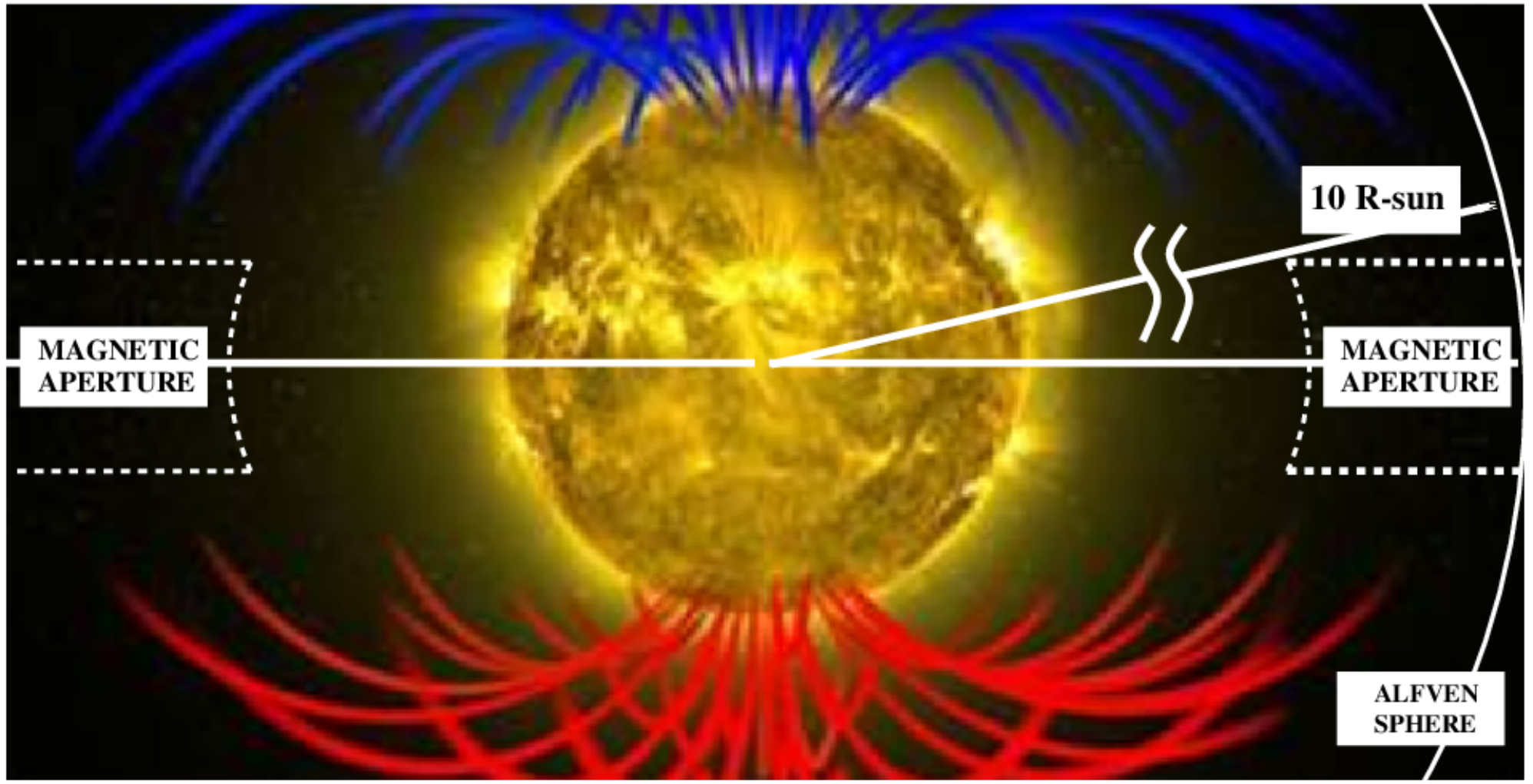}
   \caption{\label{fig:sun-exterior-magnetic-field}{\bf Top left:\ }Magnetic dipole field pattern.
     {\bf Top right:\ } Perspective view of dipole field pattern. 16 field lines are shown
     (actually +1, including the straight line from observer's view point).
     {\bf Bottom:\ } 2025 image shows magnetic fields radiating from the sun's poles. 
     Courtesy of NASA's Goddard Space Flight Center.  Superimposed is the outline of the
     aperture of the sun as a betatron particle accelerator.  Interpolated onto the photograph are
     outlines of a virtual vacuum chamber for the sun as particle accelerator.  Also indicated,
     at approximately 10 times the sun's radius, is the Alfven radius, a reference radius 
     that figures prominently in the sun's injection, acceleration, and extraction  processes.}
\end{figure}
\subsection{Solar electric fields}
The top image in Figure~\ref{fig:Parker-solar-wind} is copied from Owens and Forsythe\cite{Owens-Forsyth}.
The bottom figure breaks out the electric and magnetic field patterns. The so-called ``Parker spiral angle'',
$\theta_s$, is indicated by all the arrows in the bottom figure.  (The fact that these angles seem not quite
equal is a defect of the figure, not of the theory.)  The original version of this figure is contained
(as a simple sketch) in Parker's original paper explaining the solar wind.

The tangential electric fields result from the 22 year periodic variation of the sun's magnetic
polarity.  The tangential electric field causes the sun to act as a betatron accelerator.  The electromotive
force resulting from Faraday's law produces the tangential electric fields shown in the lower figure.
The original caption to the figure referred to the figure as\\

\noindent
``a sketch of the steady-state solar magnetic field
in the ecliptic plane. Close to the sun, in a spatial region approximately bounding the solar corona, the magnetic
field dominates the plasma flow and undergoes significant non-radial (or super-radial) expansion with height. At
the source surface, typically taken to be a few solar-radii, the pressure-driven expansion of the solar wind
dominates and both the field and flow become purely radial.  In the heliosphere, rotation of the heliospheric
magnetic field (HMF) footprints within a radial solar wind flow generates an azimuthal component of the HMF,
$B_{\phi}$, leading to a spiral geometry.  Regions of opposite HMF polarity, shown as red and blue, lines, are
separated by the heliospheric current sheet (HCS), shown as the green dashed line.  Image adopted originally from
Schatten, Wilcox, and Ness.''\\

\begin{figure}[hbt!]
   \centering
   \includegraphics[scale=0.56]{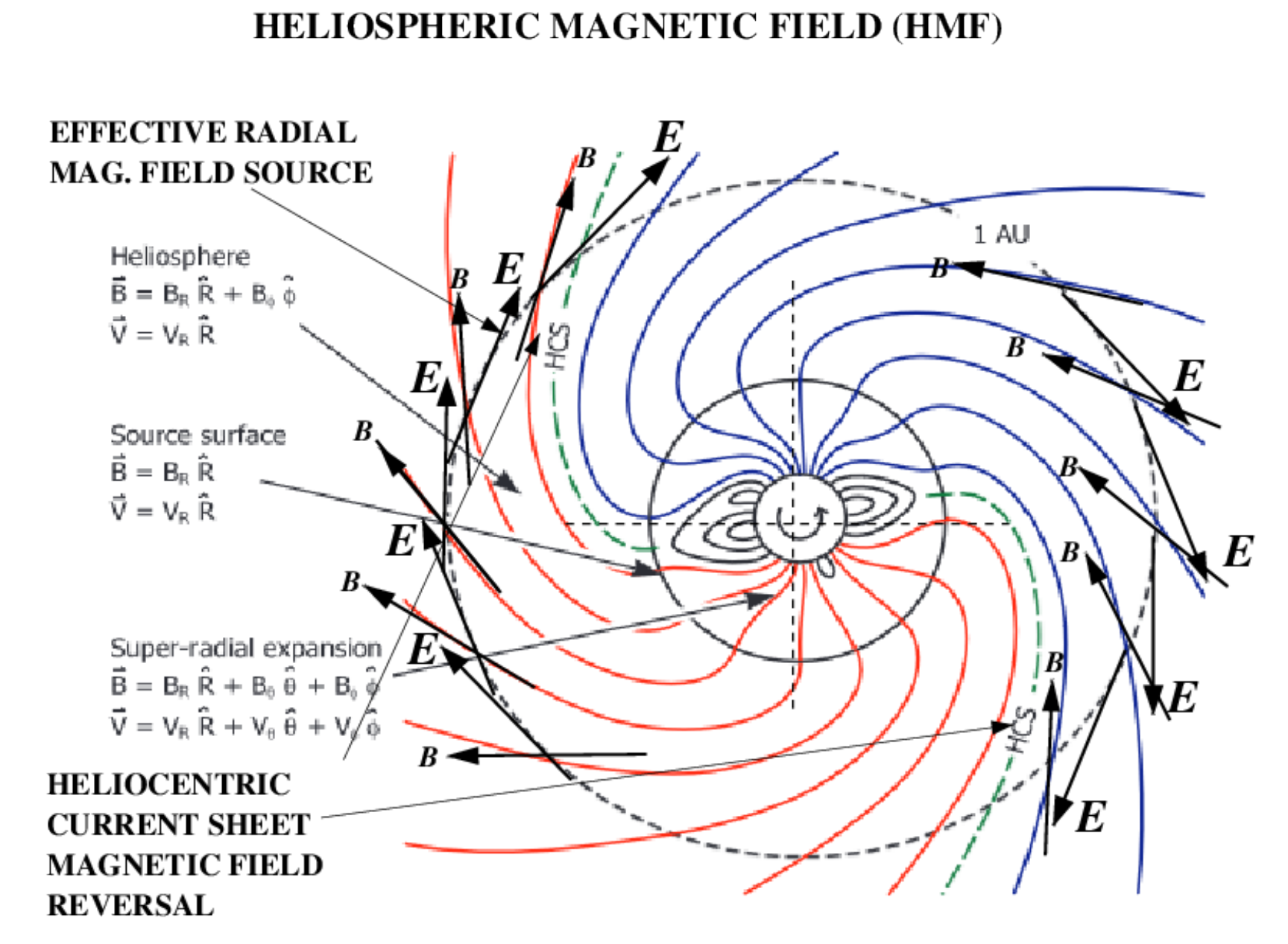}
    \includegraphics[scale=0.47]{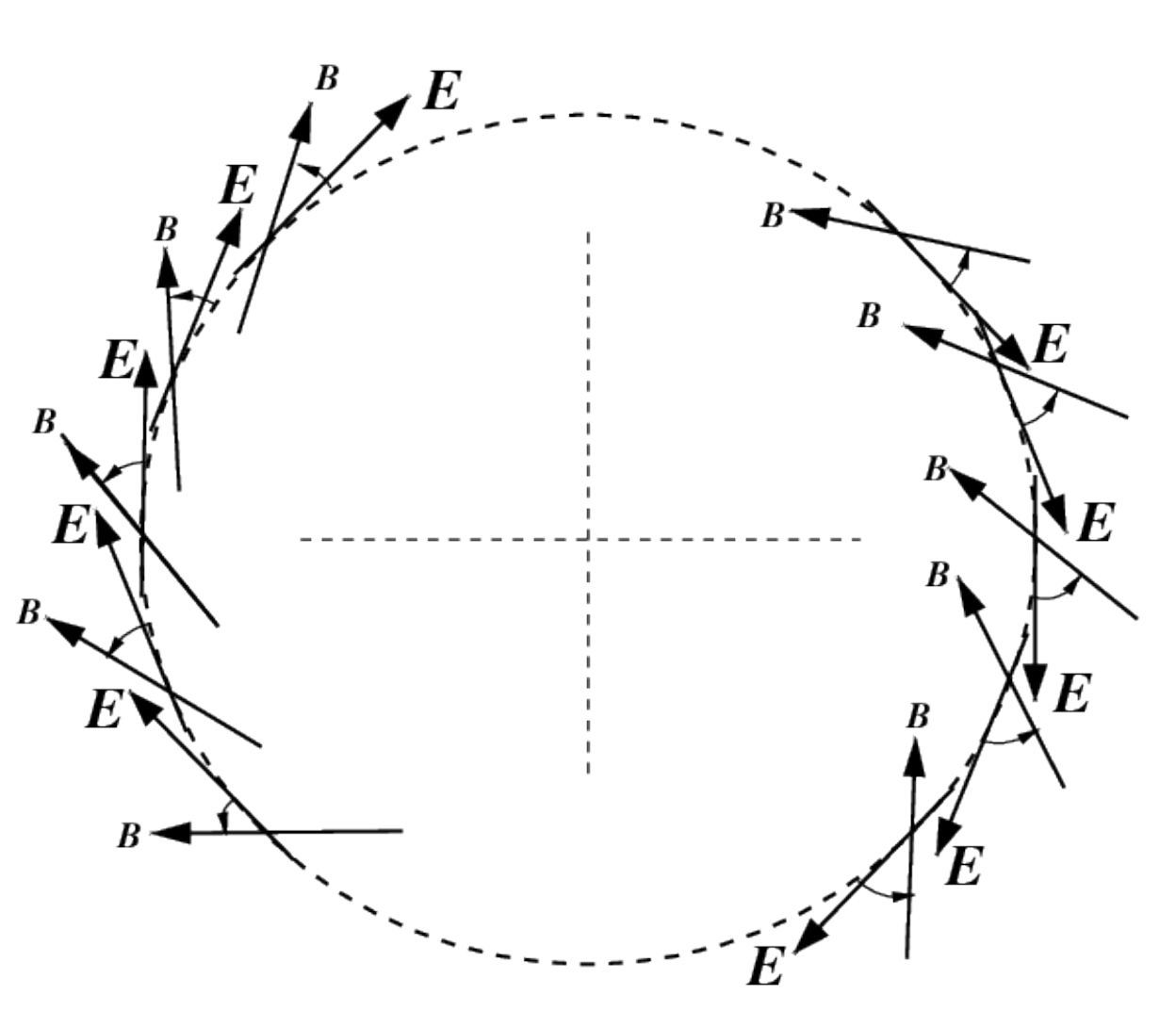}
    \caption{\label{fig:Parker-solar-wind}{\bf Top:\ }Copied and annotated figure from Owens and
      Forsythe\cite{Owens-Forsyth}, showing magnetic field directions, along with added
      electric field directions.  The electric field directions correspond to the Faraday's law
      electromotive force resulting from the 22 year period, time-varying axial solar magnetic flux.
      {\bf Bottom:\ }Copied from the top figure, the local electric and magnetic field directions are
      shown, illustrating, for example,  the magnetic field reversal across the heliocentric current
      sheet (HCS).  This figure defines the Parker angle $\theta_P$ orienting the $B$-field lines relative
      to the effective radial magnetic field source and demonstrates its constancy.''}
\end{figure}
As complicated as it is, Figure~\ref{fig:Parker-solar-wind} requires some qualification in order to
be comprehensible. Though a nuclear isotope orbit is sometimes said to coincide with a magnetic field line,
\emph{this is categorically incorrect.}  The isotope orbits are actually helical, winding around a ``guiding center'', just
as in the earth's Van Allen belts.  It is the guiding center that coincides with a magnetic field
line.

Ever since the first reported observation of cosmic rays (by Victor Hess in 1912)
their source has never been unambiguously identified.  A well known first attempt
by Fermi, has only been grudgingly accepted, even by Fermi himself.  But
Fermi's mechanism has commonly favored the view that cosmic ray production
is a galactic or extra-galactic phenomenon.

\subsection{The sun as a betatron}
The first fully successful circular particle accelerator was the ``betatron'' invented by Donald Kerst, at the
University of Chicago, just before the Second World War.  The source of the particle acceleration was
the Faraday's law electromotive force (EMF) induced by the time-varying magnetic field linked by the
toroidal vacuum chamber of the ring.

There is a very substantial changing magnetic flux associated with the 22 year periodic variation of the
suns magnetic dipole moment (MDM).  The sun's changing magnetic flux linkage makes it a natural betatron
cosmic ray particle  accelerator. 

\subsection{Faraday's law determination of the sun's maximum betatron EMF}
We wish to find the Faraday EMF encountered in a nuclear trip around the sun.
The Sun's time-dependent, 22 year periodic magnetic dipole moment variation is the
source of time variation.
\begin {equation}
 \mu_{\rm sun}(t) \approx 3.9\times{10}^{33}\,{\rm Am^2}\,\cos(\omega_{\rm sun} t),
 \label{eq:Faraday.1}
\end{equation}
where $\omega_{\rm sun} = 22\times 10^7\times 0.31536\,{\rm s}$.
The maximum time rate of change of $\mu_{\rm sun}$ is 
\begin{equation}
  \frac{d\mu_{\rm sun}}{dt} = \frac{3.9\times{10}^{33}}{22\times 10^7\times 0.31536}   \,{\rm Am^2/s}.
  \label{eq:Faraday.1t}
\end{equation}
The area of the equatorial plane of the sun is $A_{\rm sun} = \pi R_{\rm sun}^2 = 3.14\times{0.6957^2\times10^{18}}$ .
Another formula for the sun.s magnetic dipole moment is $\mu_{\rm sun} = \mu_0 I_{\rm loop} A_{\rm sun}$, or
\begin{equation}
\frac{dI_{\rm loop}}{dt} = \frac{1/\mu_0}{3.14\times{0.6957^2\times10^{18}}}\
                       = \frac{3.9\times{10}^{33}}{22\times10^7\times 0.31536}\,{\rm A/s}
\label{eq:Faraday.2}
\end{equation}
where  $\mu_0 =4\pi\times10^{-7}\,{\rm H/m}$  (Henries/meter a.k.a. T m/A)  
where $I_{\rm loop}$ is the current in a loop around the equitorial diameter having the same MDM.

For a thin ring of radius that of the sun let us take the wire radius to be $r$=1\,m, in
which case the inductance is given by
\begin{align}
L &= \mu_0 R\ln \Big(\frac{8R}{1} - 1.75\Big) = 4\pi\times10^{-7}\,0.6957\times10^9\ln(0.55656\times10^{10})\\
  &= 4\times3.159\times69.57\times22.44 \approx 2\times10^4\ {\rm H}.
\label{eq:Faraday.4}
\end{align}
The maximum instantaneous Faraday law EMF is the rate of change of the magnetic flux through the
inductor using the formula,
\begin{equation}
  \mathcal{E} = -2\times10^4\times\frac{dI_{\rm loop}}{dt}
              =  -2\times10^4\times\frac{3.9\times{10}^{33}}{22\times10^7\times 0.31536}\,{\rm A/s}
\label{eq:Faraday.3}
\end{equation}

\subsection{Disambiguation of the term ``cosmic ray''}
Protons are essentially the only hadronic particles escaping from the solar system. Furthermore, these escaping
protons can be conjectured to be being replaced (on the average) by immigrating protons from other stars.
(To the very limited extent that other light nuclear particles, such as deuterons, escape, they can
be treated in the same way as escaping protons.)
On a hierarchy of  solar, then galactic, then cosmic, sequentially increasing distance and time scales, there
can be balanced immigration and emigration of protons, on solar, then galactic, then cosmic time scales. 

As regards cosmic rays, my policy for this paper has been to discuss only cosmic ray acceleration
within the solar system, treated as a closed and isolated system.  But, to produce  a ``start-up'' mechanism,
it is useful to contemplate a modest flux of multi-GeV, extra-galactic origin, known from reliable
observations to be coming from supernova explosions.

An unfortunate consequence of my policy is that it renders the term ``cosmic ray'' \emph{hopelessly ambiguous}.
Ever since their discovery in 1912 by Victor Hess, it has been mainly assumed that all particles ``coming out of the sky''
originated outside the solar system.  This is very nearly opposite to the policy in this paper, which treats all
``cosmic-rays'' as a solar phenomenon.

Though confusing, this is not unprecedented.  It simply means that one must admit the possible existence of two
classes of particles coming out of the sky.  One is ``solar system cosmic rays'', the other  ``non-solar system cosmic rays''.
It seems certain that both classes will consist primarily, of protons, deuterons, and $\alpha$-particles,
in that order. 

Regrettably, since cosmic ray identifications never establish their source,  it must always be accepted that what
has been measured is an incoherent sum of solar and non-solar cosmic rays.

It is my prejudiced belief that most ``cosmic rays'' are ``solar''.

As an aside, it can be mentioned that golf ball sized ice balls routinely come out of the sky
during electrical storms in Texas.  Hundreds of golf ball size pockmarked automobiles guarantee that this is
not ``fake news''.  Apparently this phenomenon is caused, in the presence of electrical storms, by vertical winds, that
temporarily prevent the ice balls from falling. In the presence of water vapor, and extremely low temperature,
it is no surprise that the ice balls grow quickly.

One supposes that a similar phenomenon can occur when the solar wind holds up counter-traveling nuclear ions
long enough for partially charged atomic balls containing radioactive nuclear isotopes to join.  The real mystery
is how such balls can subsequently retain or recover semi-relativistic velocities high enough to produce
$\mu$-mesons, or other incontrovertible evidence of their nuclear history. 

\section{Astronomical bodies as stochastic accelerators}
\subsection{Rough parameter values \label{sec:Rough-values}}
During isolated collisions between two nuclear particles the ``strong'' nuclear force competes with
the ``medium strength'' electric force.  At nuclear scale the nuclear force wins; at atomic scale
the electric force wins. At astronomical scale, because of charge neutrality (cancellation
of nuclear and electron charge densities to a first approximation) the competition
shifts to gravitation versus magnetism.

As earth observers we are poorly equipped to assess any of these forces.  With magnetic compass we can
detect, in round numbers, a half Gauss earth magnetic field, which is amply strong enough for a
light needle on a decent bearing to measure the field direction---in other words negligibly small
for doing real work.  As it happens, a typical magnetic field at the surface of the sun is
not very different---only several Gauss, with the further complication of varying more or less
sinusoidally with a period of 22 years.

But, by no means, may the sun's magnetic field be ignored,
since the sun's radius exceeds the earth's radius by five orders of magnitude.

``Storage rings'' such as the Parker ring around the sun, or the
earth's Van Allen belts, store particles with momentum proportional to the ring radius.
With the sun/earth radius ratio being $10^5$, and the momentum of a 2\,MeV electron in the Van
Allen belt around the earth being possible, the momentum of an electron or proton stored in the Parker ring
around the sun could be 200\,GeV. \footnote{In this calculation an electron has been chosen to avoid needing
to use an exact relativistic formula for the deuteron in an only semi-relativistic context.
Both particle energies are large enough for their momenta to be roughly 
proportional to their energy. In both cases referring to spherical bodies as accelerators is misleading.
Actual accelerator have apertures of, perhaps 10\,cm. Astrophysical accelerators require no beam
tubes, and have nearly unlimited (one-sided) apertures.} As a matter of fact, the essence of this paper
is that protons (and other nuclear isotopes) can acquire energy far greater than 200\,GeV, in the solar system.
See Figure~\ref{fig:rpp2019-cosmic-rays}, copied from chapter 29 of the Particle Data
Group report\cite{Particle-Data-Group-Chap-29-Cosmic-Rays}.

Jupiter could also serve as a curiously powerful particle accelerator.  Its surface magnetic field is
nearly the same as the sun, and its radius almost exactly 1/10 as great. Though the radial dependence follows
an inverse cube (1/r³) law for the magnetic field strength, it is effectively constant (weakly defocusing)
for elevations small compared to Jupiter's radius.  With injection from the solar wind, in the equatorial
direction and bending by the centripetal gravitational force, the beam of charged particles can circulate
in a belt-shaped orbit directly above the equator.

In traditional accelerator coordinates, if the $x$ (horizontal) axis is chosen to point toward the center of
Jupiter, and the $y$ (vertical) coordinate points north, then the $z$ (longitudinal) coordinate points east.
The gravitational force then points along the positive  $x$ axis, and a beam of positive charges travels in
the positive $z$ direction.  Then the ${\bf v}\times{\bf B}$ cross product of velocity and
magnetic field points toward the center of Jupiter.  In this case, the gravitational and Lorentz magnetic
forces on a (necessarily positive) nuclear isotope can be said to be ``constructive'' or,
otherwise ``destructive''.

Along with the latitudinal electric field provided by the Parker mechanism, Jupiter could serve as a suitable
injector for a cosmic particle accelerator complex.  Currently, the highest energy terrestrial accelerator
(the LHC at CERN) has a maximum proton energy of 7000\,GeV, with an average magnetic field of about 6\,Tesla.

Somewhat similar considerations apply to the relative importance of gravitational
and magnetic forces in stars.  Cosmologically natural magnetic fields, either on the earth's surface or on
the surface of the sun are conveniently quoted in Gauss units.

\subsection{Circular, gravitational, proton orbits around the sun}
The major subject of this paper concerns orbits of nuclear particles around the sun for
which the bending is due, primarily, to the sun's gravitational attraction.  For the occasional ``sanity check''
of derived formulas it is useful to establish a trusted calculation to be referred back to
occasionally.  The (presumably non-relativistic) circular proton orbit around the sun, along
the sun's equator, calculated using MKS units, provides such a check. The value for the gravitational
acceleration on earth is $g_{\rm earth} = 9.8\,{\rm m/s}^2$. The needed quantities for the sun
are $R_{sun} = 6.963 \times 10^8$ m, $g_{\rm sun}= 273.8\,{\rm m/s}^2$,
$m_{\rm sun}=1.989 \times 10^{30}$ kg, and proton mass, $m_p = 1.672\,621\,925\,95\times10^{-27}\, {\rm kg}$,
all of which should be reconfirmed, before being applied.

According to Newton's various laws one has, for the centripetal force, applied to a proton
traveling on an orbit along the equator of the sun is
\begin{equation}
F_p = m_p g_{\rm sun} = m_p \frac{v_p^2}{R_{sun}}, \hbox{\quad or\quad} v_p^2 = g_{\rm sun}R_{sun}. 
\label{eq:Newtonian-eqs.1} 
\end{equation}
The proton velocity and the proton kinetic energy, $KE$, are then given by
\begin{align}
  v_p &= \sqrt{273.8\times6.963\times10^8} = 4.3663\times10^5\, {\rm m/s}, (\hbox{\ or\ } \beta_p\approx 10^{-3}),  \notag \\
  KE_p  &=  m_p v_p^2/2 = 1.67262\times10^{-27}\times 273.8\times6.963\times10^8/2 = 1.5994\times10^{-15}\, {\rm J}.
  \label{eq:Newtonian-eqs.2}
\end{align}
To convert from Joules to GeV one multiplies by $6.242 \times 10^{9}$, with the result
%
\begin{equation}
KE_p = 1.5994\times10^{-15}\times6.242\times 10^{9} = 9.984\times10^{-6}\,{\rm GeV} \hbox{\ or\ } 10.0\,{\rm keV}, 
\label{eq:Newtonian-eqs.3} 
\end{equation}
which confirms our expectation that the result would be non-relativistic.  Furthermore, it provides an easy
to remember mnemonic.  To be categorized as a ``healthy'' cosmic ray (meaning ``effectively fully relativistic''),
such protons would need to be accelerated by six orders of magnitude, to $10\,$GeV, for example.

One can also refresh one's memory by calculating the gravitational potential on the sun's equator,
using Newton's gravitational formula $V_G = -\frac{GM}{R}$.
\begin{align}
V_{G}\Big|_{\rm sun-surface} &= -\frac{GM_{\rm sun}}{R_{\rm sun}} \hfill \notag\\  
  &= -\frac{6.674\times10^{-11}\text{m}^3/\text{kg}\cdot\text{s}^2\times1.989\times10^{30}\text{kg}}{6.963\times10^8\,\text{m}} \notag\\   
  &= -1.906\times10^{11}\,\text{m}^2\,\text{s}^{-2}\\
  &\equiv -1.906\times10^{11}\,\text{J}/\text{kg}.
\label{eq:Newtonian-eqs.5} 
\end{align}
It is important to realize that ``escape velocity'' is a purely-classical concept.  The accelerating force applied
by the solar EMF acting on cosmic ray particles has no radial component, it is almost impossible for cosmic rays to
escape from the solar system. 

\subsection{Superimposed E,M, or G,M, circular bending formalism \label{sec:SuperimposedBending}}
In a storage ring with superimposed E\&M bending,
the circulation direction of a conventionally-named
``master beam'' (of whatever charge $q_1$) is assumed to be CW or, equivalently,
momentum $p_1>0$. A ``secondary'' beam charge $q_2$ is allowed to have either 
sign, and either CW or CCW circulation direction.

A design particle has mass $m>0$ and charge $qe$, with electron charge 
$e>0$ and $q=\pm 1$ (or some other integer). These values produce circular 
motion with radius $r_0>0$, and velocity ${\bf v}=v{\bf\hat z}$, where the motion
is CW (clockwise) for $v>0$ or CCW for $v<0$. With $0<\theta<2\pi$ being 
the cylindrical particle position coordinate around the ring, the angular 
velocity is $d\theta/dt=v/r_0$. 

(In MKS units) $qeE_0$ and $qe\beta c B_0$ are commensurate forces, 
with the magnetic force relatively weakened by a factor $\beta=v/c$ 
because the magnetic Lorentz force is $qe{\bf v}\times{\bf B}$. 
By convention $e$ is the absolute value of the electron charge; where it
appears explicitly, usually as a denominator factor, its purpose in 
MKS formulas is to allow energy factors to be evaluated as electron volts (eV)
in formulas for which the MKS unit of energy is the joule. 
Newton's formula for radius $r_0$ circular motion, expressed in terms of 
momentum and velocity (rather than just velocity, in order to be relativistically valid)
can be expressed using the total force per unit charge in the form
\begin{equation}
\beta pc/e = \Big(E_0 + c\beta B_0\Big)\,qr_0,
\label{eq:CounterCirc.1} 
\end{equation}
Coming from the cross-product Lorentz magnetic force, the factor $q\beta cB_0$
is negative for backward-traveling orbits because the $\beta$ factor 
is negative.

A ``master'' or primary beam travels in the ``forward'', CW direction. 
For the secondary beam, the $\beta$ factor can have either sign.
For $q=1$ and $E_0=0$, formula~(\ref{eq:CounterCirc.1}) reduces to a standard 
accelerator physics ``cB-rho=pc/e'' formula.  For $E_0\ne 0$ the formula 
incorporates the relative ``bending effectiveness'' of $E_0/\beta$ 
compared to $cB_0$.  As well as fixing the bend radius $r_0$,
this fixes the magnitudes of the electric and magnetic bend field values 
$E_0$ and $B_0$. To begin, we assume the parameters of a frozen spin ``master'',
charge $qe$, particle beam have already been established, including the signs
of the electric and magnetic fields consistent with $\beta_1>0$ and $p_1>0$.  
In general, beams can be traveling either CW or CCW.  For a CCW beam both $p$ and 
$\beta$ have reversed signs, with the effect that the electric force is unchanged, but the 
magnetic force is reversed. The $\beta$ velocity factor can be expressed as
\begin{equation}
\beta = \frac{pc/e}{\sqrt{(pc/e)^2 + (mc^2/e)^2}}.
\label{eq:CounterCirc.2} 
\end{equation}
Eq.~(\ref{eq:CounterCirc.1}) becomes
\begin{equation}
\frac{pc}{e} = \Big(\frac{E_0\sqrt{(pc/e)^2 + (mc^2/e)^2}}{pc/e} + cB_0\Big)qr_0.
\label{eq:CounterCirc.3} 
\end{equation}
Cross-multiplying the denominator factor produces
\begin{equation}
\Big(\frac{pc}{e}\Big)^2 = qE_0r_0\sqrt{(pc/e)^2 + (mc^2/e)^2} + qcB_0r_0\frac{pc}{e}.
\label{eq:CounterCirc.4} 
\end{equation}
To simplify the formulas we make some replacements and alterations, 
starting with
\begin{equation}
pc/e \rightarrow p,
\quad\hbox{and}\quad
m c^2/e\rightarrow m, 
\label{eq:Alterations.1}
\end{equation}
The mass parameter $m$ will be replaced later
by, $m_p$, $m_d$, $m_{\rm tritium}$, $m_e$, etc., as appropriate
for the particular particle types, proton, deuteron, triton, electron, helion, etc..
These changes amount to setting $c=1$ and switching the energy units from joules to electron volts. 
The number of ring and beam parameters can be reduced by forming the combinations
\footnote{Yet another font, for $\mathfrak{E}$ and $\mathfrak{B}$, has been introduced to
represent the scaled values of squares of electric and magnetic fields, without
clashing with energies or electric or magnetic fields.  It is worth noticing that
Eq.~(\ref{eq:CounterCirc.4}) depends on $B_0$ and $E_0^2$ and $B_0^2$, but \emph{not on}
$E_0$ un-squared. This influences the variety of solutions to the quartic equation.}
\begin{equation}
\mathfrak{E} = qE_0r_0,
\quad\hbox{and}\quad
\mathfrak{B} = qcB_0r_0.
\label{eq:Alterations.2}
\end{equation}
After these changes, the closed orbit equation has become  
\begin{equation}
p_m^4 -2\mathfrak{B}p_m^3 + (\mathfrak{B}^2-\mathfrak{E}^2)p_m^2 - \mathfrak{E}^2m^2=0,
\label{eq:AbbrevFieldStrengths.3}
\end{equation}
an equation to be solved for either CW and CCW orbits.  The absence of a term linear in $p_m$
suggests the restoration, using Eq.~(\ref{eq:Alterations.2}), of the explicit form of 
$\mathfrak{B}$ in the coefficient of the $p_m^3$ term to produce;
\begin{equation}
p_m^4 - 2cB_0(qr_0)p_m^3 + (\mathfrak{B}^2-\mathfrak{E}^2)p_m^2 - \mathfrak{E}^2m^2 = 0,
\label{eq:AbbrevFieldStrengths.3-rev} 
\end{equation}
The product factor $(qr_0)$ can be altered arbitrarily without influencing any essential conclusions.
This and other properties can be confirmed by pure reasoning, based on the structure of the equation,
or by explicit partially-numerical factorization of the left hand side.

These considerations have removed some, but not all of the sign ambiguities introduced by the quadratic 
substitutions used in the derivation of Eq.~(\ref{eq:AbbrevFieldStrengths.3-rev}). 
The electric field can still be reversed without altering the set of solutions of the equation. 
Note that this change cannot be compensated by switching the sign of $q$, which also reverses the 
magnetic bending.  The most significant experimental implication is that it is not only positrons,
but also electrons, that can have orbits identical to (usually positive in ordinary practice) baryons.

We can contemplate allowing the signs of $E_0$ or $B_0$ to be reversed for observational purposes,
such as interchanging CW and CCW beams, or replacing positrons by electrons.

Fractional bending coefficients $\eta_E$ and $\eta_m$ can be defined by
\begin{equation}
\eta_E = \frac{qr_0}{pc/e}\,\frac{E_0}{\beta},\ 
\eta_M = \frac{qr_0}{pc/e}\,cB_0,
\label{eq:BendFrac.2}
\end{equation}
neither of which is necessarily positive.  These fractional
bending fractions satisfy
\begin{equation}
\eta_E + \eta_M = 1\quad\hbox{and}\quad
\frac{\eta_E}{\eta_M} = \frac{E_0/\beta}{cB_0}.
\label{eq:BendFrac.2p}
\end{equation}
The ``potency's'' of magnetic and electric bending are in the ratio $cB_0/(E_0/\beta)$ because  the electric
field is stronger than the magnetic by the factor $1/\beta$ as regards bending charge $q$ onto an orbit with 
the given radius of curvature $r_0$. The curious parenthetic arrangement of
Eq.~(\ref{eq:AbbrevFieldStrengths.3-rev}) is intended to aid in the demonstration that, when expressed in term
of spin tunes, the  ``potency's'' of magnetic and electrically induced MDM precessions are in the same ratio as
the bending potencies. 

\subsection{Co-magnetometry}
This section, which discusses spin dependence, is superfluous, in the sense that there is little or no available
data concerning cosmic ray spin dependence. It is included to advertise the importance of spin measurement
in laboratory-based storage rings, for the precise determination of resonance energy widths, 
and to make the point that spin dependence could, in principle, enter into the interpretation of cosmic ray
observations.  

For particles at rest ``co-magnetometry'' in low energy ``table-top particle traps'' has been essential; 
especially for the \emph{direct} measurement of anomalous magnetic  dipole moments (MDMs),
storage ring technology with beam
pairs that can counter-circulate simultaneously in a storage ring with superimposed electric and magnetic bending
is required. In this context the term ``mutual co-magnetometry'' can be used to apply to ``beam type pairings''
for which both beams have frozen spins.  

In an idealized electromagnetic storage ring, the
fields are ``cylindrical'' electric ${\bf E}=-E_0{\bf\hat x}r_0/r$ and, 
superimposed, uniform magnetic ${\bf B}=B_0{\bf\hat y}$.
The bend radius is $r_0>0$. Terminology is useful to specify the relative
polarities of electric and magnetic bending:
Cases in which both forces cause bending in the same sense will be called
``constructive'' or ``frugal'';  Cases in which the electric and magnetic
forces subtract will be referred to as ``destructive'' or ``extravagant''.

There is justification for the ``frugal/extravagant'' terminology. 
Electric bending is notoriously weak (compared to magnetic bending) and
iron-free (required to avoid hysteretic effects) magnetic bending is also 
notoriously weak. As a result, an otherwise-satisfactory laboratory configuration can be
too ``extravagant'' to be experimentally feasible.

For a particle with spin circulating in a (horizontal) planar magnetic storage ring,
its spin axis precesses around a vertical axis at a rate proportional to the particle's 
anomalous magnetic dipole moment, $\mathcal{G}$.  For an ``ideal Dirac particle'' (meaning $\mathcal{G}=0$) 
\emph{in a purely magnetic field} the spin precesses at the same rate as the momentum---pointing always 
forward for example.
Conventionally the spin vector's orientation is specified by the in-plane angle $\alpha$ between 
the spin vector ${\bf S}$ and the particle's momentum vector ${\bf p}$ (which is tangential, by definition). 
For such a ``not-anomalous'' particle the spin-tune $Q_M$ 
(defined to be the number of $2\pi$ spin revolutions per particle revolution) 
therefore vanishes, in spite of the fact that, in the laboratory, the spin axis has actually precessed 
by close to $2\pi$ each turn.  

In general, particles are not ideal; the directions of their spin vectors deviate 
at a rate proportional to their anomalous magnetic moments, $\mathcal{G}$, and their spin tunes differ from 
zero even in a uniform magnetic field.  Note also, that a laboratory electric field produces a magnetic 
field in the particle rest frame, so a particle in an all-electric storage ring also has, in general, a 
non-vanishing spin tune $Q_E$. Along with $\mathcal{G}$ and $Q$, all of these comments apply equally to the 
polarization vector of an entire bunch of polarized circulating particles.  

By convention, in the Bargmann, Michel, and Telegdi, BMT-formalism, the orientation of the spin vector 
${\bf S'}$ is defined and tracked in the rest frame of the circulating particle, while the 
electric and magnetic field vectors are expressed in the lab. The spin equation of motion 
with angular velocity $\pmb{\Omega}$ is 
\begin{equation}
\frac{d{\bf S'}}{dt} = {\pmb{\Omega}}\times{\bf S'},
\label{eq:BMT.1} 
\end{equation}

with orbit in the horizontal $(x,z)$ plane assumed, where
\begin{align}
{\pmb\Omega}
 &=
-\frac{q}{\gamma mc}\,
\bigg(\Big(\mathcal{G}\gamma\Big)cB_0 + \Big(\big(\mathcal{G} - \frac{1}{\gamma^2-1}\big)\gamma\beta^2\Big) \frac{E_0}{\beta}\bigg)\,{\bf\hat y} \notag\\
 &\equiv
-\frac{q}{\gamma mc}\,\bigg((Q_{M})cB_0 + (Q_E)\,E_0/\beta\bigg)\,{\bf\hat y},
\label{eq:BMT.2} 
\end{align}
This equation serves to determine the ``spin tune'', which is defined to
be the variation rate per turn of $\alpha$, as a fraction of 
$2\pi$. Spin tunes in purely electric and purely magnetic rings are given by
\begin{equation}
Q_E = \mathcal{G}\gamma - \frac{\mathcal{G}+1}{\gamma},
\quad
Q_M = \mathcal{G}\gamma,
\label{eq:BendFrac.7}
\end{equation} 
where $\gamma$ is the usual relativistic factor.
Note that the sign of $Q_M$ is the same as the sign of $\mathcal{G}$, which is positive for
protons---proton spins precess more rapidly than their momenta in magnetic fields. 
Deuteron spins, with $G$ negative, lag their momenta in 
magnetic fields.  With $\mathcal{G}$ positive, $Q_E$ increases from -1 at zero velocity, eventually switching sign
at the ``magic'' velocity where the spins in an all-electric ring are ``globally frozen'' relative 
to the beam direction.  When a particle spin
has precessed through 2$\pi$ in the rest frame it has also completed one full revolution
cycle from a laboratory point of view; so the spin-tune is a frame invariant quantity. 

In a celestial storage ring with predominantly gravitational bending, to represent a physical limitation on
the magnetic bending force at radius $r_0$, there can be a go, no-go condition such as 
$$|\eta_M/\eta_G| < 1/3. $$
The resulting magnetic force dependence on direction causes an $\eta_M>0$ (call this ``constructive'') 
or $\eta_M<0$ (``destructive'') perturbation to shift opposite direction orbit velocities (v) of the same radius, 
one up in radius and one down, resulting in two stable orbits in each direction.  For stored beams, any further 
$\Delta\eta_M \ne 0$ change causes beam velocities to ramp up in kinetic energy ($KE=\mathcal{E}-mc^2$) 
in one direction, down in the other.\footnote{In this paper there are several different types of ``energy''.
The most important of these is the total relativistic particle energy $\mathcal{E}$, with font chosen to be
calligraphic, in order to avoid clashing with $E$, the symbol for electric field.}

Depending on the sign of magnetic field $B$, 
either the lighter or the heavier particle bunches can be faster, ``lapping'' the slower bunches 
periodically, and enabling ``rear-end'' nuclear collision events. (The only longitudinal complication 
introduced by dual beam operation in laboratory storage rings is that the ``second'' beam  needs to be
injected with accurate velocity, directly into stable RF buckets.)

\section{Part 1: Solar particle acceleration; galactic particle injection?}
\subsection{``Start up'' and  ``topping up`` solar injection}
As an accelerator, the sun needs to have an injector, at least for ``start up''. Commonly, with
terrestrial accelerators, the injector is also available for occasional ``topping up``.
As conceived in this paper, the start-up capability is required, but ``topping up'' will crtainly be unnecessary.
The sun itself can replenish intermediate energy particles from its own solar wind.

There seems to be no credible direct way in which particles, say multi-MeV scale protons, can be accelerated
from the solar wind up to the GeV energy range.  However, once started up, the solar accelerator is
self-sufficient, until it is not; for example once every 11 years, while the sign of magnetic dipole component 
is reversing. Within these periods a self-sufficient cosmic ray atmosphere will have been established, which
probably has enough ``inertia'' to recover from the magnetic field reversals every 11 years.

\subsection{Arbitrary nuclear particle type~$A$ notation}
To simplify the formulas, and to emphasize that the dominant centripetal force is gravitational,
and proportional to the nuclear mass, we will use $A$, as in $A=Z+N$, as subscript for (nuclear) particle~2.
This means, to adequate precision, that $m_A=Am_p$ or that the rest mass of particle-2, is $A$, when
measured in GeV units.  To simplify the treatment of ultra high energy performance we are generalizing the
formalism to a single nuclear particle for each value of of mass index, namely $A=Z+N$. In other words,
there is a unique particle type $A$ for any integer value of $A$, in spite of the fact that particle names
retain their traditional identification with $Z$; e.g. for beryllium  Be, implies that $A=9$ and $Z=4$.

There are grounds, based on electrodynamics and general relativity, for excluding photons from the class of
``particles'' that can follow the orbits under consideration.  \footnote{This comment is based on conversation
with Saul Teukolsky. With the photon mass being exactly zero,  we are saved from the embarrassment of
establishing the  value of the $\gamma_2/m_{\rm photon}$ ratio in the vanishing mass limit.}
\footnote{With $G_n=6.67258\,m^3\,s^{-2}\,kg^{-1}$,
$M_{sun}=1.989\times10^{30}kg$, and conversion\ factor\ from\ Joule\ to\ GeV$=6.242\times10^{9}$,
the limiting radius is $r_{2-sun-lim} = 1.0631\times10^{10}\,{\rm m}$. For the six proton energies
plotted in Figure~\ref{fig:zoomed-AMS-PRL-protons-2011-2019}, the limiting radius values, in units of
$10^{10}$ meters, are
8.085,
1.973,
1.518,
1.366,
1.269, and
1.177.
}
However, electrons and positrons,
are covered, even though their small masses make them more responsive to magnetic fields.

The radius $r_A$ of a unique circular orbit around the center of the sun for a particle $A$,
with relativistic gamma-factor $\gamma_2$, a standard radius  value is defined by
\begin{equation}
  r_{\rm A-sun-lim_1} = \frac{G\,M_{\rm sun}}{2c^2} = 738.24\times\gamma_2 \,{\rm m}.
\label{eq:Newton-kinetics-revised-b-bis}
\end{equation}
This value is less than the actual radius of the sun for $\gamma_2=1$.  This result no longer depends on
the mass of (nuclear) particle~2.  Such an orbit would be internal to the sun, making it unphysical.  To find
the smallest particle energy consistent with circulation outside the sun, based entirely on the sun's
gravitational attraction, the ratio of these two results yields
\begin{equation}
  \gamma_{\rm 2-sun-lim}  = \frac{R_{sun}}{r_{\rm A-sun-lim_1}} = \frac{0.6957 \times 10^9}{738.24} = 0.943\times 10^6.
\label{eq:gamma2-sun-lim}
\end{equation}
By chance, since the rest mass of a proton is 0.938\,GeV, this means that only protons with energy
greater than $10^6$\,GeV can circulate stably in a circular gravitational orbit around the sun,
without the possible aid of some (presumably magnetic) centrifugal, centrifugal radial force pointing away
from the sun.

\subsection{Who ordered cosmic ray positrons and anti-protons?}
According to the data in Figure~\ref{fig:AMS-p-e-pos-pbar-asymptotes}, there are
(negative) anti-proton cosmic rays present in the solar system.  They are not abundant, but even
one confirmed anti-proton would needs to be explained.  Certainly, if protons can be accelerated
clockwise (CW) then anti-protons can be accelerated CCW.  The same goes for electrons and
positrons. In short, the sun would then be a colliding beam storage ring. This would be doubly
welcome, as it would help to answer the question ``Who ordered $\bar{p}$ and $e^+$ cosmic rays?''.

The relative abundances of $p$ and $\bar{p}$'s, on the one hand. and $e^-$ and $e^+$'s
on the other, have now been measured.  Along with careful calculation of the most likely QED
production channels, checking these abundance ratios would provide luminosity-independent
consistency checks of the proposed solar betatron acceleration mechanism.

By referring to Figure~\ref{fig:rpp2019-cosmic-rays}, one sees that the highest energy
cosmic ray protons just barely fail to meet $\gamma_{\rm 2-sun-lim}$ condition evaluated in
Eq.~(\ref{eq:gamma2-sun-lim}.  The data shows that all other
cosmic ray nuclear isotopes fail, but also just barely, to meet the same condition. This
statement depends on the abscissa label of the figure, which is ``Kinetic Energy Per Nucleus [GeV]''.

But it is more important that there must be a mechanism in nature capable of accelerating
nuclei to, say, $10^5$\,GeV/nucleus, in the context of the present paper.  To meet
this ``start-up'' condition one can, or must, assume that some particles with energies
in excess of $10^5$\,GeV/nucleus are entering the solar system from the solar galaxy, depending
upon the magnetic dipole field, shown, for example, in Figure{\ref{fig:sun-exterior-magnetic-field}}\,.

There is some leeway here, in that the start-up gravitational centripetal force has been neglected,
based on a radius ten times the solar radius, and the centripetal gravitational force varies
with inverse square law dependence on radial distance.  

\subsection{The Alfven-sphere}  
There is another issue influencing the sun as accelerator.  It concerns the Alfven sphere, with
radius
\begin{equation}
  {\rm R_{Alfven}} \approx 10\times {\rm R_{sun}},
\label{eq:R-Alfven}
\end{equation}
or 4\% of the distance to the earth; a characteristic radius of the magnetohydrodynamic field surrounding the sun.
See Figure~(\ref{fig:Crammmer-Solar-Wind}), copied from S.R. Cranmer, et al.\cite{Cramer-Alfven-surface},
shows features at elevations from the sun's surface out to 100 times the sun's surface, which is closely
related to the range of nuclear and electron orbit radii being discussed in the present paper.

Copying from that paper,\\
\noindent
``It is known that the strongest forces acting on the plasma undergo a
transition from being mostly magnetic, near the Sun, to mostly hydrodynamic –
i.e. depending on gas-pressure gradients and nonlinear inertial gas flow terms –
far from the Sun. A common way to quantify this transition is to locate the Alfven
surface, the place where the radially increasing solar-wind speed [u] exceeds the
radially decreasing Alfven speed [V A ].
``.
One sees from the figure that the Alfven radius (though significantly variable as
a function of time) is approximately 10 times the radius of the sun.

NASA's Parker Solar Probe (PSP) encountered the specific magnetic and particle conditions
at 18.8 solar radii that indicated that it had penetrated the Alfven surface;
Averaging these figures, one expects a value $r_{\rm Alfven}$ in the range of 10 t0
20 times $R_{\rm sun}$ for the Alfven radius.
This will become significant in the context of discussing particle injection onto the sun as accelerator,
for which the Alfven} physics will be important.  But the existence of stable circular orbits around the
sun is only significant if there is an efficient injection mechanism, and the Alfven physics will be
germane to this paper.

Understanding of the solar wind has been expanding continuously ever since its most clear elucidation
by Parker 70 years ago. The abstract to a 2023 paper by Crammer et al.\cite{Crammer-Alfven} reads as
follows\\
\noindent
``
The solar wind is the extension of the Sun’s hot and ionized corona,
and it exists in a state of continuous expansion into interplanetary space. The
radial distance at which the wind’s outflow speed exceeds the phase speed of
Alfvenic and fast-mode magneto-hydrodynamic (MHD) waves is called the Alfven
radius. In one-dimensional models, this is a singular point beyond which most
fluctuations in the plasma and magnetic field cannot propagate back down to
the Sun. In the multi-dimensional solar wind, this point can occur at different
distances along an irregularly shaped “Alfven surface....Combined with recent
perihelia of Parker Solar Probe, these studies seem to indicate that the Alfven
surface spends most of its time at heliocentric distances between about 10 and 20
solar radii. It is becoming apparent that this region of the heliosphere is
sufficiently turbulent that there often exist multiple (stochastic and time-dependent)
crossings of the Alfven surface along any radial ray. Thus, in many contexts,
it is more useful to make use of ... a complex “Alfven zone” rather than one closed
surface.\\ 
``
\\
\noindent
Figure~\ref{fig:Crammmer-Solar-Wind} displays a combination of theoretical behaviors
in the “Alfven zone”. As well as in the original caption, this figure is explained in
the new caption.
%
\begin{figure}[hbt!]
\includegraphics[scale=0.8]{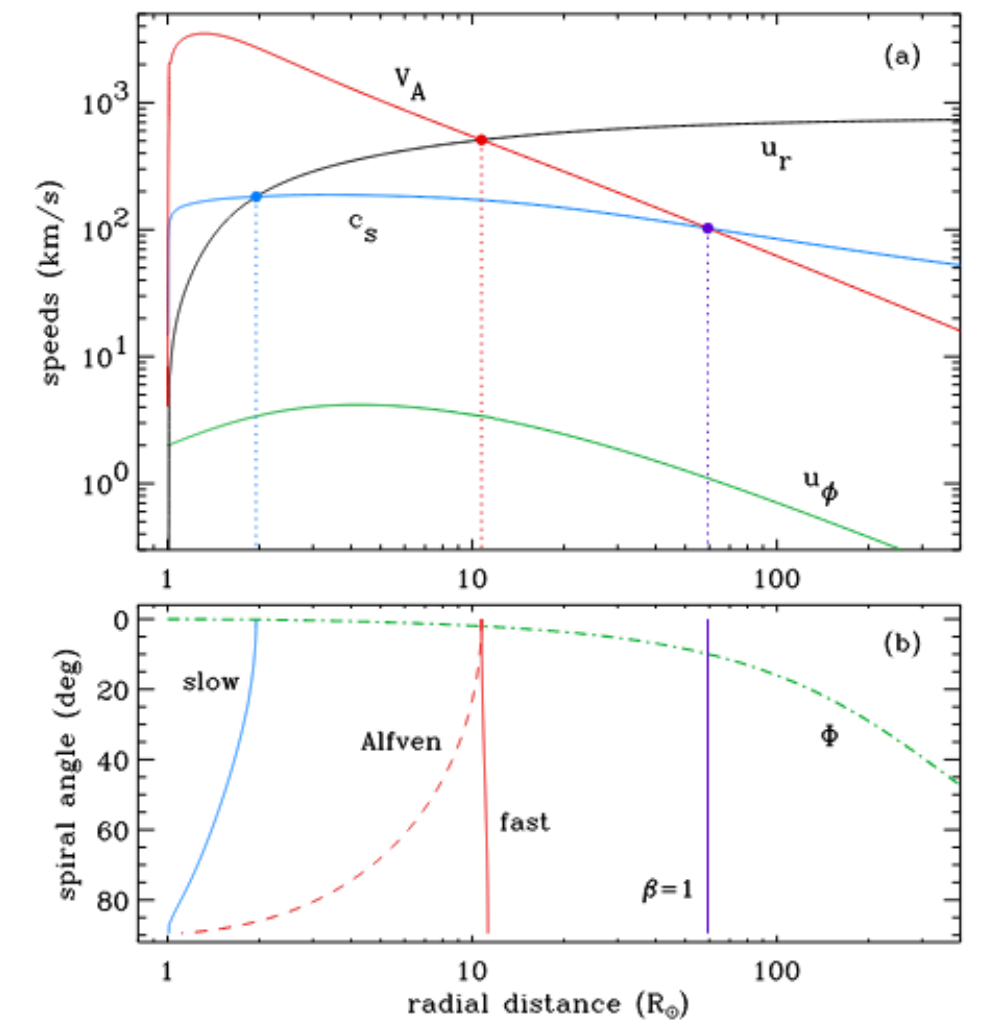}
\includegraphics[scale=0.4]{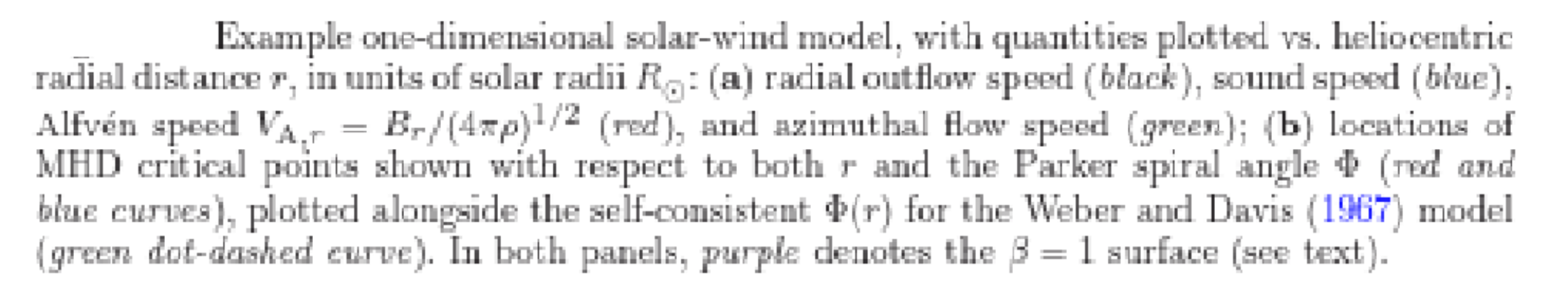}
\caption{\label{fig:Crammmer-Solar-Wind}Figure copied with original caption from reference\cite{Crammer-Alfven},
showing, in particular, the radius of the Alfven surface, located at approximately 10 times the solar radius.
Note: $\beta$ in this figure is not velocity as fraction of the speed of light.  
}
\end{figure}
%
There is hardly anything in the two captions to Figure~\ref{fig:Crammmer-Solar-Wind} that is not relevant
for the present paper.  Here we emphasize the ``(stochastic and time-dependent)'' characterization
of the region which will be referred to as the ``Alfven zone'' in the present paper.

Every solar cosmic ray has to make its way through the Alfven zone frequently and periodically from
its birth to its death.  It is predominantly these passages that can contribute stochastic noise to the
solar cosmic ray equilibrium.  Its evolution outside this region, to be described below, is governed by
deterministic Hamilton-Jacobi evolution. It remains hard to understand how all particle type energies,
for hadrons, leptons, and their anti-particles, can evolve identically in the fully-relativistic limit.
But, at least, it is clear that this behavior is governed by systematic passages through the Alfven region
of the sun.

\subsection {Magnetic bending at star and planet equators}
The magnetic dipole moments common to most stars and planets provide a magnetic field with magnetic dipole
pattern.  The magnetic field to be emphasized in this section is aligned with the rotational axis and
uniform close to the equatorial plane.  By symmetry, external to the sun, this field parallel to the axis
of rotation of the sun is sufficiently uniform for large aperture treatment.  The ``natural'' cyclotron
orbits are circles of radius greater than the solar radius, and centered on the axis in the equatorial
plane.  Any cosmic ray orbit discussed  in this paper follows a helix of increasing or decreasing radius,
slowly drifting ``out of plane'', or `` vertically'', in accelerator terminology, away from, or toward
the equatorial plane.  

The Parker solar wind provides another essential field component.  It is the latitudinal (or ``tangential''
in Frenet-Serret accelerator terminology) component of electric field, shown in the lower left corner of
Figure~\ref{fig:Parker-solar-wind}. Though small at the surface of the sun, this electric field, which is
caused by rotation of the sun's magnetic moment, becomes substantial at the surface labeled ``effective radial
magnetic field source (ERMFS)'' in the figure. This tangential electric field can serve the same role as
the RF cavities in terrestrial circular accelerators or storage rings.

When referring to a star or planet as an accelerator, these circles are slowly expanding or
contracting, nearly-closed orbits. Since free space is a fairly good vacuum, and only a quite
small number of turns expected, no vacuum chamber is required. In a linearized sense then, the
ring aperture is the infinite half-space external to this sphere; or, rather, to the nonlinear
dynamically limited portion of this half-space. Unlike any functioning terrestrial accelerator,
there is no significant focusing, other than the weak geometric focusing of closed (except for precession)
elliptical orbits.  As a consequence nonlinear bending will surely be unimportant. 

To make this kind of celestial accelerator better yet one could wish for focusing to better preserve beam
emittances; otherwise known as ``Courant invariants''. These were  named after ``Ernest'', who was the the
son of ``Richard'', co-author of ``Courant and Hilbert'', with the ``Hilbert'' of Hilbert space fame.
Their colleagues in Gottingen, Germany, include far too many to list famous physicists and mathematicians,
including Riemann in 1859. 

Helical Alfven guiding centers could supply such focusing if they lay in the equatorial plane. But
the only important Alfven guiding centers are aligned with magnetic fields perpendicular to the
equatorial plane.  Any useful net beam acceleration must, therefore, be accomplished in a fairly small
number of turns, or possibly just a fraction of one turn.

Though some protons are constantly being lost, they remain predominant, since they are constantly being
regenerated by nuclear collisions.  Note, though, that it is only low energy protons, not high energy
cosmic rays that are being regenerated.

\subsection{``Magnetic rigidity'' and ``fine tuning'' formulation}
For interpreting graphs published by International Space Station (ISS) authors it is important
to understand ``magnetic rigidity''. 

A universal nuclear mass $m_U$ can be defined as the mass of a carbon-12 nucleus,
divided by 12.  To express the mass of a nuclear isotope in the form $A m_U$ the mass of Z electrons
must have been subtracted from the atomic mass.  In accelerator physics jargon, the definition of
``magnetic rigidity'' of a nuclear isotope (Z,A) species, on a circular arc of radius $\rho$ in magnetic
field $B$ is referred to as ``B-rho'' where, in MKS units,
\begin{equation}
B\rho=\hbox{\ magnetic\ rigidity} = (A/Z)m_U\gamma\beta c/e.
\label{eq:Brho}
\end{equation}
We are mainly interested in order of magnitude precision, limiting discussion to
protons, for which (Z, N, A)=(1,0,1),and ``$\alpha$-material'', for which
(Z, N, A)=(Z, N=Z, A=2Z).\footnote{The ``$\alpha$-material'' approximation is accurate for a
few low $Z$ cases, but becomes increasingly less valid as $N$ exceeds $Z$ above helium,
carbon and oxygen and beyond.}
This terminology may seem to imply that any $\alpha$-material nucleus
can, for some purposes, be treated the same as any other, irrespective of their particular
Z-values.  This is intentionally and precisely what is being implied. 

Specific to magnetic bending; all such isotopes have the same curvature while in the same
magnetic field.  Within the Parker solar wind model, all such particle orbits have
the same curvature while in the same magnetic field.  

Also, a standard mass $m_U$ is defined for which, quoted as an energy, $m_Uc^2=1$\,GeV,
Also, the proton rest energy is 0.938\,GeV, not very different from $m_U$. 

In many papers describing results from the ISS and elsewhere, a parameter also referred to
as ``rigidity'' is defined as
\begin{equation}
R = (pc)/(Ze) = B\,r_L,
\label{eq:Alfven-proton-0}  
\end{equation}
which differs from what accelerator physicists refer to as ``magnetic rigidity''. Evidently, $\rho$
and $r_L$ are two names for the same thing.  Also, if, one refers to the ISS quantity as
``rigidity per charge'' in units of eV, as a momentum in units of GV, one obtains the
same result as an energy in units of GeV in accelerator physics terminology.

The radius of the sun is approximately 696,340 kilometers or $0.6963\times10^{9}$\,m.
For an example needed later in the present paper, one can determine the energy in GeV for
a proton traveling along a circular orbit at a (minimal) value of the Alfven radius distance
from the center of the sun given by $\rho = 0.7\times10^{10}$\,m, or approximately 10 solar radii,
for a (maximal value) magnetic field of 10 gauss, i.e. 0.001\,T.

Eq.~(\ref{eq:Brho}) and Eq.~(\ref{eq:Alfven-proton-0}), after appropriate substitution of
symbols, become
\begin{equation}
  \rho = \frac{(m_Uc\gamma_p\beta_p}{B\,e} \hbox{\quad or, for proton momentum, \quad}
  \frac{p_pc}{e} = B\rho.
\label{eq:Alfven-proton}
\end{equation}
One sees, from the units, that what is actually being determined by ``B-rho'' is the
particle momentum, in this case $p_p$, for the proton, expressed in convenient units,
in a formula that is valid both relativistically and classically. This is the reason why the
abscissa horizontal label of ``energy dependence'' in ISS is correctly expressed in GV units,
over the full range of energies.  In the fully relativistic regime GV and GeV can be used
interchangeably, for a particle having charge equal in magnitude to the charge of an electron.

The magnetic field at the Alfven radius varies significantly, ranging from 1\,gauss to 10\,gauss,
correlated with the 22\, year periodicity of the sinusoidal solar magnetic
field.  Neglecting the gravitational force of the sun on the proton, (which is substantially
smaller) we have calculated the proton energy for a proton circulating around the sun,
at a radius ten or twenty times the radius of the sun to be in a range from 1 to 10\,GeV.

It is only because the start-up bending is magnetic, that positive and negative particles can
counter-circulate in the same ring.  Later, both beams have been sufficiently accelerated for
the magnetic bending to ``saturate'', which enables the bending to be gravity-dominated, without
losing either of the counter-circulating beams.  If true, this is a remarkable instance
of ``fine-tuning'' in nature.

The data in Figure~\ref{fig:rpp2019-cosmic-rays} shows ISS-measured cosmic rays for energies
per nucleus from $10^1$ to more than $10^5$\,GeV/nucleus.  The gravitational force of the
sun is proportional to $1/rho^2$ and the centripetal force needed to hold high momentum
particles in circular orbit is  proportional to $1/rho$. The radius of the solar system
is not less than $10^5$\,AU, which is six orders of magnitude greater than the Alfven radius
of the sun.  It seems, therefore, that cosmic rays of energy as high as $10^{11}$\,GeV
can be held captive in the solar system.

\subsection{${\rm G}\&{\rm M}$ solar bending and electric acceleration}
The sun's magnetic dipole moment field magnetic field, alone, given by Eq.(\ref{eq:mag-dipole}),
and normal to the sun's equatorial plane, is capable of bending low energy nuclear isotopes and
electron's into counter-circulating orbits above the equator of the sun.  During an ``injection phase''
the kinetic energies of all of these particles  are being accelerated by the ``longitudinal'' electric
field component shown, for example in Figure~\ref{fig:Parker-solar-wind}.

The actual situation is further complicated by magnetic forces, which ``saturate'' as the
speed approaches the speed of light. While still non-relativistic, the magnetic deflection strengths
track proportionally with increasing momentum.  Once a particle's speed has become essentially equal
to the speed of light, it is only the gravitational force that can provide the bending force required
for further increase in momentum.  One has to anticipate an intermediate situation during which the magnetic
bending has become negligible.  This might be referred to as the end of the injection phase.

Later, for cosmic ray particles remote from the sun, the magnetic bending may eventually becomes significant
again.  Though weak, this magnetic bending will ``corrupt'' the otherwise predictable relativistic Keplerian
cosmic ray particle trajectories.  To the extent these weak magnetic fields are known, they can be
treated as perturbations of ideal Hamilton-Jacobi-Kepler orbits.

\begin{figure}[hbt!]
\centering
\includegraphics[scale=0.50]{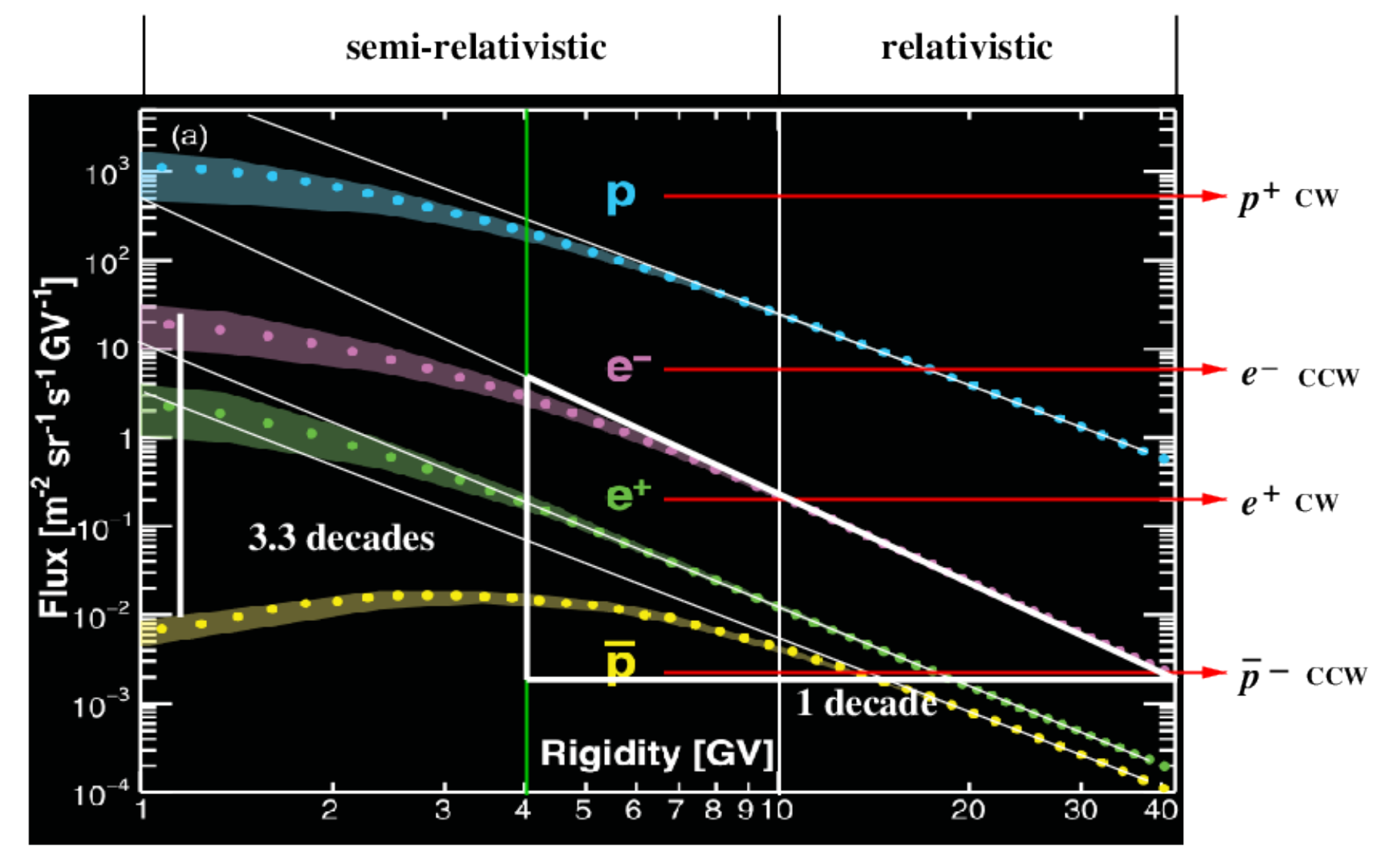}
\caption{\label{fig:AMS-p-e-pos-pbar-asymptotes}Figures copied from
reference\cite{Agular-AMS-p-e-pos-pbar}, showing rigidity dependence
(also known as momentum over charge dependence) and relative abundances of protons, electrons, positrons,
and anti-protons measured by the AMS detector in the ISS, interpreted as four separated beams of protons,
electrons and their anti-particle being accelerated by the sun, with superimposed gravitational and
magnetic bending, and Parker electric field acceleration. For the asymptotic extrapolation, 
$ {\rm Flux[db]} = -10\ {\rm log} 10^{3.3} = -33 {\rm\ db/dec}.$  The protons have become fully
relativistic only above 10\,GeV, while the electrons have become fully relativistic midway through
the region labeled ``semi-relativistic.
}
\end{figure}

Fortunately, there is splendid experimental ISS-AMS data bearing on the relative importance of
magnetic and gravitational bending of cosmic rays.  Shown in
Figure~\ref{fig:AMS-p-e-pos-pbar-asymptotes}, are energy spectra, expressed as rigidities in
GV units, for protons, electrons, positrons and anti-protons.
\footnote{GV units are appropriate for momentum $pc$, or rather for ``momentum by charge'', where
it is convenient to describe the charge as ${\rm Z}$, in units of the proton charge ${\rm e}$,
with the numerical value of $e$ being 1.}
For the superposition of magnetic and gravitational bending we need to derive the analogs
of Eqs.~(\ref{eq:BendFrac.2}) which apply to the superposition of magnetic and electric bending.
A gravitational/magnetic Lorentz force law analog is,
\begin{align}
{\bf F}_{GM}(r) &= -(Gm_1)\frac{\gamma_2m_2}{r^2}\hat{\bf r} + qc\hat{\pmb\beta}\hat{\bf z}\times B(r)\hat{\bf y}\\
            &= \Big((Gm_1)\frac{\gamma_2m_2}{r^2} + q\beta cB(r)\Big)\ (-{\hat\bf r}),
\label{eq:F-GM.1}
\end{align}
where ${\bf\hat x}={\bf\hat r}$, ${\bf\hat y}={\bf\hat B}$ and ${\bf\hat z}={\bf\hat v}$,
form a right-handed triplet of mutually perpendicular unit vectors.\footnote{As elsewhere in this paper,
the factor $c$ in Eq.~(\ref{eq:F-GM.1}) is superfluous, since its
value is 1; its presence serves as reminder concerning the velocity dependence of the Lorentz force law.}
In the gravitational term the product $(Gm_1)$ has been combined; the reason for this is that, while ${\rm G}$
itself, is poorly known, the product of the Sun's mass $M_{sun}$ and the gravitational constant G
is known with value known to 10 decimal places, which, for our
purposes can be expressed adequately as
$$ ({\rm GM_{sun}}) \approx 1.3271244 \times 10^{+20}\,{\rm m}^3{\rm s}^{-2}.$$
This product is known as the ``Sun's standard gravitational parameter''; it has a universal symbol
which will not be needed, and is therefore not shown.  Note that, with these physical dimensions,
division by a velocity squared, such as $c^2$, produces a length in meters.

As already stated more than once, particle~1 is heavy, particle~2 light.  Note also that $m_2c^2$ is
the rest energy in GeV.

We wish to mimic the $\eta_E$ and  $\eta_M$ partitioning defined in Eq.~(\ref{eq:BendFrac.2}).
Unfortunately, in the ${\rm G}\&{\rm M}$ case we are not at liberty to choose the magnitudes
of either the gravitational or magnetic force, nor the sign of the magnetic field; they are
provided by nature. 

We therefore define $\eta^{**}_G$ and $\eta^{**}_B$, where overhead $^{**}$ or $^{*}$ symbols
provide warning that the fractions do not sum to 1.  Bending fractions are then
defined by.
\begin{equation}
\eta^{**}_G = (Gm_1)\frac{\gamma_2m_2}{r^2}, \quad \eta^{**}_M = q\beta cB(r).
\label{eq:BendFrac.GM}
\end{equation}
Unlike a laboratory storage ring, which has a fixed bending radius, an astronomical
accelerator has no such constrained radius.  However, like a laboratory-based ${\rm E}\&{\rm M}$
ring, beams of a given type, say protons, can rotate either CW or CCW.  If there are also
anti-protons, i.e. $\bar{p}$-type, the gravitational force will be the same for both beam types,
but the magnetic forces will be opposite, centripetal for one type, centrigugal for the
other.

\emph{A priori}, we had no reason to expect to see $\bar{p}$'s at all.  Yet the AMS detector
results in Figure~\ref{fig:AMS-p-e-pos-pbar-asymptotes} show, roughly, one $\bar p$ for every $10^5$
protons. Clearly the $p$'s and $\bar p$'s are traveling stably in opposite directions.
This means that, for nuclear particles, the gravitational bending is predominant in both cases.

 We also had little \emph{a priori} reason to expect to see electrons, let alone positrons.
 The AMS detector shows, roughly, one positron for every 50 electrons, and
 one anti-proton for every ten thousand protons, though these ratios only stabilize in the
 fully relativistic regime.

Clearly the electron's and protons need to be streaming initially in opposite directions
This means that, even for light (i.e. lepton) charged particles, the magnetic bending is more
important for electrons than for protons. 

In passing one notes that electrons and protons traveling in opposite directions around the sun
turn the sun into a ``colliding beam storage ring''.  This, no doubt, is how positron's and
anti-protons enter the picture. Since this physics will be trivial once the cosmic ray generation
is understood semi-quantitatively, this mechanism is not discussed further in the present paper. 

In short, Figure~\ref{fig:AMS-p-e-pos-pbar-asymptotes}, suggests that all charged particles are
being accelerated, presumably while temporarily ``captured'' in a gravitational storage
ring.  Furthermore, there is ample data provided by the same figure to establish the relative
strength of the magnetic and gravitational bending forces more quantitatively.  I, personally,
can see no reason to doubt that the acceleration in question is provided by the external fields,
gravitational and magnetic, of the sun. For nomenclature one can express the sun as a
${\rm G}\&{\rm m}$ ring, meaning that the bending is predominantly gravitational.

\section{Correlation: cosmic ray intensity with solar oscillation phase}
\subsection{Magnetic field imposed complications}
The primary importance of magnetic fields in the solar system is not inside the sun or planets;
it is in the free space external to these bodies.  There is a strong tendency
for massive atoms to fall back into the sun or one of the planets. In equilibrium, it is predominantly
protons, electrons, and alpha material (predominantly $\alpha$-particles and deuterons) that continue
to circulate freely and longer, in more or less isotropic directions.  Meanwhile electrons and nuclear
particles will have merged into atoms that are captured in the atmospheres of planets.

It is predominantly higher momentum protons, deuterons and $\alpha$-particles that continue to circulate
freely. In equilibrium these particles will have gravitated toward the most massive objects in
the solar system, namely either the sun or Jupiter.

Quoting from reference~ \cite{Agle-Juno}, ``NASA’s Juno mission to Jupiter made the first definitive
detection beyond our world of an internal magnetic field that changes over time, a phenomenon called
secular variation. Juno determined the gas giant’s secular variation is most likely driven by the
planet’s deep atmospheric winds.''\cite{Agle-Juno}  Also `` The field rotates with the approximately
9 hour rotational period of the planet.''.

When produced in standard terrestrial alpha radioactive decay, alpha particles generally have a kinetic
energy of about 5 MeV and a velocity in the vicinity of $0.04\,c$. At this speed the time of flight of an
alpha particle from Jupiter to the sun is approximately 0.0007/0.04 = 0.02 years, or about 7 days.  At higher
energy or lower mass this time would, of course, be less.

By way of contrast, with the diameter of the solar system being about one light year, a fully-relativistic
particle traveling to the sun and back (presumably following a highly eccentric elliptical orbit) would
have taken a year or so to get to the sun and back. This means, for example, that the sun's magnetic fields
may have reversed, with ten percent probability, during the longest possible duration of a solar-based
cosmic ray particle.  For purposes of this paper, this is taken to mean, even though the cosmic ray
intensities are correlated with the sun's orientation reversal, that the cosmic ray acceleration process
does not need to be restarted every 11 years. Though important cosmic ray features may vary in phase with
the sun's magnetic polarity, the cosmic ray ``atmosphere'' and the solar wind are less in lock step
with each other than is true for a laboratory accelerator being refilled from its injector. 

Nevertheless, there remains a strong correlation between cosmic ray intensity with the phase of
the sun's 22 year periodic variation. \emph{This correlation is plausibly explained in the ``standard model'' of
extra-galactic cosmic ray  production, by the deflection of remotely generated cosmic rays being caused
by solar magnetic fields they encounter, after they have entered the solar system.}

Understanding this correlation is also challenging for a theory assuming that cosmic rays are purely
of solar origin. It is hard to understand the mechanism which keeps the cosmic ray ``atmosphere''
synchronized with the solar wind.  Without such synchronism it is hard to understand how
cosmic ray flux intensities can be steadily replenished.

One is tempted to assume that no such synchronism exists. This would mean, however that partially accelerated
cosmic rays would be as likely to return to the sun ``out of phase'', meaning they would be decelerated.

\emph{The problems are of very different character, however.}
\emph{Nevertheless. the presence of substantial magnetic fields in the free space within the
solar system seriously disrupts any simple, periodic, closed form theory of solar cosmic ray production.}

All this makes it important to learn what is known about the correlation between cosmic ray intensity
correlation with phase of the sun's 22 year oscillation period.
The ``open solar flux'' (OSF) represents the variable heliospheric magnetic flux. The ``galactic cosmic
ray'' (GCR) measure represents cosmic ray intensity near Earth.  Quoting Koldobsky\cite{Koldobskiy}, ``The
flux of \emph{galactic} cosmic rays (GCRs) outside the heliosphere is generally assumed to be constant at 
time scales shorter than a hundred thousand years.'' as well as ``The GCR flux variability is known to be
delayed with re-respect to solar activity, leading to the so-called “hysteresis” effect of the phase shifts
in the development of the 11-year solar cycle in both indices.  Here the \emph{galactic} qualifier has been
emphasized to acknowledge that correlation between cosmic ray intensity variation presents serious problems
for both solar-based and galactic-based cosmic ray generation.

Again quoting Koldobsky\cite{Koldobskiy}, ``All of these three indices appear highly coherent at a timescale
longer than a few years with persistent high coherence at the timescale of the 11-year solar cycle. The GCR
variability is delayed with respect to the inverted SSN (solar sunspot number) by about eight 27-day Bartels
[a measure of solar activity rotation cycles] is 1/2 of a year on average, but the delay varies greatly with
the 22-year cycle, being shorter or longer around positive A+ or negative A-.''

\begin{figure}[hbt!]
\centering
\includegraphics[scale=0.47]{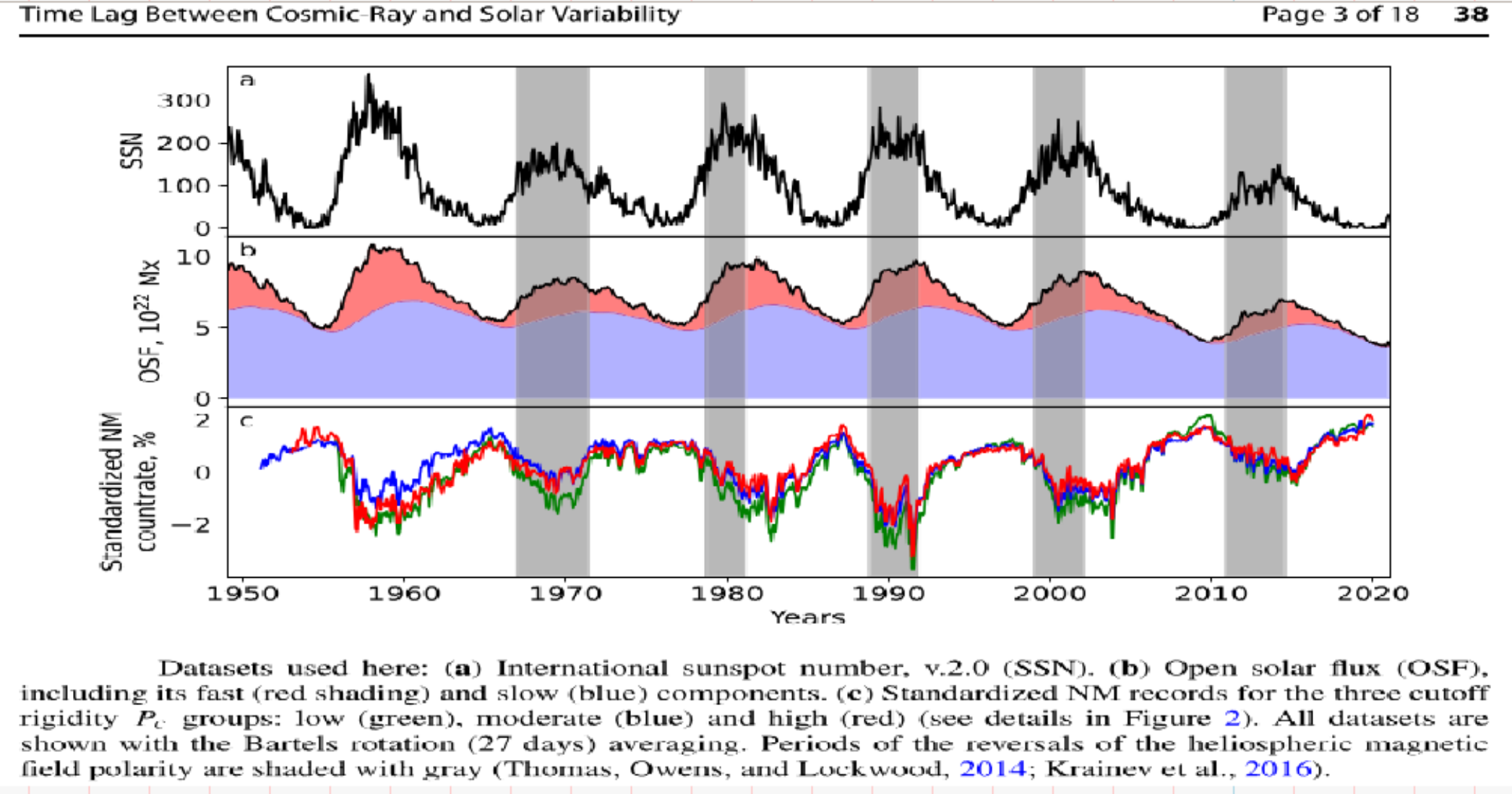}
\caption{\label{fig:Cosmic-ray-solar-cycle}Time Lag comparisons, 1950-2020, of sunspot number (SSN),
open solar flux (OSF) and cosmic ray measure (NM). {\bf (a)} Notice that, unlike the upper two solar plots,
the cosmic ray measure remains within a $\pm 2$-percent range. As regards signs, the upper two solar plots
remain faithfully in phase. So also do the cosmic ray deviations. But maximum solar flux coincides with reduced
cosmic ray rates and minimum solar flux coincides with increased cosmic ray rates. {\bf (b)} But notice also
from the top two figures, that the OSF is a superposition of an AC signal of significantly variable amplitude,
or different frequency component, and a DC signal. {\bf (c)} In a DC sense, it might be said that, on the
average, the OSF and the cosmic ray flux are proportional.  {\bf (d)} Though opposite in phase, so also are
the SSN's and the cosmic ray deviations.}
\end{figure}

Time Lag measures of SSN, OSF, and NM from 1950 to 2020 are platted in Figure~\ref{fig:Cosmic-ray-solar-cycle},
along with the original caption. Frequency compositions are commented upon in the new caption.
When produced in standard terrestrial alpha radioactive decay, alpha particles generally have a kinetic
energy of about 5 MeV and a velocity in the vicinity of $0.04\,c$. At this speed the time of flight of an
alpha particle from Jupiter to the sun is approximately 0.0007/0.04 = 0.02 years, or about 7 days.  At higher
energy or lower mass this time would, of course, be less.

The ``open solar flux'' (OSF) represents the variable heliospheric magnetic flux. The ``galactic cosmic
ray'' (GCR) measure represents cosmic ray intensity near Earth.  Quoting Koldobsky\cite{Koldobskiy}, ``The
flux of \emph{galactic} cosmic rays (GCRs) outside the heliosphere is generally assumed to be constant at the
time scales shorter than a hundred thousand years.'' as well as ``The GCR flux variability is known to be
delayed with re-respect to solar activity, leading to the so-called “hysteresis” effect of the phase shifts
in the development of the 11-year solar cycle in both indices. Here the \emph{galactic} qualifier has been
emphasized, if only to stress that the present paper is concerned only with solar cosmic rays.

The long-term variability of GCRs is routinely monitored by a global network of ground-based neutron monitors
(NMs) that are sensitive to the nucleic component of the cosmic-ray-induced atmospheric cascades and located,
in worldwide network around the globe since the 1950s.
\begin{figure}[hbt!]
\centering
\includegraphics[scale=0.40]{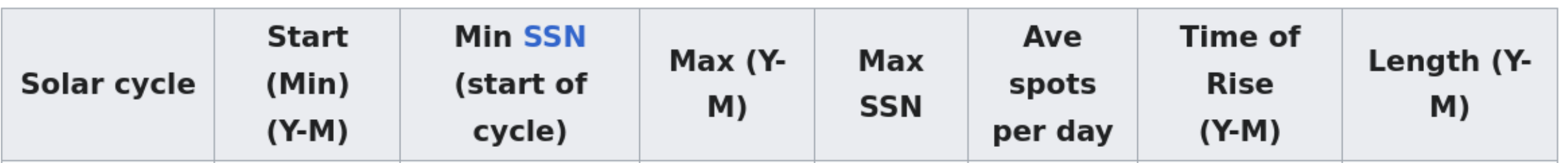}
\includegraphics[scale=0.41]{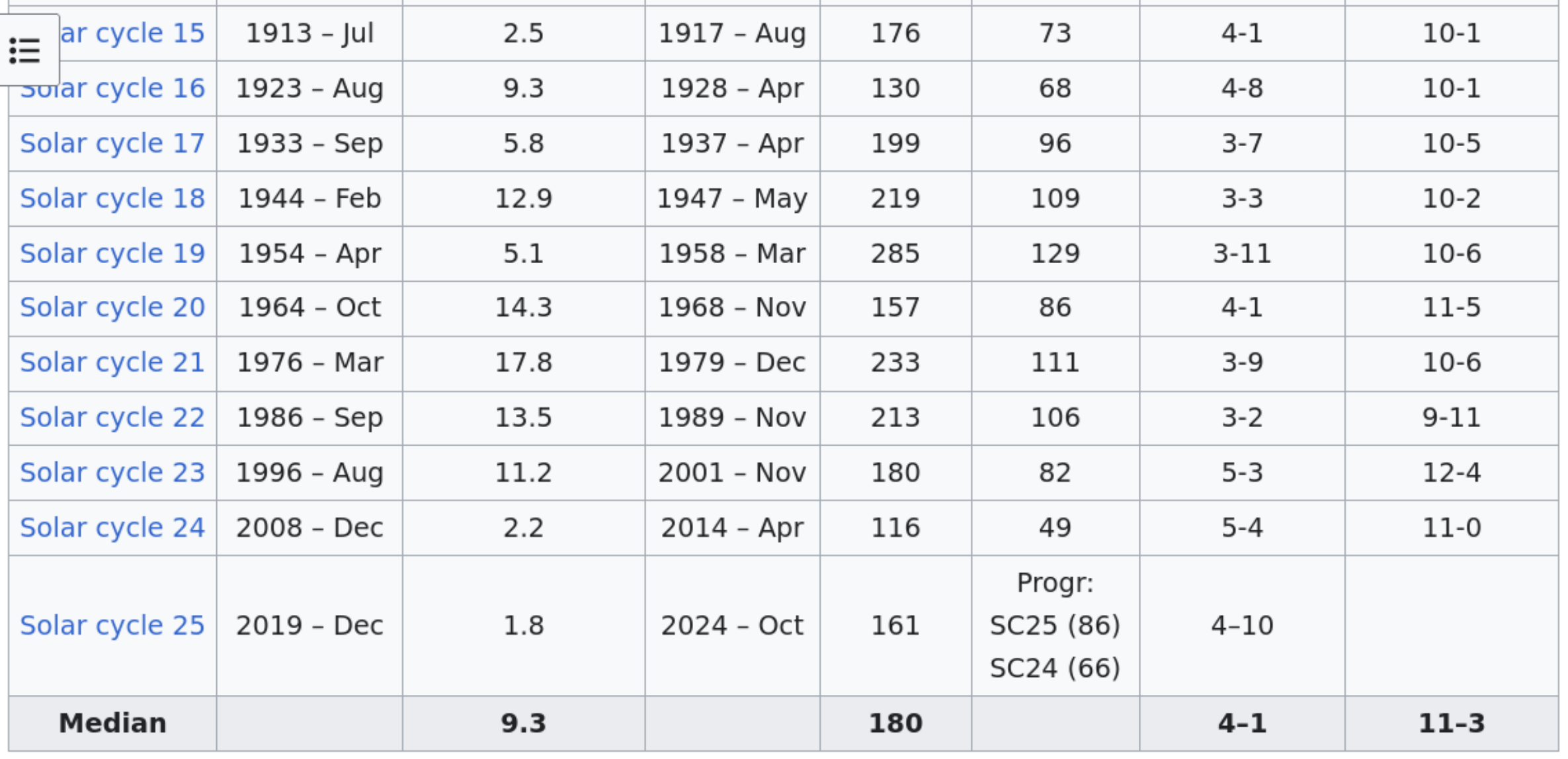}
\caption{\label{fig:Sunspot-Number}This data, which describes tne ``recent'' variation of the solar magnetic
reversal cycle is alluded to in this paper's abstract. Though basically not understood, a conjecture of the
present paper is that a significant fraction of solar cosmic rays are actally produced by Jupiter, which satisfies
the same betatron coditions as the sun. This would ameliorate what would otherwise be an embarrassingly too great
cosmic ray time dependence synchronized with the suns's magnetic field reversals.
}
\end{figure}

\subsection{Systematic particle acceleration}
Figure~\ref{fig:sun-exterior-magnetic-field}, shows a toroidal region surrounding the sun with transverse
area greater than  the sun itself, in fact semi-infinite, shaped much like the magnetic field of a ground
based particle accelerator such as the LHC.  Even though, eventually, the main centripetal deflecting force is
gravitational, provided by the mass of the sun, there is a vertical magnetic field, not unlike the field of a
weak focusing storage ring, though decreasing as $1/r^3$ radially

As shown in Figure~\ref{fig:Parker-solar-wind}
there is also a ``constant'' (modulo fluctuations and 22 year period sinusoidal variation) ``longitudinal''
electric accelerating field.  Charged particles with velocities directed more or less parallel to the electric field,
will be accelerated systematically.  In this sense the sun, as an accelerator, serves as its own injector.

Though much like ground-based storage rings, there are, of course, many features distinguishing
the Sun-SR from ordinary storage rings.  \emph{The most important of these differences is that, unlike a ground-based
accelerator, the bending field does not ramp up to match the increasing momenta of the candidate cosmic ray
charged particles.} Less important, the acceleration is being treated as  ``DC'' rather than ``AC-RF''. In fact, 
discussed in Appendix~\ref{sec:Kulsrud} can be a nonlinear plasma process capable of bunching the cosmic ray
particles, much as in a laboratory Alvarez linear accelerator, while their energies are low.
To begin with we need only discuss the acceleration to high energy of single particles being accelerated
without bunching.

Our model, therefore, initially treats each isotope particle individually and incoherently. Once the
acceleration of a single particle has been understood, it is not difficult to understand how beam bunching
acquired at low energy can be retained all the way up to high energies.  The same comment
applies to all charged particle types, including leptons and anti-particles.

It is not immediately obvious how anti-particles come into the picture.  Certainly there are neither positrons
nor anti-protons initially emerging from the sun in the solar wind.  Their source should not be hard to understand
once counter-streaming high energy electrons and proton bunched beams have bean understood, but this discussion
can be deferred until later.

Concentrating initially on a single particle, say a 100\,MeV proton, and plausibly describing its acceleration
to, say $10^{10}$\,GeV, it is easy to envisage a large number of such protons being bunched into an
arbitrarily long continuous trains of charged particle bunches.  It is also easy to visualize such trains,
like Amtrak trains, behaving sporadically.  This requires the start-up procedure to be repeated frequently
and more or less reliably.  The duration of even the most successful train could not be longer than 11 years,
which is the longest time the sign of the sun's fields remain the same.  But this would be a ludicrously
too long over-estimate, based on sporadic solar flares, electric storms, and all.

Each individual, initially 100\,MeV, proton once accelerated enough to first escape the sun, follows a
relativistically valid, highly elongated precessing Kepler orbit which, eventually, but inexorably, brings
it back periodically through the sun's atmosphere, presumably in a zone for which $R_{\rm Alfven}$ is the
typical radius.  Here it interacts strongly with the sun's solar wind, possibly to an extent that the proton
is continues onto an orbit much the same as before, except for having increased energy. See, for example,
Figure~\ref{fig:Munoz-Pavic2}.

While completing spiral turns around the sun, at radius approximately equal to the Alfven radius,
these semi-circular arcs match the highly-elongated, highly-eccentric, precessing, ellipsoidal relativistic orbits
having one focus at the center of the sun. Each such orbit includes more-or-less semi-circular orbit injection
and extraction segments.  Every other incipient cosmic ray orbit is similar. And the perihedral radii of the 
precessing ellipses are close to the so-called ``Alfven radius $R_{\rm Alfven}$.  All this is illustrated in figures
 \ref{fig:Munoz-Pavic2} and \ref{fig:Theoretical-analysis-of-AMS-data}. 

These assumption reduce the problem to mechanics that can be treated by the Hamilton-Jacobi model, as
described in Appendix~{\ref{sec:Hamilton-Jacobi}}.  During their lengthy periods remote from the sun, orbit
wandering due to unknown magnetic fields needs to be treated as stochastic perturbations; for example as
described in reference`\cite{Talman-Mechanics-Chapter-16}.

Our Hamilton-Jacobi model of planetary orbits resembles quite closely the Sommerfeld relativistic refinement
of the 1920, pre-quantum mechanics, Bohr theory of atomic structure.
It was Jacobi who developed the method of separation of variables in Hamilton's partial differential
equation (PDE) to perform a ``canonical transformation'' to produce momentum variables that are, themselves,
constants of the motion.  As well as satisfying special relativity, this is the same mathematics
as has been used ever since 1925, to solve the Schr\"odinger equation in quantum mechanics (QM). All that
was missing from modern QM at Bohr-Sommerfeld time was the introduction of non-commuting variables and
Heisenberg uncertainty.

Anyway, my proposed model of cosmic ray orbits of elementary particles around the sun is much the
same as Sommerfeld's model of electrons orbiting a nuclear atom.  The first ``constant momentum
component'' in the Jacobi treatment of orbits is the particle energy $\mathcal{E}$.
Bohr's quantized atomic energy energy level model began with his requirement for energy levels
to correspond to classical adiabatic invariants.  Our application is much the same, even though
the physical application is very different.  \footnote{Bohr's application was consistent with his requirement
for the frequencies of radiated photons to correspond to the differences of energies between
two energy levels, rather than being harmonics of a single energy level.} 

For our purposes the adiabatic invariance of the orbital energy of a single isotope, such as a proton,
is an ``orbit element'' in Lagrangian terminology. This means that its value is decoupled from the
three other canonical momenta, which therefore remain constant while the particle energy
increases, during its periodic transit through the Alfven zone of the sun.  Averaged over days or
months an equilibrium solar atmosphere of cosmic rays develops.

The duration of a ``long accelerator run'' in the Sun-SR cannot exceed 11 years; any established
``beam'' will not be able to survive this regular reversal of the sun's magnetic field (and with
it the reversal of acceleration directions).  However the decay lifetime of the solar atmosphere
is presumed to be long compared to 11 years.  This is consistent with the time spent during acceleration,
being long compared to time spent in a more-or-less stable atmosphere of cosmic rays, many of which
had  previously been accelerated to quite high energy, though never high enough to escape the
solar-system.

As with land-based storage rings, the details of solar storage ring start-up are bound to be quite
messy.  For example, start-up may be synchronized with unpredictable sun-spot activity.  Let us,
therefore, skip to a conjectural low energy start-up condition, with specified proton beam of definite
energy and current, circulating stably while passing periodically through the Alfven zone
having radius about ten times the radius of the sun.

It is only the ``longitudinal'' velocity component of circulating cosmic ray particles that is
accelerated.  Large transverse velocity components will typically cause the fractional acceleration
of such particles to be quite limited, especially for large vertical ``out-of-plane'' velocity
component particles.  Furthermore, the acceleration will not necessarily always be positive. But
this depends on the beam bunching which will favor positive acceleration but, for now, is being
neglected.

There will, however, be a substantial difference of injection efficiencies.  Particles approach the sun
with equal probability of missing to the left or to the right.  But the accelerating efficiency
will be half and half, depending on their sign.  The centrifugal bending to establish this condition
is produced by the ``destructive'' superposition of gravitational and magnetic bending.  Being magnetic,
this will be effective one way for positive charges, the other for the other sign.

While the velocities remain non-relativistic, the radial magnetic force will scale up proportionally
with the particle momentum, supporting the orbit radius to be greater than the radius of the sun.

\section{ Part 2: The sun as betatron cosmic ray factory}
\subsection{Balancing magnetic and gravitational force expressions}
In SI units, with $B_{\rm sun}(r)$ being the longitudinal magnetic field component at radius $r$ above
the sun's equator,
\begin{equation}
F_E=qeE_0, F_M=qe\beta c B_{\rm sun}(r) \hbox{\ and\ } F_{G,p} = (Gm_1)\frac{m_pc^2}{r^2}
\label{eq:force-expressions}
\end{equation}
represent electric, magnetic, and gravitational forces.

It has been highly advantageous for the electric and magnetic forces to be ``commensurate'', as this has
enabled the ``bending fraction'' formalism.  For the sun as ``circular accelerator'', we now wish to
exchange the electric bending fraction $\eta_E$ with a gravitational bending fraction, $\eta_G$.  The
fact that both electrical and gravitational radial dependencies are ``inverse square law'', makes this
very natural, but it requires the gravitational law expression to closely match the other two laws.

A natural way to do this is to limit the generality of the Newton law to be specific to the sun,
which (not counting Jupiter) is
the only celestial accelerator we are (currently) trying to describe.  As an accelerator, the
sun has a  ``characteristic'' bending radius
$$r_0 = R_{A-sun-lim} = 6.6 \times 10^{9}\,{\rm m} \qquad\qquad\qquad\qquad\qquad\qquad \inlineeqno, $$
which is roughly ten times the sun's radius.  We therefore rearrange Eqs.~(\ref{eq:force-expressions}) to become
$$F_E=qeE_0, F_M=qe\beta c B_{\rm sun}(R_{A-sun-lim}) \hbox{\ and\ } F_{G,{\rm sun}} = \big(Gm_{\rm sun}\big)\frac{m_A}{R^2_{A-sun-lim}}.
\qquad\qquad  \inlineeqno $$
In this form the gravitational notation is quite general, since, as well as including all nuclear particles,
``A'' includes all nuclear anti-particles, as well as electrons and positrons.  In other words, any non-zero mass,
stable, fundamental particle is included.  In practice, this gravitational force law excludes massive objects such
as meteorites or planets, which can never acquire relativistic velocity.

In order to refine the discussion of the electron, proton, positron, anti-positron cosmic ray distributions
shown in Figure~(\ref{fig:AMS-p-e-pos-pbar-asymptotes}) it is convenient to rearrange the ratio of
bending fractions,
\begin{align}
\frac{\eta_M^*}{\eta_G^*} &=  q_2e\beta_2 cB_{\rm sun}(R_{\rm A-sun-lim})\,\frac{R^2_{\rm A-sun-lim}}{GM_{\rm sun}m_2} \notag\\
   &=  q_2\beta_2 \Big(\frac{R^2_{\rm A-sun-lim}}{GM_{\rm sun}}\Big)\frac{e}{m_2}\,cB_{\rm sun}(R_{\rm A-sun-lim}), 
\label{eq:fractional-bending-ratios}
\end{align}
where the overall sign of the ratio of forces is determined by the initial $q_2\beta_2$ and final $cB_{\rm sun}$
products, for which the sign depends on the sign of the sun's magnetic field and on the signs and rotation directions
of the four particles under discussion. See the annotations on Figure~(\ref{fig:AMS-p-e-pos-pbar-asymptotes}.
Notice also that the fractional bending ratio for electrons and protons only become the same for fully relativistic
motion, as $\beta_2\rightarrow 1$.  From the figure it is clear that the magnetic and gravitational forces are
constructive in all cases. It is (presumably) also true, that the longitudinal Parker electric field causes
positive acceleration in all cases.

The only unknown quantity is the Sun's surface equatorial magnetic field $B_{\rm sun}$, which needs to be established
empirically, either by alternate measurement method or by curve fitting based on the present theory.
This bending has become negligible in the relativistic limit, but one faces the complication that the electron orbits
at low energy depend on radial position $r = R_{\rm sun} + \delta r$, which is shrinking rapidly in the low energy and
semi-relativistic phases. Recall, however, that close to the sun, to first order in $\delta r/R_{\rm sun}$, the radial
magnetic and gravitational force components fall off proportionally.  This reduces the sensitivity of the radial
magnetic to the gravitational bending ratio and reduces the need for explicit knowledge of the radial variation of
$cB(r)$. 

\subsection{Complication due to superimposed magnetic bending}
What remains to be done is to explain the curves in Figure~\ref{fig:AMS-p-e-pos-pbar-asymptotes}, in order to
estimate the actual external gravitational and magnetic fields of the sun by matching the observed centripetal
and longitudinal accelerations. Toward this end, we return to the task of refining the gravitational and magnetic
bending fractions; 
\begin{align}
\eta^*_G &\approx \Big(\frac{GM_{\rm sun}}{R^2_{\rm sun}}\Big)m_2\Big(1-2\frac{\delta r}{R_{\rm sun}}\Big),
\label{eq:BendFrac.etaG}\\
\eta^*_M &\approx q_2\beta_2 cB_{\rm sun}R_{\rm sun}\Big(1-2\frac{\delta r}{R_{\rm sun}}\Big),
\label{eq:BendFrac.etaG}
\end{align}
where a reference radius $R_{\rm sun}$, the radius of the sun, and the magnetic field 
at the surface is $B_{\rm sun}$, deviation $\delta r$ has been
introduced and the radial position $r$ is expressed as
           $$r=R_{\rm sun}+\delta r = R_{\rm sun}\Big(1 + \frac{\delta r}{R_{\rm sun}}\Big),$$
where only the leading power has been retained.

Expressed in this way, it can be seen, to a first approximation, that the dependence on
radial deviation ``$\delta r$'' of the bending fractions cancels in their ratio.
Though the bending fractions still do not sum to 1, their individual meanings are useful approximations,
with roughly correct ratio.  So, there is some justification in retaining the separate expressions
for $\eta^*_G$ and  $\eta^*_M$ as convenient linearized approximation of the individual bending fractions
while treating their ratio as known and fixed.

Apart from the questionable approximations, the only unknown in the bending ratio is the magnetic field,
which can be expressed in terms of the sun's surface magnetic field;
\begin{equation}
  B_{\rm sun}(r) = B_{\rm sun}(R_{\rm sun}) \frac{R_{\rm sun}^3}{r^3},\quad\hbox{yielding}
\label{eq:radial-sun-mag-fld}
\end{equation}
\begin{equation}
  \frac{\eta^*_M}{\eta^*_G}(r)
  \approx q_2\beta_2(t)\,\frac{cB_{\rm sun}(R_{\rm sun})}{GM_{\rm sun}m_2}\,\frac{R^3_{\rm sun}}{r^3},
\label{eq:Magnetic-Gravitationl-bending-ratio}
\end{equation}
\emph{but only briefly, while there is significant bending, say of electrons, resulting from the sun's magnetic dipole moment field.}
The factor $\beta_2(t)$ causes the relative M/G bending ratio to be greater for electrons than anti-protons
while the electrons have become relativistic but the anti-protons have not.

We anticipate finding four solutions, one for each of the four curves plotted in
Figure~(\ref{fig:AMS-p-e-pos-pbar-asymptotes}). 
It is essential to realize that the points plotted in Figure~(\ref{fig:AMS-p-e-pos-pbar-asymptotes})
are not at all the energies of solutions of Newton's equation.  They represent the detected particle
fractions at each energy compared to a nominal low energy fraction.

At present there is no theoretical calculation for these four processes.  It is not
easy to explain the low energy curve shapes.  But, empirically, the high energy curve
shapes explain themselves, in two ways; they are all straight lines on a log-log plot,
and, as can be best established with pencil and ruler, the straight lines are all parallel.  

Trusting the data plotted in Figure~(\ref{fig:AMS-p-e-pos-pbar-asymptotes}), a serious challenge
is to account for the ratios of the four exhibited intensities.
By first order Taylor expansion about the sun's radius, Eq.~(\ref{eq:BendFrac.etaG} has been written to
linearize the bending fraction deviations in terms of radial deviation $\delta r$.  Because
Figure~\ref{fig:AMS-p-e-pos-pbar-asymptotes} has log-log dependencies, it seems natural to replace the
linear factors in Eq.~(\ref{eq:BendFrac.etaG}) by negative exponentials.

The analytically evaluated orbits in Figure~\ref{fig:Theoretical-analysis-of-AMS-data}
correlate an energy measured in ISS apparatus apparatus near the earth with
a unique orbit that will eventually pass close to the sun.  The six cases shown span the range from
partially relativistic, to almost fully relativistic velocity.

\subsection{Relativistic Kepler orbits}
\begin{figure}[hbt!]
\centering
\includegraphics[scale=0.60]{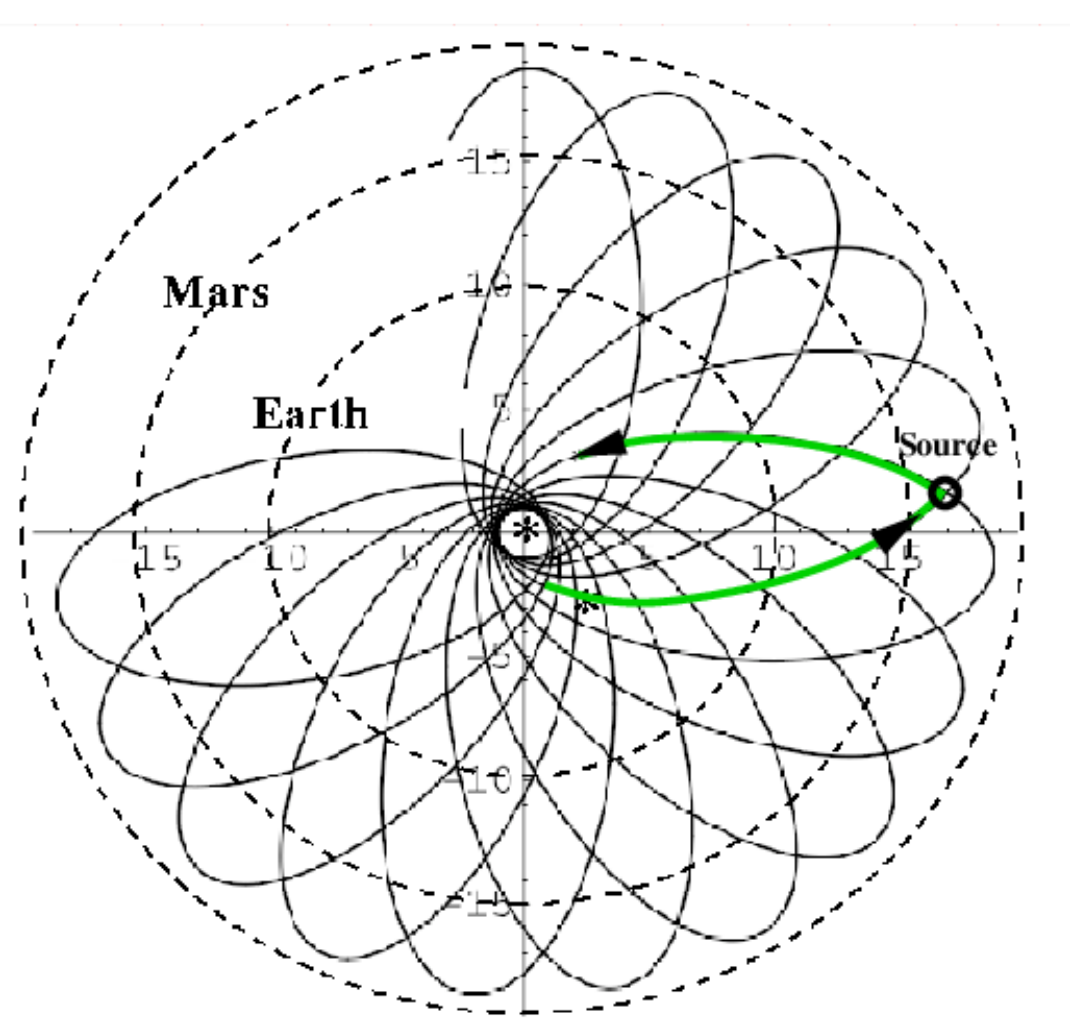}
\caption{\label{fig:Munoz-Pavic2}Figure copied from a paper by Munoz and Pavic\cite{Munoz-Pavic} with
  a solitary, precessing elliptical orbit representing a single, relatively low energy, cosmic ray
  cosmic ray orbit.  This particle is intended to stand in for an entire atmosphere of cosmic rays.
  In such a distribution there would be balanced in-flow and out-flow through spherical surfaces
  centered on the sun. The red arc beginning at ``Source'' represents a ``balancing orbit'' passing
  down through the earth's orbit.}
\end{figure}

Using special relativity Newton's ellipse planetary formalism is readily modified to respect special
relativity.\cite{Munoz-Pavic}.  The leading effect is the in-plane precession of the orientation
of the Newtonian elliptical orbit, under appropriate conditions on angular momentum $L$,
is described by the formulas 
\begin{equation}
\gamma = \frac{\mathcal{E} + k/r}{mc^2}, \quad 
\label{eq:munoz-Pavic.1}
\end{equation}
and 
\begin{equation}
r = \frac{\lambda}{1 + \epsilon \cos(\kappa\theta)}, \quad
\lambda = \frac{k}{\mathcal{E}}\Big[\Big({Lc}{k}^2\Big) -1 \Big], \quad
\epsilon = \frac{Lc}{k}\sqrt{1 - \Big(\frac{mc^2}{\mathcal{E}}\Big)^2\kappa^2}
\epsilon
\label{eq:munoz-Pavic.2}
\end{equation}
This introduction is too abbreviated to follow without referring to the Munoz and Pavic
paper. Its purpose here is primarily as excuse to display the orbit shape description introduced
initially by Newton:
       $$r = \frac{\lambda}{1 + \epsilon \cos(\kappa\theta)}$$
For our conditions, this formula makes
understandable a typical relativistic orbit such as the one shown in Figure(\ref{fig:Munoz-Pavic2}).
As visualized in this paper, once remote from the sun every cosmic ray particle orbit shape
is ``geometrically similar'' to this one, though not so regular, due to unknown magnetic field
perturbations.

Close to the sun, the orbits more nearly resemble the (pseudo)-orbits shown in
Figure~(\ref{fig:Theoretical-analysis-of-AMS-data}, as continued to enter as well as escaping
the sun.
\subsection{Voyage to the sun and back}
\begin{figure}[hbt]
\includegraphics[scale=0.30]{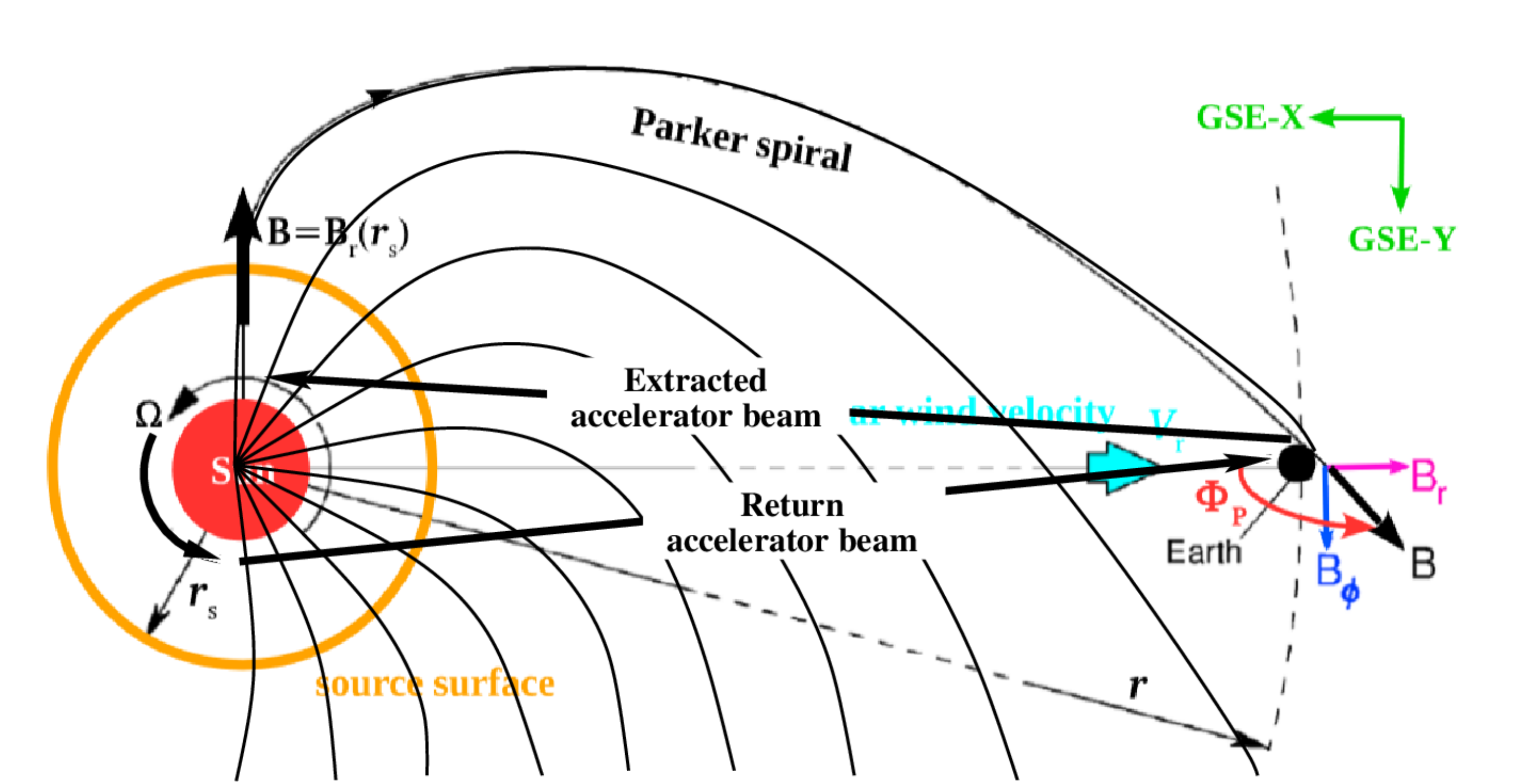}
\caption{\label{fig:FNAL-sun-apparatus}Experimental ``thought apparatus'' sending a Mars-based
  Tevatron beam on a ``voyage to the sun and back''.}
\end{figure}

Figure~\ref{fig:FNAL-sun-apparatus} is related to the red arcs in Figure~\ref{fig:Munoz-Pavic2}.
It represents a thought experiment employing a FNAL Tevatron accelerator situated on Mars and aimed
toward the sun.  This is not unlike the actual FNAL experiment, with Tevatron neutrinos \emph{bent down
by 7.5 degrees}, and aimed at an underground detector site in South Dakota.

To match the red arcs in Figure~\ref{fig:Munoz-Pavic2} the Tevatron would have to be situated on Mars,
at the point labeled ``Source'' (with the beam \emph{bent up  by, say, 5 degrees}, to reduce beam damage
caused by interactions with the Mars atmosphere).
Comparing Figure~\ref{fig:FNAL-sun-apparatus} with Figure~\ref{fig:Munoz-Pavic2}, one sees that the
Tevatron is not pointed correctly, It would need to be pointed significantly to the right, in order
for the return beam to return to the FNAL site on Mars.  Also, beam steering caused by the sun's
magnetic field would need to be taken into account.

\clearpage

\subsection{``Improving'' the performance of solar particle acceleration}
Continuing our effort to design, or rather to understand the design, of cosmic accelerators,
an essential ingredient is the longitudinal electric field component $E_{\parallel}$ that is
causing the beam particle energy ${\mathcal{E}_2}(t)$ to increase with time;
\begin{equation}
\mathcal{E}_2(t) = \mathcal{E}_{2\,\rm inj} + q_2E_{\parallel}t.
\label{eq:Accel.1}
\end{equation}
At this point we make a few significant observations:
\begin{enumerate}
\item
  We have been assuming the Parker longitudinal electric field component $E_{\parallel}$ is constant,
  causing the beam energy to increase monotonically.  This would seem to violate the
  E\&M requirement that the line integral of static longitudinal electric field must vanish.
\item
  The task seems, therefore, to be impossible. But wait; the Sun's electric field is not
  constant; it oscillates with a period of 22 years.  Perhaps, or even probably, this adequately
  overcomes the item~1 constraint. In any case, the direction of stable particle rotation around
  the sun certainly must reverse every 11 years. During reversals, when the magnetic field vanishes
  temporarily, the launching of cosmic rays into space must stop, at least temporarily.
\item
  It is impractical, however, to suppose that the entire process of populating the solar system with
  cosmic rays needs to be repeated every 11 years.  There are too many empirical observations that
  would refute this possibility.  It must, therefore, be the case that the vast majority of cosmic
  rays represent a nearly-constant-in-time equilibrium in which the replenishment of the equilibrium
  is briefly interrupted ever 11 years.
\item
  We are striving to accelerate a beam, say a proton beam, from an energy, of say $10^{-2}$\,GeV,
  to an energy of, say, $10^6$\,GeV, which is just slightly higher than the highest
  energy cosmic ray proton energy plotted in Figure~(\ref{fig:rpp2019-cosmic-rays}).
  This is eight orders of magnitude acceleration.
\item
  Previously, in Section~(\ref{sec:Rough-values}), based on the known magnetic field at the
  surface of the sun it was estimated that protons could be accelerated from 2\,MeV 
  to an energy of 200\,GeV in a single revolution, an acceleration
  by a factor of five orders of magnitude. Though impressive, this is not even close to
  eight orders of magnitude. We are still missing three orders of magnitude.
\item
  The whole point of a ground based particle accelerator is to increase the energy a little
  bit each turn, for many turns. All we need is for the cosmic ray beam to stay captured
  by the sun for a thousand optimal turns or, at least, to make many thousands of less than
  optimal turns around the sun.  This would seem to be easy, considering that beams
  traveling at (almost) the speed of light have been stored in ground based accelerators,
  with tiny apertures, for day-long runs.
\item
  In actuality, in real life, the situation is more complicated than has so far been assumed.
  However, we need not require 100 percent beam capture, nor perfect beam flux retention
  through the acceleration process.  According to Figure~(\ref{fig:AMS-rpp2019-cosmic-rays}),
  the extracted beam flux at maximum energy is less by 12 orders of magnitude at maximum compared
  to minimum energy.
\end{enumerate}

\section{International Space Station (ISS) cosmic ray results}
\subsection{Properties of the Alpha Magnetic Spectrometer (AMS-02)}
The quality of the cosmic ray data from the ISS over the last twenty years is outstanding; a treat
to pore over and understand.  The data in Figure~(\ref{fig:AMS-p-e-pos-pbar-asymptotes}) alone
provides enough data to confirm or refute a significant fraction of the cosmic ray theory presented
in this paper.  But the critical experimental data most needed to test my solar cosmic ray picture is
quite weak at present.  Rather than scanning all cosmic ray experimental data, this section explains why
the AMS magnetic spectrometer data is blind to the data that would provide the most critical test of  
the solar cosmic ray picture.

By design, the AMS-02 spectrometer is optimized to provide universal coverage of cosmic rays aimed more or
less in the direction pointing toward the sun from the location of the ISS which, for this purpose, is the
same as the location of the earth. The acceptance cone of the AMS, though large compared to laboratory
spectrometers, is conical, with conical acceptance angle of approximately 0.5 steradians, which is quite
small compared to the $4\pi$ needed for full angular coverage.

Measured in space within the solar system, incident cosmic ray directions are more or less isotropic,
with diﬀerential energy distribution falling with power law, $\mathcal{E}^{\gamma}$ spectrum, with
``spectral index''$\gamma \approx 3.0$. The several plots in this paper of this energy distribution,
as measured in the ISS, agree well with this behavior, modulo small ``kinks'' to be discussed in the sequel.

My solar cosmic ray picture is consistent with this isotropic, power law energy distribution.  In this
picture, should the AMS-02 apparatus have been pointed away from, rather than towards, the sun, it would
be observing much the same cosmic ray distributions as at present. Unfortunately (in this respect) the AMS-02
spectrometer was not designed for isotropic coverage. In fact, the 17 radiation lengths thick electromagnetic
calorimeter (ECAL), present to  measure the energy released by electromagnetic particles, terminates the
AMS-02 beam line.  Along with the electronic vetoing of sideways traveling cosmic rays, the AMS-02
apparatus is, by design, insensitive to the angular distribution prediction of the present paper.

An apparatus ``perfect'' for for testing solar cosmic ray generation, would have a pair of AMS-02
spectrometers, pointed in opposite directions, one towards, the other away from the sun.

\subsection{ISS-measured isotope cosmic ray survival vs energy}
Many cosmic ray detection experiments have been performed aboard the International Space
Station (ISS): PAMELA, AMS02, ATIC, CREAM, CALET, DAMPE, NUCLEON, etc. For example, the AMS02
experiment is responsible for the splendid data in Figure~\ref{fig:rpp2019-cosmic-rays}.
Energy spectrum measurements from some or all of these experiments are shown in
Figure~\ref{fig:AMS-rpp2019-cosmic-rays}, in the form of a lower energy data set in the
bottom figure, as well as a combination of the same lower energy data and a higher energy
data set in the top figure.

Nearly perfect negative exponential energy dependence is shown in these plots.
Because these are log-log plots, the pure exponential dependence can be quantified
in simple decibels per decade terms, as indicated by the inset calculations.
Though not calculated very carefully, the slopes are sufficiently equal to suggest that
all these data sets have the same origin, either solar or galactic, but certainly not both.
But, looking more carefully, there seem to be ``kinks'' in the plots of individual
nuclear isotopes.

Since much has been made of the proton energy dependence in this region,
this is pursued in the subsequent figures and discussion, centering on
Figure~\ref{fig:Lipari-Vernetto-proton-cosmic-ray-spectrum2-mod}, which is copied from Lipari
and Vernetto reference~\cite{Lipari-Vernetto}, who performed multi-parameter curve fitting
algorithms devoted entirely to analyzing the energy region from $10^2$ to $10^5$,\,GeV. In order
to enhance variation through this region they factor out the average slope to amplify the ``kink''
region visually.

In one sense the Lipari-Vernetto fitting procedure was a good idea. It shows that all seven
experiments, presumably analyzed independently, agree quantitatively as to the existence and
magnitude of the kink effect under study. On the other hand, the expanded energy dependence,
is highly misleading. Sorting this out is, however, hugely instructive, and is the subject of
much of the following discussion.

The analytic representation of the exponential decay of voltage as a function of time in an RC circuit
is elementary. It is the unambiguous solution of a linear differential equation. It meets the
requirements required to be a Markov process, in that the past is constantly being forgotten.
The decay of detection probability as a function of energy, though less elementary, seems to be
essentially similar, with the exception of an occasional kink. A simple physical interpretation for
this kink is that there is a sudden discontinuity in the proton energy.  Once again, this could not
be said to be ``predictive''. It could only be referred as ``descriptive''.

In the days of the ``Star Wars'' movie, the clever description of such
a phenomenon, coined by Etienne Forest, was ``beam me up, Scottie'' propagation, which is, of course,
impossible.  What is possible is that, over time, the energy changes but the position does not. This
phenomenon is most commonly observed in circular accelerators as the beam passes through an RF cavity,
treated as having zero length.  This is a ``sudden'' approximation, of the sort that is commonly
quite accurate.

As an experimental interpretation, this ``proves'' that a significantly large fraction of the protons
being detected have been accelerated, in principle with either sign, though in practice always
positive. as they stay in contact with the sun.  This can only have happened during temporary
circulation around the sun.  Once escaping the sun, the energy density has deviated slightly
prom its previous value, causing a small kink in the data.

The data plotted in Figure~\ref{fig:Lipari-Vernetto-proton-cosmic-ray-spectrum2-mod}
is a spectral distribution, energy spectrum in this case, for which ``local curve fitting'' is not
at all appropriate.  Circuit currents ``decay'', but they never ``un-decay''.  Particle energies
can change and their energy densities must change correspondingly.

As with the Heisenberg uncertainty principle, the effect of an instantaneous impulse is spread over a
large range of energies. What makes the kink in the proton energy distribution most persuasive is that
all seven experiments noticed the same impulse. Since the protons are fully relativistic, their time
and their orbit radius known, the fractional revolution around the sun can be determined.  Then, by their
energy change, which seems to have been, by a factor of one hundred, according to the kink, or by a
factor of $10^5$ in Figure~\ref{fig:Lipari-Vernetto-high-energy-CR-spectrum2}. In passing, one notes
that the longitudinal Parker electric field component could therefore be said to have been measured
approximately.

It is the nature of the vertical scale in these figures that brings Heisenberg uncertainty into the
picture.  Unlike a particle decay curve, which ptots survival probability versus energy (and is necessarily
monotonic) the horizontal scale is energy, while the vertical scale is energy density.  If every
particle acquires the same sudden energy increase then the effective energy density does not change during
this time interval. 

\subsection{ISS measured cosmic ray fluxes}
The closely matching ``shapes'' of ISS plots argues that all the cosmic ray data is either solar or
galactic, but not both. There has, however, been a curious kink in the cosmic ray energy distribution,
observed persuasively by seven independent experiments on the ISS have been subject to the same impulse.
\begin{figure}[hbt!]
\centering
\includegraphics[scale=0.75]{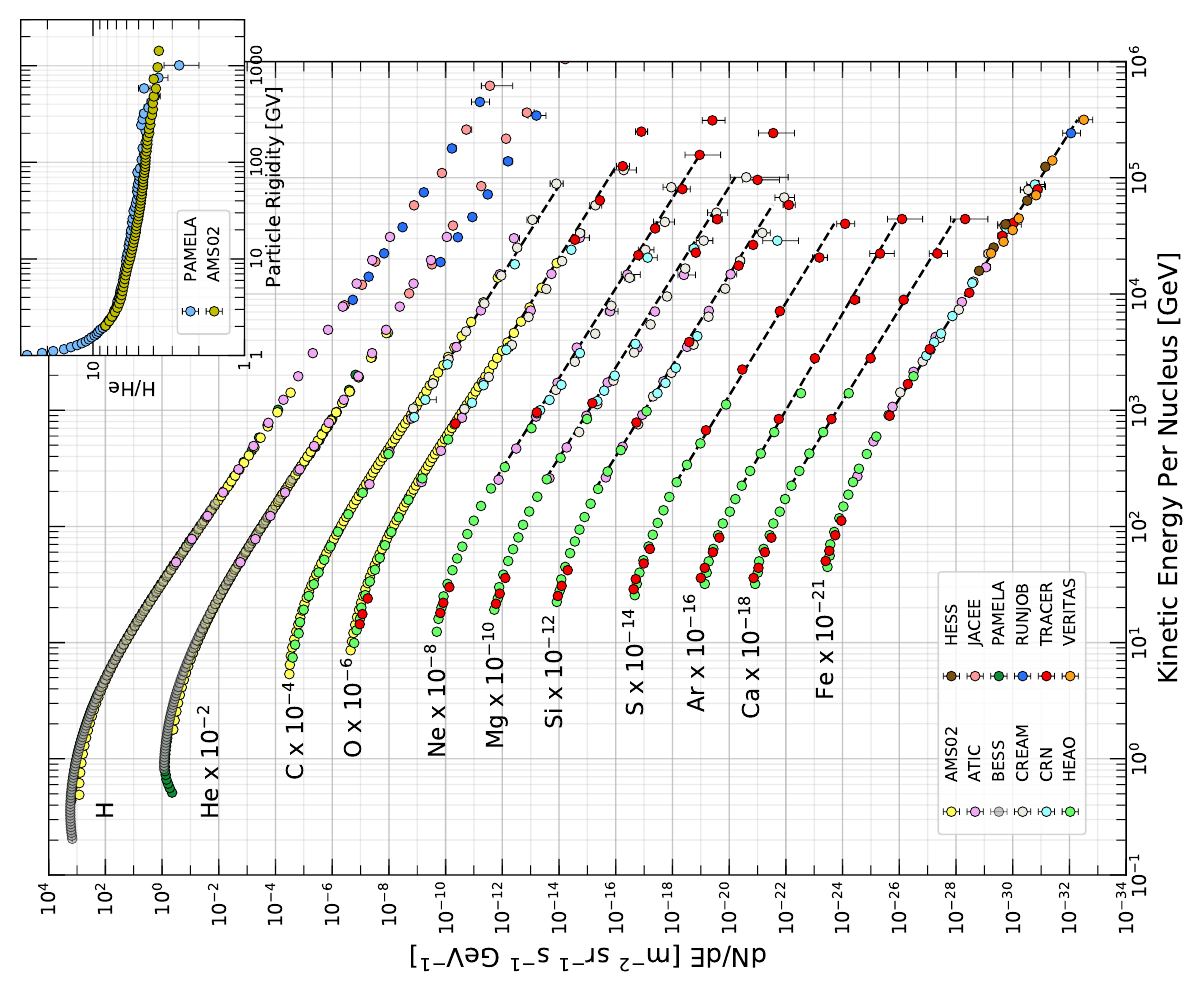}S
\caption{\label{fig:rpp2019-cosmic-rays}
Figure copied from Particle Data Group, Chap-29, Cosmic-Rays.
showing energy dependence and relative abundance ratios measured by AMS within the ISS.  This plot
is superficially misleading by suggesting it is representative of all atomic nuclei.  In fact, the
``light elements'' (Li, B), and especially Be, abundances are not shown; presumably because 
these nuclei are so anomalously missing as cosmic rays.
} 
\end{figure}

The helical orbits in Figure~\ref{fig:Theoretical-analysis-of-AMS-data} are measured
by AMS to have the same energy, augmenting by one, the number shown of protons indicated as
already present in the bin labeled 500.  There are numerous points to make about orbit tracking
in the AMS detector, and related to Figures~{\ref{fig:Theoretical-analysis-of-AMS-data}.
and Figure~\ref{fig:zoomed-AMS-PRL-protons-2011-2019}.  Let us say they correspond to
one of the six events labeled in Figure~\ref{fig:zoomed-AMS-PRL-protons-2011-2019}
To understand the figures it is essential to have available the original paper\,\cite{M.Agular},
while assessing the following points:

\begin{enumerate}
\item
  The tracking measurement is made near the earth, while the orbits are accurately reconstructed near the sun.
  No matter how carefully the orbit has been aimed, it is clear that, during the fitting process, the orbits
  have been constrained near the sun, based on the measured proton energy.
\item
  Each orbit does not represent a single proton, nor even a single beam of protons. It is an orbit
  constructed from points detected in a particular energy range over many years, and binned accordingly.
  Clearly the jagged nature of the orbit reflects fluctuation effects that are prominent at low proton energy,
  but which become less pronounced with increasing energy (as expected from unknown magnetic fields encountered)
  Already by 1\,GeV, with proton velocity $\beta_p=0.87$, the orbits are nearly perfect circles or seemingly,
  shrinking helices.
\item
  One can see from these figures how measured flux per bin as ordinates (in figures like
  Figure~\ref{fig:AMS-rpp2019-cosmic-rays}) can be correlated with matching bin energy as abscissas.
\item
  Though these figures are naturally analyzed as helices shrinking as they approach the sun, it may be
  possible that some of the points are arriving at the sun, and some are leaving the sun.
\item
  This confuses the interpretation of these reconstructed particle orbits as representing an equilibrium
  distribution of more or less isotropic protons and other cosmic ray particles.
\item
  In equilibrium there need to be approximately the same number of cosmic rays going either way,
  to or from the sun.  However the AMS detector acceptance is limited to a narrow solid angle of
  directions pointing toward the sun.
\item
  Forward and backward orbits are spatially identical, but their time evolutions are opposite.
\end{enumerate}

\section{Markov model of cosmic rays as accelerator beam particles}
\subsection{Measured solar cosmic ray distributions}
Continuing the interpretation of proton cosmic ray data, Figure~\ref{fig:zoomed-AMS-PRL-protons-2011-2019}
is a \emph{tour de force} plot of proton flux density measurements, measured continuously from 2011 until 2019
(and to this day, but data not yet published).  \footnote{ISS data for all nuclear isotopes 
is displayed in Figure~(\ref{fig:rpp2019-cosmic-rays})}.  To display such a vast amount of data, they
bin the data day by day, producing and plotting a new point every day--eventually for the 22 year period
of solar magnetic field oscillation, but, for now, just for 10 years.  

The ISS is situated close to the earth. For the comfort of the astronauts, the orientation of the
ISS cannot be rotating very rapidly, and it must also remain more or less upright, with axis pointing
toward the sun. This is important also for the cosmic ray detection apparatus,  which provides precision
particle tracking, primarily over a quite small solid angle, as imposed by the limited
magnetometer mass. 

Though the cosmic rays are more or less isotropic, the detection apparatus is oriented to
emphasize incoming cosmic rays aimed more or less toward the sun.  Fixed relative to the
spacecraft, it is challenging to establish the cosmic ray orbits relative to, say, a coordinate
system with origin at the center of the sun, and orientation fixed relative to the sun's orientation..
It is likely true, also, that most of the ISS experiments cannot distinguish whether events are
headed towards or away from the sun. Even the AWS detector is short enough to make it
challenging to establish the sign difference between ``entrance'' and ``exit'' times. In fact,
the final particle detector is a ``thick'' calorimeter, intended to establish the total energy of
every detected event.

According to the present paper, as illustrated in ~\ref{fig:Munoz-Pavic2}, cosmic rays are expected
be nearly isotropic at the ISS location.  In particular the distribution should be more or less
forward/backward symmetric at the ISS.  But only those traveling in the forward direction are detected.
In a future space mission this limitation could, perhaps, be rectified, but the current data
cannot confirm or contradict the expected cosmic ray isotropy.  

This section discusses the orbitry illustrated in Figures~\ref{fig:Theoretical-analysis-of-AMS-data},
and ~\ref{fig:zoomed-AMS-PRL-protons-2011-2019}.

There is a procedure in differential geometry, using  ``Cartan moving frame'' formalism,
that reconstructs spatial orbits in one frame of reference (fixed, say, to the sun),  based on kinetic
formulas and directions in a relatively-moving frame, say the ISS. With momenta inferred from energy,
and presuming Newton's gravitational law to be valid, a single point on the cosmic orbit can be
established remotely, from an observation point near the earth.

Of the six orbits established (and
labeled at the bottom of the upper plot, in Figure~\ref{fig:zoomed-AMS-PRL-protons-2011-2019}. the
highest momentum ``proton'' follows a circular orbit closest to the sun, captured at least temporarily by the
gravitational attraction toward the sun. As can be seen by the labeling of the figure, what is
plotted along the radial axis, is the measured proton flux through the magnetic spectrometer in the
ISS. As already stated, this is all very challenging.  Yet there is a further challenge.

Even though the Cartan reference frames have been established, there is a time delay issue, resulting
from the time delay equal to the transit time of the protons from the ISS to the sun, which is a distance
of approximately 1\,AU, or $1.5\,\times10^{11}$\,m. Traveling at the speed of light the
time delay is about eight minutes.  For a 1\,GeV proton, $\gamma$ is approximately 2.5,
and the speed is about 0.87\,c.

This is a large enough difference to account for the ragged 1\, GeV proton ``orbit'' shown in
Figure,~\ref{fig:zoomed-AMS-PRL-protons-2011-2019}.  It can be seen, however, in the same figure, that
the 10\,GeV orbit around the sun is very nearly closed.  Note also, that this same orbit matches almost
exactly the white circular orbit, labeled ``$r_2$ limit circle'', which is the circle that almost fully
relativistic charged particles first follow.

As protons are further accelerated by the sun's electromotive force (EMF), they acquire the precessing
elliptical shapes shown in Figure~\ref{fig:Munoz-Pavic2}.  There is, of course, no evidence of this in
Figure~\ref{fig:zoomed-AMS-PRL-protons-2011-2019} which takes account of no such acceeleration.
Because of their eccentric, processing, elliptical-shaped orbits, which have become ``cosmic rays'',
spend little time near the sun but, with increasingly long periods, their energies continue to be
increased by every passage by the sun. This continuing acceleration can be treated perturbatively,
as discussed in Section~\ref{sec:Hamilton-Jacobi}.  

It is especially important to realize that any particle  acceleration is orthogonal to the radial
direction from the center of the sun. This is what permits ultrahigh energy particles to remain
captured in the solar system.  Because of their systematically increasing momenta particles will
follow spirals of increasing transverse radius corresponding to their increased momenta, but
without direct radial acceleration.

This is one respect in which non-relativistic mechanics is misleading. There is no such thing
as ``escape velocity'' once the velocity is already maximal.

As already mentioned, the Cartan orbit reconstruction is time independent.  If an
orbit respects CPT invariance, the Cartan tracking cannot distinguish between forward and
backward track following.

With these considerations, it seems plausible to suppose that the orange curve, defined by ragged
data points, can be interpreted as either the entering-to, or the exiting-from, curve of the
1\,GeV protons.  The more-nearly fully relativistic 10\,GeV orbit, is much less ragged, presumably
due to its reduced sensitivity to magnetic bending.

\begin{figure}[hbt!]
\centering
\includegraphics[scale=0.6]{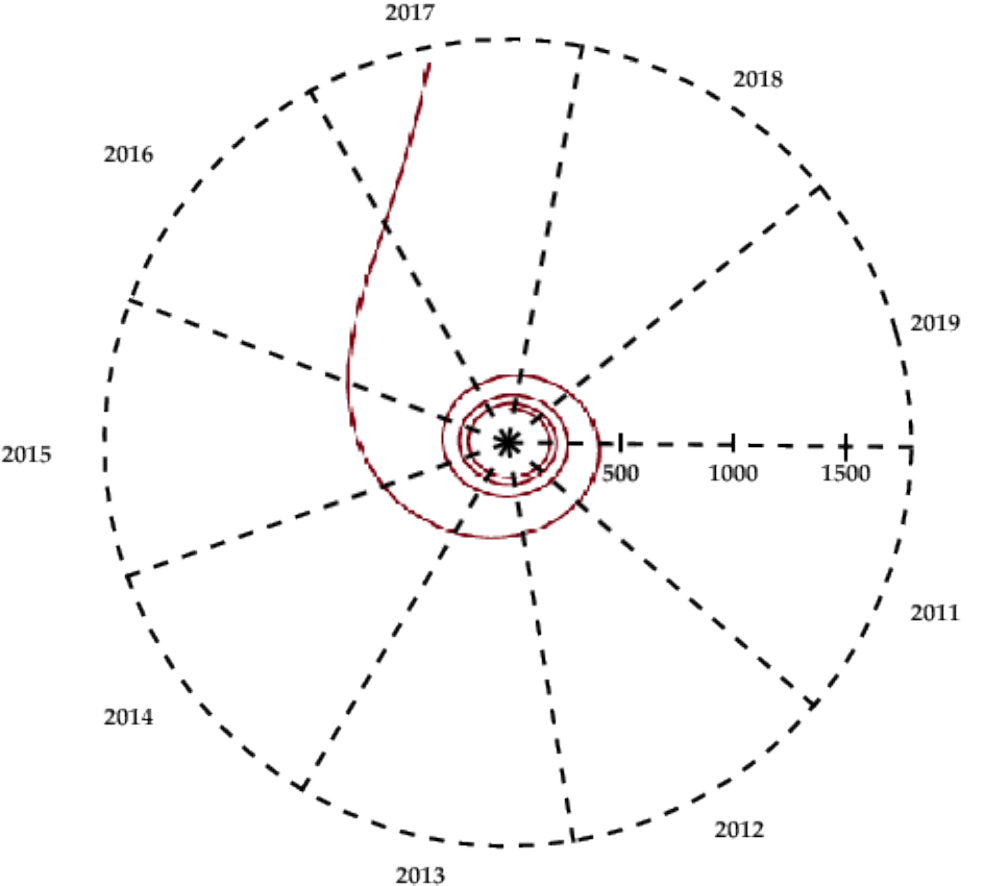}
\includegraphics[scale=0.6]{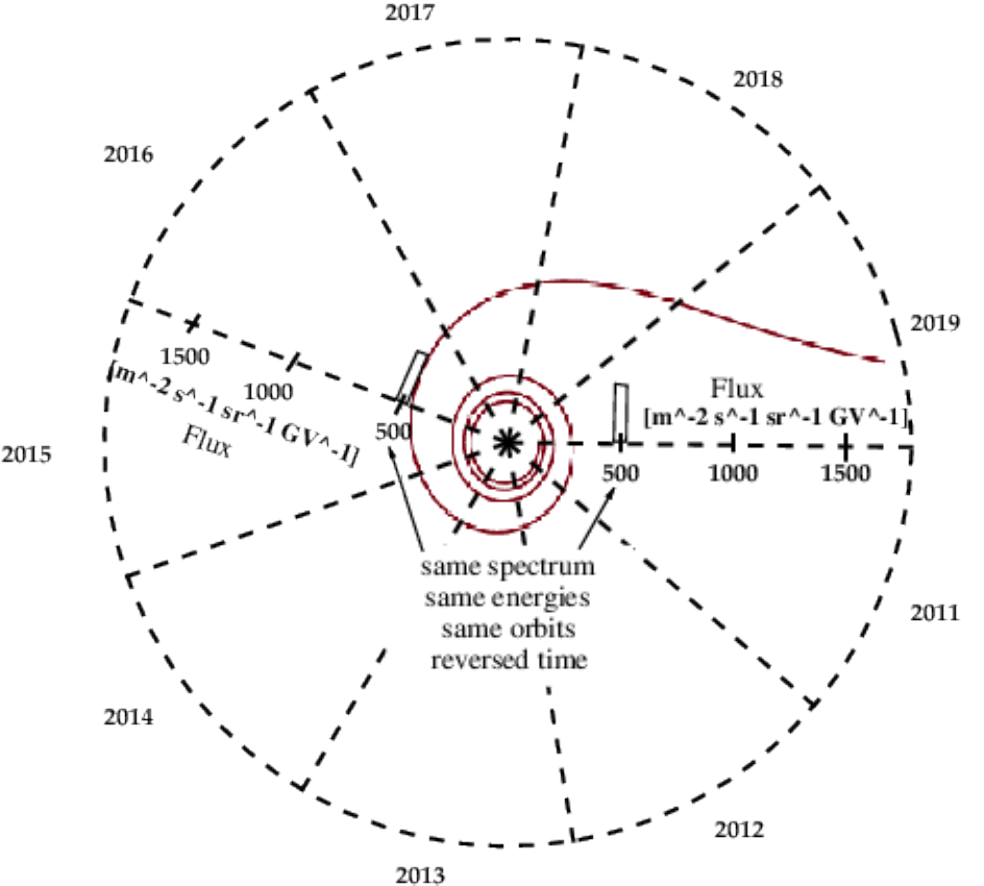}
\caption{\label{fig:Theoretical-analysis-of-AMS-data}{\bf Top:\ }Calculated (pseudo)-orbit for
a proton departing from  AMS for the sun in mid 2017. {\bf Bottom:\ }Calculated (pseudo)-orbit
(assuming negligible magnetic field in both cases) for proton departing from 
the sun for AMS in mid 2019.\smallskip\\
\noindent
The spiral orbits are calculated analytically 
lituus-modified spirals, (introduced by Galileo, who died the same year
that Newton was born) as the orbit of a mass falling onto a massive sphere,
which we, unlike Galileo's assumed Earth, are taking to be the sun.  The lituus shape
modification allows adjustment of the direction of the asymptotic straight line passing
through the ISS. It is possible, but frustrating, to count the number of revolutions, even
in this, almost the simplest possible case.  In any case the orbit need not be permanently
captured; out-of-plane velocity components cannot be seen in this projection.\smallskip\\
\noindent
As with all orbits observed by the ISS, these orbits do not represent single particles.
They represent binned and sorted reconstructions with inferred energies having reliably
small statistical sampling error, but no way of distinguishing the direction of travel.}
\end{figure}
\begin{figure}[hbt!]
\centering
\includegraphics[scale=0.75]{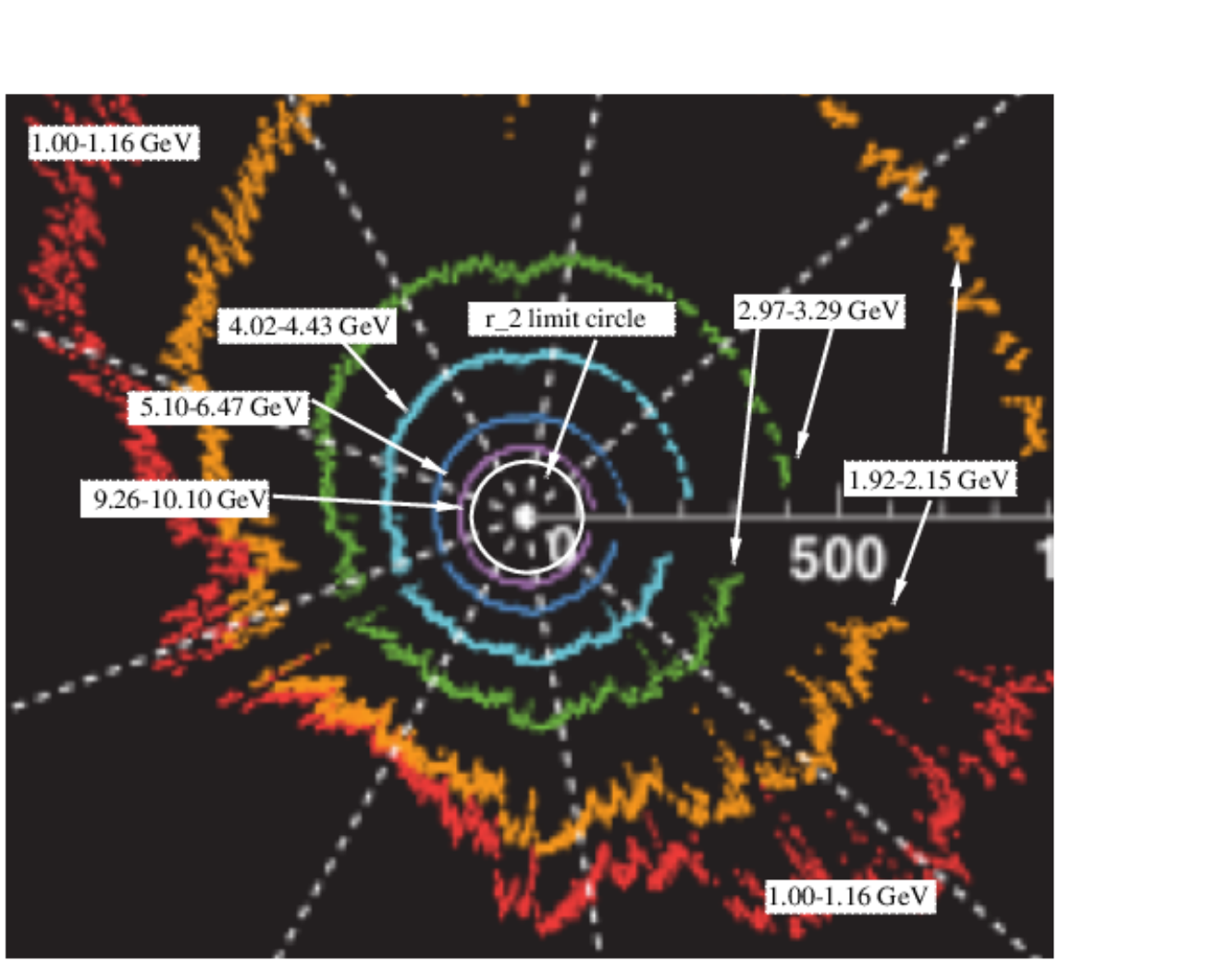}
\caption{\label{fig:zoomed-AMS-PRL-protons-2011-2019}
Central portion of a figure from M. Agular, et al.~\cite{M.Agular}. showing the daily AMS
proton fluxes for six typical rigidity
bins from 1.0 to 10.10 GV measured from May 20, 2011 to October 29, 2019 which
includes a major portion of solar cycle 24 (from December 2008 to December 2019).
The AMS data cover the ascending phase, the maximum, and descending phase to the minimum of
solar cycle 24.)}
\end{figure}
\begin{figure}[hbt!]
\centering
\includegraphics[scale=0.48]{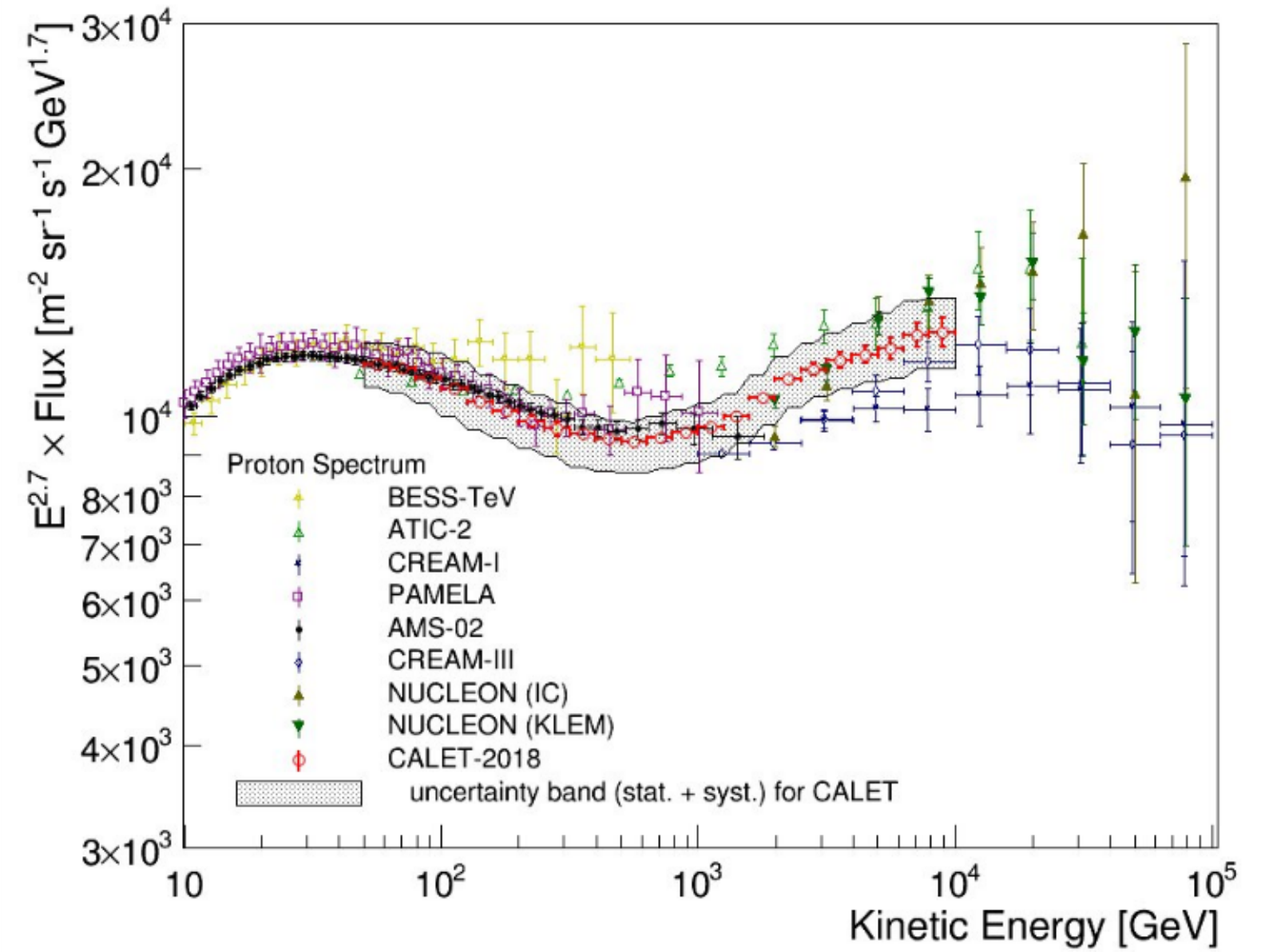}
\includegraphics[scale=0.67]{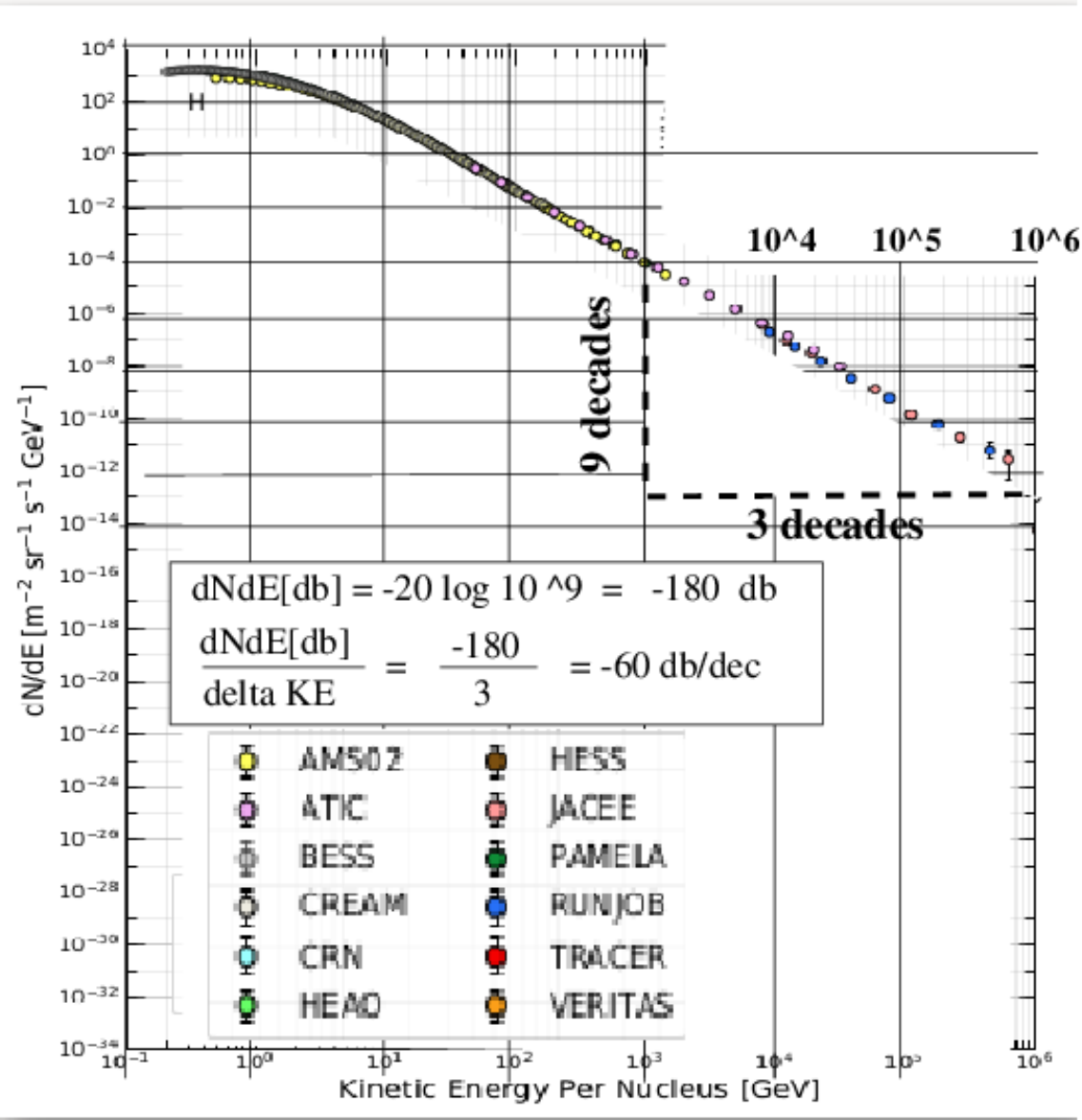}
\caption{\label{fig:AMS-rpp2019-cosmic-rays}ISS measured flux of cosmic ray protons,
plotted as function of proton energy:
{\bf Top:\ }${\rm dN/dE}$ flux dependence, with best fit exponential factor ${\rm KE}^{2.7}$ scaling;
{\bf Bottom:\ }unscaled data, with best fit exponential decay (expressed as db/decade).
This slope of (approximately) 3 decades per decade only matches the fully relativistic data,
which is more or less consistent with the $\mathcal{E}^{2.7}$ energy scaling in the upper figure.
This proton data is the same as the more complete data shown in
Figure~(\ref{fig:rpp2019-cosmic-rays}).
}
\end{figure}

\begin{figure}[hbt!]
  \centering
\includegraphics[scale=0.52]{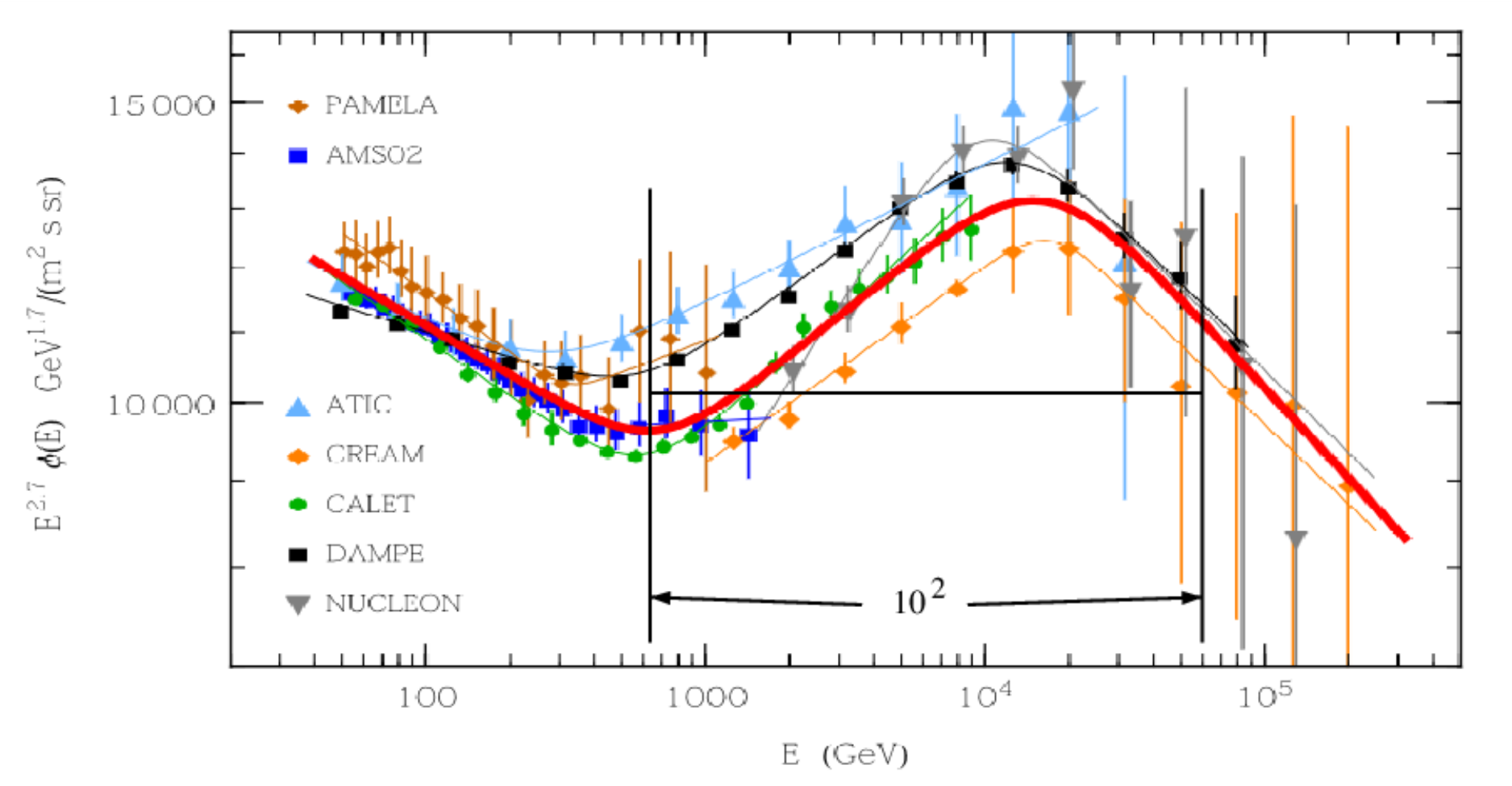}
\caption{\label{fig:Lipari-Vernetto-proton-cosmic-ray-spectrum2-mod}This figure has been
copied from Lipari and Vernetto\cite{Lipari-Vernetto}. This data shows that all
seven experiments on the ISS were subjected to the same impulse, or that all protons were
subjected to the same impulse when near the sun. In the present paper this is interpreted
as acceleration of protons from 600\,GeV, to 60,000\, GeV, while briefly circulating around
the sun.  What looks like a slope discontinuity, i.e. kink, in the particles per energy spectrum,
corresponds to a substantial increase in the energy of individual particles, with little change
in particles per unit of energy.  This vaguely resembles ``Heisenberg uncertainty''. While
seemingly ``at rest'', though actually performing multiple turns around the sun, an energy
alteration over a long time is perceived to have been sudden.  A ``discontinuous''
change in energy registers as a ``kink'' in particles per unit energy. 
}
\end{figure}
\begin{figure}[hbt!]
\centering
\includegraphics[scale=0.47]{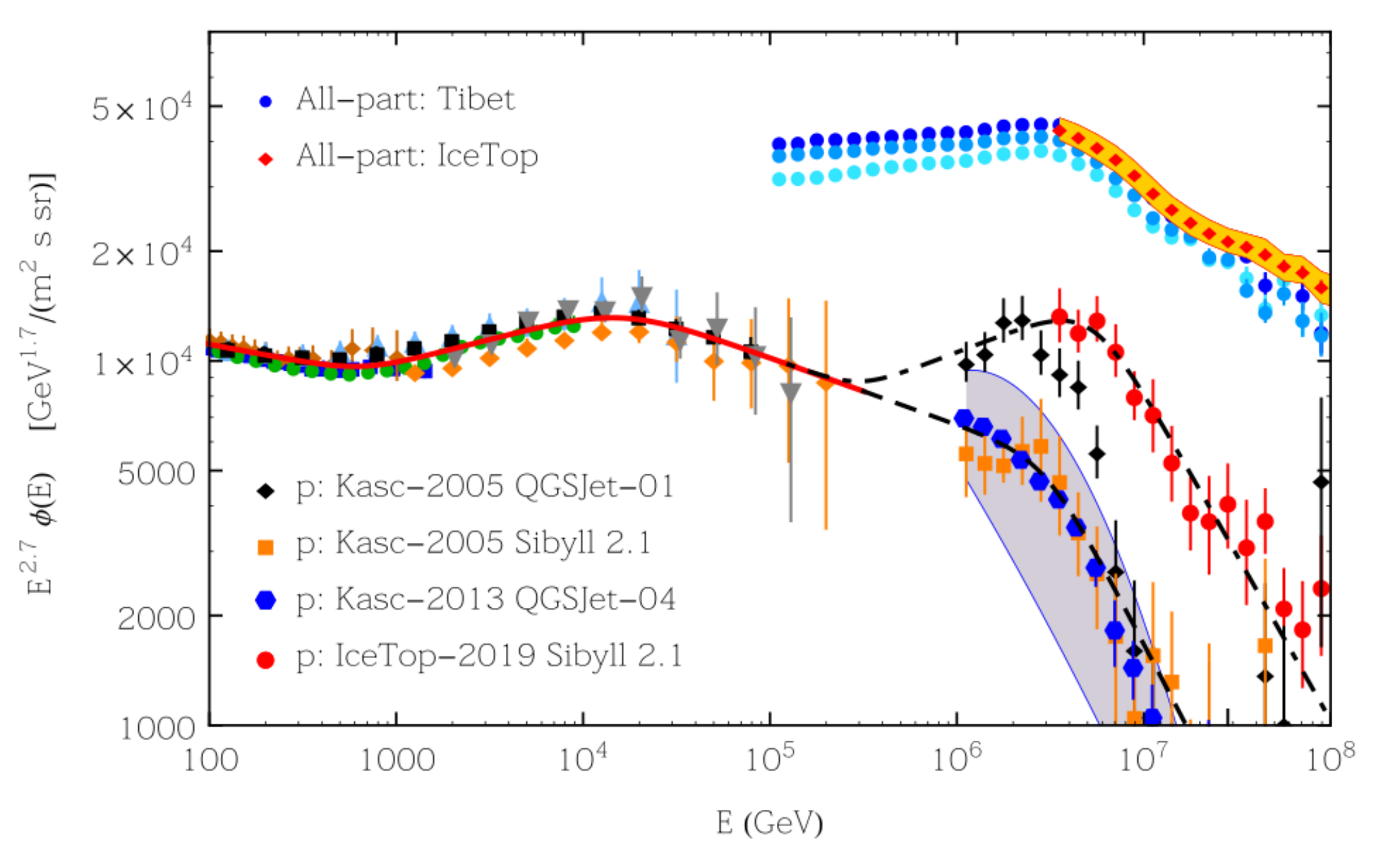}
\caption{\label{fig:Lipari-Vernetto-high-energy-CR-spectrum2}
Also copied from Lipari and Vernetto\cite{Lipari-Vernetto}, the data in this plot,
increase the acceleration range, starting from 100\,GeV and running
almost to $10^7$\,GeV.}
\end{figure}

\subsection{Original and acceleration-imposed Schottky noise}
There has always been an intimate connection of statistical physics with accelerator physics.
Particle beams of a single species have statistically distributed distributions in six dimensional
phase space.  Though these distributions limited the performance of early (linear) accelerators,
the statistics was quite simple because the only way of improving beam precision was by
collimation, necessarily reducing the counting rates. 

The advent of circular accelerators changed this situation dramatically. With long term stability
understood, and  good vacuum achieved, the same beam could impinge repeatedly on the same target,
hour after hour.  Though eliminating the previous problem, this introduced a new statistical problem,
namely beam quality reduction caused by the stochastic scattering of every beam particle in the
target material, causing stochastically increased ``emittance growth'' and corresponding loss of particle
phase space density, \emph{that would otherwise be conserved by  Liouville's theorem.}  This, too,
was reasonably easy to understand, but could only be reduced by reduced target thickness and
counting rate.  Unlike the original beam spreads this spread is ``induced'' by the beam
acceleration mechanism.

In the sun as beam accelerator, both of these noise sources are present and both can, in principle,
be measured.  This is why Markov chain treatment is appropriate for diagnosis of the sun as
accelerator. Markov processes are processes with ``no after effect'' meaning the past is always
forgotten; the future is based only on present conditions and future sources of noise.  

In the present context, all of this discussion could not be said to be ``predictive''.  It could
only be referred as ``descriptive'', based on the experimental data. Most of the following
discussion is subject to the same qualification.

A more serious application of stochastic processes in accelerator physics came with the
introduction of detectors sensitive to ``Schottky noise'', both ``longitudinal'' caused by arrival
time spread, and ``transverse'' caused by coupling of longitudinal and transverse coordinates.
Then came stochastic cooling, which comes as close as possible to serving as a ``Maxwell demon'' 
violating the second law of thermodynamics.  For the sun as accelerator, \emph{neither of these topics
is applicable}, since the observer has no control over the external signals influencing the
statistical properties of the beams being accelerated.

\subsection{``Liouvillean'' cosmic ray high energy flux dependence on energy}
Figures such as Figure~(\ref{fig:AMS-p-e-pos-pbar-asymptotes}) and
Figure~(\ref{fig:rpp2019-cosmic-rays}) provide information concerning the energy dependence of
the flux of various particle types.\footnote{Technically ``rigidity'' is ``momentum per charge'',
not ``energy'', but, at fully-relativistic energy this distinction has become inessential.}
From these plots one can infer the energy dependence of 
abundance ratios of pairs of particle types.  Superficially the abundance ratios themselves show
no strong energy dependence. (i.e. the ``shapes'' of the curves are similar).  Especially striking
is the fact that, once relativistic, these ratios seem to be ``frozen'', independent of energy
as shown by the identical high energy slopes, independent of particle type.

At the same time, a figure such as Figure~(\ref{fig:rpp2019-cosmic-rays} shows abundance ratios
varying over 20 orders of magnitude, all with the same flux vs energy dependence, at fully
relativistic energy.  Simple description of this behavior is that, for fully relativistic particles,
the acceleration mechanism is ``the same'' in all cases, independent of charged particle type.
Once again, this could not be said to be ``predictive''. It could only be referred as ``descriptive''.

A universal feature of radiation from sources of small size, is that the intensity
(expressed as the flux in this case) varies inversely as the square of distance from the source.
This represents conservation of energy. Plotted on log-log paper the slope would be two decades
of flux for every one decade of energy.
\footnote{
Especially in accelerator x-ray sources, this radiation property is expressed as ``brightness''
or as ``brilliance'', two quantities that are technically different but, for present purposes,
are much the same~\cite{Talman-Accelerator-Accelerator-X-ray-Sources}.}
However, Figure~(\ref{fig:rpp2019-cosmic-rays}) represents the distributions in energy, not
radius.  The data itself, for example Figure~(\ref{fig:AMS-p-e-pos-pbar-asymptotes}), shows 
3.3 decades of flux per decade of ``rigidity'', meaning ``momentum per charge'', with momentum
and energy. It is this data that  converges to parallel straight lines at high energy.

We are familiar with straight lines on log-log plots representing the semi-permanent survival of
long-lived, but unstable, nuclear isotopes. Starting with equal numbers of a stable and an unstable
isotope, the plot of their surviving fractions would be identical only until the last of
the unstable particle had decayed.

One can account for this by noting that only stable nuclei are displayed and unstable nuclei
are both rare and not plotted.  We know, for example, that there are few enough Be nuclei to justify
their not being included.

Something still seems surprising about Figure~(\ref{fig:rpp2019-cosmic-rays}. The data seems to show
in some mechanical sense, that all charged elementary particles are ``identical''.  Looking closely
at the straight line fits to C, O, Ne, Mg, Si, S and Ar, it can be seen, in almost every case, that
there are actually kinks in the data which seem to be to be matched by two, rather than one, broken
fit line.  This is consistent with the requirement, following the curves from right to left, for them
to roll over to be more or less horizontal at the left edge of the graph.  But why should the asymptotic
extrapolations be exactly parallel in the fully relativistic limit?  Perhaps this was imposed by
the choice of energy scaling factors?

The absence of lithium, and (especially) beryllium, and boron, has always been ascribed to the
ultra-short $10^{-18}$\,s lifetime of ${}^8{\rm Be}_4$; its absence from the solar wind is discussed
in reference~\cite{Talman-solar-cosmic-ray-1}. 

It is important to be careful about possible misinterpretation of ``kinks'' in the nuclear isotope
distribution plots in the final several figures of the paper.
Figure~(\ref{fig:Iterated-Markov}), copied from Roman and Bartoletti,~\cite{Roman-Bertolotti},
``A master equation for power laws'', describes statistical mathematics representation of
``kinks in power laws'' such as are observed in those figures.  This is sketched in the caption to
that figure.

A more physical explanation of the kinks observed in cosmic ray energy distributions, is provided
by the Heisenberg uncertainty principle, as explained in the caption to
Figure~\ref{fig:Lipari-Vernetto-proton-cosmic-ray-spectrum2-mod}.

In our model of the sun as accelerator there are periodic sources of ``noise'', encountered
while individual particles pass by the sun.  In the high energy limit, where all particle
energy fits are parallel, the same stochastic fit applies to every particle type, but
the incoherence of the perturbation requires the iterated fit to more nearly fit the
measured data, in spite of the quite large energy range of the individual fits. 
\begin{figure}[h!]
\centering
\includegraphics[scale=0.50]{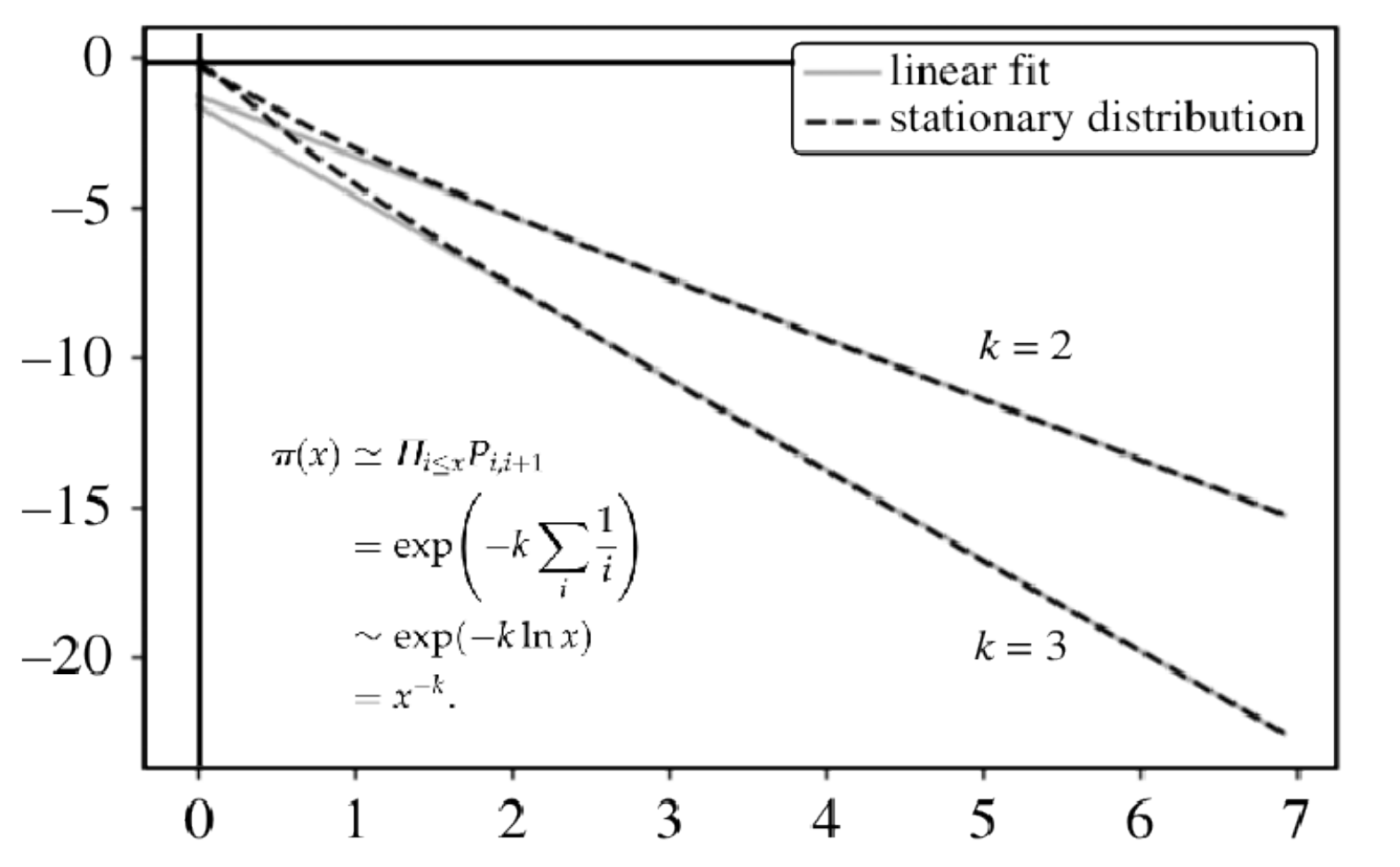}
\caption{\label{fig:Iterated-Markov}Copied from Roman and Bartoletti \cite{Roman-Bertolotti}
``A master equation for power laws'', this figure shows how an iterated Markov
stochastic distribution with one parameter (in this case, k=2) can match a different
Markov process with parameter (k=3). Applied to our case, periodic perturbation
of one Markov process, over a large (but finite) range of abscissas, can be accurately
matched to a different Markov process.  As explained in the text, this could account
for the kinks in Figure~(\ref{fig:rpp2019-cosmic-rays}), while continuing to treat
the overall process as Markovian. In this way one can understand (as mathematics)
why there need to be kinks at low energy but (as physics) there has to be an explanation
for all fits having exactly the same slopes in the high energy limit, as shown in
Figure~(\ref{fig:rpp2019-cosmic-rays}.
}
\end{figure}

\subsection{Systematic increase in solar cosmic ray particle energies}
A effort has been made so far, based on ``established physics'', meaning, primarily,
special relativity, to account for observed cosmic ray properties. This section contradicts nothing
described so far.  Rather, it reformulates the astrophysics of cosmic rays starting from a Hamilton-Jacobi
(H-J) wave-like interpretation of Newtonian gravity.  This is in contrast with Einstein's general relativistic
formulation of cosmology.

The Jacobi partial differential equation reformulation of Hamilton's classical mechanics is reviewed in
Appendix~A, \emph{Hamilton-Jacobi Wave Mechanics}, which has been generalized to account for special relativity,
and patterned after Schr\"odinger wave mechanics, but not encumbered by quantum mechanical complications,
such as failure of commutation and Heisenberg uncertainty, to be sufficient for understanding cosmic ray
generation.

There is one experimental cosmic ray feature that complicates our understanding of cosmic rays.
It was hinted at in the previous ``Disambiguation of the term `cosmic ray' sub-section'' of this paper.  
Now it is returned to in Figure~\ref{fig:Munoz-Pavic2} which exhibits a class of particle orbits in an
appreciably large central portion of the solar system.  The figure also  serves as a reminder that, from
the earth's perspective, the solar and galactic systems are both ``astronomically large''.

Just a single celestial orbit is plotted in Figure~\ref{fig:Munoz-Pavic2}, but in its various phases
and amplitudes it can represent a multiplicity of different particle orbits constituting the 
distribution of ``light'' (meaning atomic and nuclear) material in the solar system.
An important observation is that there are as many particles aimed away-from, as aimed towards the earth.
This satisfies a major requirement of equilibrium in the cosmic ray atmosphere.  Plotted in 1950,
Figure~\ref{fig:Munoz-Pavic2} takes proper account of special relativity to explain the dependence of
particle velocity on distance from the sun.  Plotted here it does not yet account for the stochastic,
more or less monotonic, increase of single cosmic ray particle energies during their brief but repetitive
circuits around the sun.

\subsection{Hamilton-Jacobi description of accelerating particle orbits}
To begin with, it is important to qualify the meaning of ``unperturbed'' in this section.
The basic understanding of terrestrial particle accelerators is said to be ``exact'', even though
engineering defects, and only partially understood ``second order'' effects persist, that
cause ``unperturbed'' to be somewhat of an exaggeration.

For our solar accelerator, the meaning of the term ``unperturbed'' needs to be qualified far further,
even to the point of making the term misleading.  Nothing concerning injection can be said to be
``exact'' even for laboratory-based accelerators. However, the term applies quite nicely to
the subsequent acceleration and extraction processes.  Anything said about injection into
or extraction from, the solar accelerator, can only be conjectural.  

For the sun as accelerator, the acceleration phase is described by closed-form Hamilton-Jacobi
formalism that provides unambiguous answers that can be correct or incorrect, depending on the
truth of the assumptions and the accuracy of the parameters. This is the only sense intended 
as the meaning of ``unperturbed'' in this section.

Our model of solar cosmic rays is based on charged particles following relativistic Keplerian orbits.
Not yet described, however, is a quantitative calculation of the gradual acceleration of particles during
the brief sections of their orbits as they temporarily circulate around the sun.  For this calculation,
a Hamilton-Jacobi partial differential equation description of the relativistic mechanics seems to be
appropriate.  The special relativistic Kepler orbit treatment is already ``time-dependent'' by virtue
of the time dependence implicit in $\gamma_2(t)$ variation; we are now introducing the further time
dependence associate with acceleration by the sun's longitudinal electric field.

Formally, this new time dependence could simply be accounted for by a more detailed evaluation
of $\gamma_2(t)$.  But this is easier said than done.  A perturbative treatment seems needed in order to
calculate the further dependence on time of the orbits during the brief time they are circulating
around the sun.

This (more or less periodic) time dependent acceleration is different for different particle types. 
It can best be treated as the perturbation of otherwise energy-conserving relativistic Keplerian orbits.
The unperturbed H-J description of these orbits is described in Appendix~A, \emph{Hamilton-Jacobi Wave Mechanics.}

Calculations like this have become routine in Celestial Mechanics (again with the simple generalization,
in our case, from non-relativistic to special relativistic mechanics). In fact, the most general treatment is
due to Lagrange himself, in his ``Lagrange Planetary Equations'', which introduced adiabatic invariant
``Lagrange brackets'', which were later inverted into ``Poisson brackets'', and later yet, by Dirac, into
quantum commutation relations.  Much of this is described in references\cite{Talman-Mechanics-Chapter-14}
and \cite{Talman-Mechanics-Chapter-16}.

\section{An experimental test of solar cosmic ray production}
Many aspects of solar cosmic ray astrophysics have been addressed in this paper.
At most one of the extra-galactic or the solar source models can be essentially correct.
Based purely on presently available experimental data, one cannot decide which
is more nearly correct.

An experimental test is proposed here which may be workable and might be expected to
favor one or the other of the purely solar or the purely extra-galactic cosmic ray source
possibilities.  This proposal places two identical but anti-parallel AMS spectrometers
on a next generation ISS space ship.\footnote{This is not as easy as it sounds, since the
solar panels for the ISS, always shield the ISS from the sun, possibly  impeding detection
of cosmic rays coming directly from the sun.}  Comparison of the symmetries of the cosmic
ray distributions measured by these two detectors should help to distinguish between
solar-based and extra-galactic-based cosmic ray models.

For the extra-galactic model, nearly perfect initial cosmic ray isotropy is assumed; currently
observed asymmetry is ascribed to shielding by magnetic fields encountered by cosmic rays after
their entry into the solar system.  For the solar-origin cosmic ray model, there will also be
observed asymmetry.  But the details of these two cosmic ray anisotropies will be very different.

What has not yet been provided in this paper, is a self-consistent modeling of existing cosmic
ray data, based on formulas derived in this paper.  Ingredients for such a model have been described,
including the H-J description of individual particle orbits, dominated by their brief but repetitive
encounters with the sun.  Concerning these orbits while remote from the sun, the absence of coupling
between particle energy and the other three H-J separation momenta helps reduce uncertainty
associated with orbit wandering caused by unknown large scale magnetic forces.   

Comprehensively measured abundance-ratios have been displayed, but not yet carefully accounted
for using theoretical methods described in the paper. Some of these comparisons should be
quit easy; especially p/p-bar and $e^+/e^-$ energy-dependent ratios, which are especially challenging
for extra-galactic sources.  

\section{Acknowledgments}
Special acknowledgments are owed to Peter Wittich, Anders Ryd, Lawrence Gibbons, Maxim Perelstein,
Liam McAllister, Eanna Flanagan, Dave Chernoff, Saul Teukolsky, Nils Deppe, Nigel Lockyer,
Csaba Csaki, Mike Niemack, Ritchie Patterson, Jared Maxson, and Leonard Gross.  My son John
has also made valuable contributions.

\clearpage

\appendix

\bigskip
\centerline{\bf APPENDICES}
\section{Relativistic energy and momentum    \label{sect:Relativistic-energy_and-momentum}}
\subsection{The relativistic principle of least action}
It is straightforward to generalize the principle of least action 
in such a way
as to satisfy the requirements of relativity while at the same time leaving
non-relativistic relationships (e.g. Newton's Law) valid when speeds are
small compared to $c$.
Owing to the homogeneity of both space and time,
the relativistically generalized action $S$ cannot depend on the
particle's coordinate 4-vector $x^i$. Furthermore it must be a
relativistic scalar since otherwise it would have directional 
properties, forbidden by the isotropy of space.

The action of a free particle 
(i.e., one subject to no force) is
\begin{equation}
S = (-mc)\int_{t_0}^t ds 
  = (-mc^2)\int_{t_0}^t \sqrt{1 - \frac{v^2}{c^2}}\,dt,
\label{eq:Relint.nine2} 
\end{equation} 
where the invariant interval $ds$ is the proper time multiplied by $c$.
As always, the dimensions of $S$ 
are momentum$\times$distance or,
equivalently, as energy$\times$time.
Though the first expression for
$S$ is manifestly invariant, the second depends on values of
$v\equiv|\dot{\bf x}|$ 
and $t$ in the particular frame of reference in which Hamilton's 
principle is to be applied. \emph{A priori} the
multiplicative factor could be any constant, but it will be seen 
shortly why the factor has to be $(-mc^2)$. 
The negative sign is significant. It
corresponds to the seemingly paradoxical result
that the free particle
path from position $P_1$ to position $P_2$ {\it maximizes} 
the proper time taken. Comparing with the standard definition of the
action in terms of the Lagrangian, it can be seen that the free particle 
Lagrangian is
\begin{equation}
{L}({\bf x},\dot{\bf x}) = -mc^2 \sqrt{1 - \frac{|\dot{\bf x}|^2}{c^2}}.
\label{eq:Relint.ten2} 
\end{equation} 
As always, the Lagrangian has the dimensions of an energy.

In Lagrangian mechanics, once the Lagrangian is specified, the equations
of motion follow just by ``turning the crank''. Slavishly following
the Lagrangian prescriptions, the momentum ${\bf p}$ is \emph{defined} by
\begin{equation}
{\bf p} = \frac{\partial {L}}{\partial \dot{\bf x}}
      = \frac{m{\bf v}}{\sqrt{1 - v^2/c^2}}.
\label{eq:Relint.eleven2} 
\end{equation} 

\noindent
For $v$ small compared to $c$, this gives the non-relativistic result
${\bf p}\simeq m{\bf v}$. 
This is the relation that fixed the constant factor in
the initial definition of the Lagrangian.
From this equation one obtains the Hamiltonian $H$ and hence 
the energy $\mathcal{E}$ by
\begin{equation}
H = {\bf p}\cdot{\bf v} - {L}
      = \frac{m c^2}{\sqrt{1 - v^2/c^2}}.
\label{eq:Relint.twelve2} 
\end{equation} 

\noindent
For $v$ small compared to $c$, and the numerical value of $H$
symbolized by $\mathcal{E}$, this gives 
\begin{equation}
\mathcal{E} \simeq \mathcal{E}_0 +\frac{1}{2} mv^2,
\label{eq:Relint.thirteen2} 
\end{equation} 
which is the classical result for the kinetic energy,
except for the additive constant  $\mathcal{E}_0=mc^2$,
known as the rest energy. An additive constant like this has no effect
in the Lagrangian description.
From Eqs.~(\ref{eq:Relint.eleven2}) and 
(\ref{eq:Relint.twelve2}) come the important identities
\begin{equation}
\mathcal{E}^2 = {\bf p}^2c^2 + m^2 c^4,\quad
{\bf p} = \frac{\mathcal{E} {\bf v}}{c^2}.
\label{eq:Relint.fourteen2} 
\end{equation}
For massless particles like photons these reduce to $v=c$ and
\begin{equation}
p = \frac{\mathcal{E}}{c}.
\label{eq:Relint.fifteen2} 
\end{equation} 
This formula also becomes progressively more valid for a massive 
particle as its total energy becomes progressively large
compared to its rest energy.
As stated previously, $m$ is the ``rest mass'', a constant
quantity, and there is no
question of ``mass increasing with velocity'' as occurs in some
descriptions of relativity, such as the famous ``$\mathcal{E}=mc^2$'', which
is incorrect in modern formulations.

Remembering to express it in terms of ${\bf p}$, the 
relativistic Hamiltonian is given by
\begin{equation}
{H} ({\bf p}) = \sqrt{p^2c^2 + m^2c^4}.
\label{eq:Relint.sixteen2} 
\end{equation} 

\subsection{4-vector notation}
It can be seen that ${\bf p}$, as given by Eq.~(\ref{eq:Relint.eleven2}), and $\mathcal{E}$, 
as given by Eq.~(\ref{eq:Relint.twelve2}), are closely related to the 4-velocity $u^i$. We
define a momentum 4-vector $p^i$ by
\begin{equation}
p^i = mu^i
 = 
\frac{m}{\sqrt {1 - v^2/c^2}}\,
\begin{pmatrix} c \\ {\bf v} \end{pmatrix}
 =
\begin{pmatrix}\mathcal{E}/c \\ {\bf p}\end{pmatrix}.
\label{eq:Relint.seventeen2} 
\end{equation} 
We expect that $p^ip_i$, the scalar product of $p^i$ with itself should,
like all scalar products, be invariant. 
The first of Eqs.~(\ref{eq:Relint.fourteen2}) shows this 
to be true;
\begin{equation} 
p^ip_i = \mathcal{E}^2/c^2 - p^2 = m^2c^2.
\label{eq:Relint.eighteen2} 
\end{equation} 
Belonging to the same 4-vector, the components of 
${\bf p}$ and $\mathcal{E}/c$ in different
coordinate frames are related according to the Lorentz transformation.

\subsection{Forced motion}
If the 4-velocity is to change, it 
has to be because force is applied to the particle.
It is natural to define the 4-force $G^i$ by the relation
\begin{equation}
G^i = \frac{dp^i}{ds/c}
    = \gamma\,\bigg(\frac{d\mathcal{E}/c}{dt},\frac{d{\bf p}}{dt}\bigg)^T
    = \bigg( \frac{{\bf F}\cdot{\bf v}/c}{\sqrt {1 - v^2/c^2}},
              \frac{{\bf F}}{\sqrt {1 - v^2/c^2}} \bigg)^T 
\label{eq:Relint.twentythree2} 
\end{equation} 
where 
\begin{equation}
{\bf F} = \frac{d{\bf p}}{dt}
\label{eq:Relint.twentyfour2} 
\end{equation} 
is the classically defined force. Since this formula is valid both
relativistically and non-relativistically it is least accident-prone
3D-form of Newton's law. 
The energy/time component $G^0$ is 
related to the rate of work done on the particle by the external 
force.
Note that this component vanishes
in the case that $ {\bf F}\cdot{\bf v} = 0 $, as is true,
for example, for a 
charged particle in
a purely magnetic field.

\section{Hamilton-Jacobi wave-like particle mechanics  \label{sec:Hamilton-Jacobi}}
For simplicity this appendix discusses the simplest non-trivial application
of Hamilton Jacobi wave-like mechanics, namely time independent, single particle mechanics.
\footnote{For the present paper the single particle restriction will be strictly respected,
but the time independence will need to be violated. Time dependence will be incorporated
by exploiting the H-J separation of variables formalism; just like the Schr\"odinger,
quantum mechanical treatment of the hydrogen atom.  This will account for an occasional,
once per cycle, brief energy acceleration. See Figure~(\ref{fig:Munoz-Pavic2}).
This can account for an (adiabatically changing)
periodic external accelerating force, while holding the other orbit elements constant.}
In this case a so-called
''complete integral'' of Hamilton's partial differential equation exists.  It is conventionally
referred to as the ``principal integral'' or the ``action function''  $S(q_i, \alpha_i, t)$,
in terms of which  Hamilton's first order differential equations can be obtained from the H-J
partial differential equation.
\begin{equation}
\frac{\partial S}{\partial t} + H(q_i, \frac{\partial S}{\partial q_i};\ t) = 0.  \quad i=1..3,
\label{eq:KeplerHJ.1}
\end{equation}
Hamilton's differential equations can then be expressed in the form
\begin{equation}
  \dot p_i = \frac{\partial S}{\partial q_i},\quad  \dot \alpha_i
           = -\frac{\partial S}{\partial \alpha_i}, \quad i=1..3,
\label{eq:KeplerHJ.2}
\end{equation}
where $\alpha_1,\ \alpha_2,\ {\rm and}\ \alpha_3$ are (adjustable) constant parameters.\

Using polar coordinates, the Lagrangian for a fundamental particle of variable mass
     $m_{\beta}\equiv m(\beta) \equiv m\gamma$, where $\gamma\equiv 1/\sqrt{1-\beta^2}$, and
where, to begin with, the velocity $v=\beta c$ is assumed to be non-relativistic, but eventually highly
relativistic, say $\beta=0.98$ velocity, moving in three dimensions in the field of an inverse
square law central gravitational force, and/or a transverse magnetic force, and/or a longitudinal
electric force.
\begin{equation}
L = 
\frac{1}{2}m\gamma(\dot r^2 + r^2\dot\theta^2 + r^2\sin^2\theta\,\dot\phi^2) - \frac{K}{r},
                \hbox{\ where\ } K=\frac{Ze^2}{4\pi\epsilon_0 r}.
\label{eq:KeplerHJ.1}
\end{equation}
The canonical momenta are
\begin{equation}
p_r = m\gamma\dot r, \quad
p_{\theta} = m\gamma\dot\theta, \quad
p_{\phi} = m\gamma\sin^2\theta\,\dot\phi,
\label{eq:KeplerHJ.2}
\end{equation}
and the Hamiltonian is
\begin{equation}
H = \frac{p^2_r}{2m\gamma}
  + \frac{p^2_{\theta}}{2m\gamma^2} 
  + \frac{p^2_{\phi}}{2m\gamma^2\sin^2\theta}
  + \frac{K}{r}.
\label{eq:KeplerHJ.3}
\end{equation}
For nuclear particles $K$ and $Z$ are both positive.
Looking for a solution by separation of variables, the time-independent H-J
equation is\footnote{As elsewhere in this paper,
the conventional symbol $E$, for ``energy'' has been replaced by $\mathcal{E}$.}
\begin{equation}
\mathcal{E} = \frac{1}{2m\gamma}
\Bigg(
\bigg(\frac{dS^{(r)}}{dr}\bigg)^2
+
\frac{1}{r^2}\,\bigg(\frac{dS^{(\theta)}}{d\theta}\bigg)^2
+
\frac{1}{r^2\,\sin^2\theta}\,\bigg(\frac{dS^{(\phi)}}{d\phi}\bigg)^2
\Bigg) + \frac{K}{r}.
\label{eq:KeplerHJ.4}
\end{equation}
Since $\phi$ does not appear explicitly we can separate it
immediately in the same way $t$ has already been separated;
\begin{equation}
S = -\mathcal{E}t + \alpha_3\phi + S^{(\theta)}(\theta) + S^{(r)}(r) .
\label{eq:KeplerHJ.5}
\end{equation}
Here $\mathcal{E}$ is the first ``new momentum'' of Jacobi and $\alpha_3$ is the second; it is interpretable as
the value of a conserved azimuthal angular momentum around the $z$ axis because
\begin{equation}
p_{\phi} = \frac{\partial S}{\partial\phi} = \alpha_3
\label{eq:KeplerHJ.6}
\end{equation}
is constant.  Substituting this into Eq.~(\ref{eq:KeplerHJ.4}) 
and multiplying by $2m\gamma r^2$
yields
\begin{equation}
2m\gamma \mathcal{E} r^2 - 2mK\gamma - r^2\bigg(\frac{dS^{(r)}}{dr}\bigg)^2
=
\bigg(\frac{dS^{(\theta)}}{d\theta}\bigg)^2
+
\frac{1}{\sin^2\theta}\,\bigg(\frac{dS^{(\phi)}}{d\phi}\bigg)^2
=
\alpha^2_2 ,
\label{eq:KeplerHJ.7}
\end{equation}
where the equality of a pure function of $r$ to a pure function
of $\theta$ implies that both are constant; this has permitted
a third Jacobi parameter $\alpha_2$ to be introduced.
The physical meaning of
$\alpha_2$ can be inferred by expanding $M^2$, the square 
of the total angular momentum; 
\begin{equation}
M
= 
\sqrt{(m\gamma r^2\dot\theta)^2 + (m\gamma r^2\sin\theta\,\dot\phi)^2} 
=
\sqrt{p^2_{\theta} + \frac{p^2_{\phi}}{\sin^2\theta}}
=
\alpha_2 .
\label{eq:KeplerHJ.9}
\end{equation}
From the interpretation of $\alpha_3$ as the $z$ component
of $\alpha_2$ it follows that
\begin{equation}
\alpha_3 = \alpha_2\,\cos{i} .
\label{eq:KeplerHJ.9p}
\end{equation}
%
Determination of the other terms in $S$ has been
``reduced to quadratures'' since Eqs.~(\ref{eq:KeplerHJ.7}), gives
expressions for $dS^{(\theta)}/d\theta$
and $dS^{(r)}/dr$ that can be re-arranged to yield
$S^{(\theta)}(\theta)$ and $S^{(r)}(r)$ as indefinite
integrals;
\begin{align} 
S_2 =\ &  
-\int^{\theta}\sqrt{\alpha_2^2-\frac{\alpha^2_3}{\sin^2\theta'}}\,
      d\theta',                       \notag \\
S_3 =\ &  
\int^r\sqrt{2m\gamma \mathcal{E} - \frac{2m\gamma K}{r'}-\frac{\alpha^2_2}{{r'}^2}}\,dr'.
\label{eq:KeplerHJ.9pp}
\end{align}
Instead of using $\mathcal{E}$ as the first Jacobi ``new momentum''
it is conventional to use a function of $\mathcal{E}$, namely
\begin{equation}
\alpha_1 = \sqrt{\frac{-K^2m\gamma}{2\mathcal{E}}}, \quad
\mathcal{E} = \frac{-K^2m\gamma}{2\alpha^2_1} .
\label{eq:KeplerHJ.10}
\end{equation}
Like $\alpha_2$ and $\alpha_3$, $\alpha_1$ has dimensions
of angular momentum.  The semi-major axis $a$
and the orbit eccentricity $\epsilon$ are given by
\begin{equation}
a = -\frac{\alpha^2_1}{Km\gamma} ,\quad
1-\epsilon^2 = \bigg(\frac{\alpha_2}{\alpha_1}\bigg)^2,
\label{eq:KeplerHJ.11}
\end{equation}
with inverse relations
\begin{equation}
\alpha^2_1 = -Km\gamma a ,\quad
\alpha^2_2 = -(1-\epsilon^2)Km\gamma a .
\label{eq:KeplerHJ.12}
\end{equation}
Combining results, the ``action integral'' of the H-J equation is
\begin{equation}
S = \frac{m\gamma K^2}{2\alpha^2_1}\,t
  + \alpha_3\phi
- \int^{\theta}_{\pi/2}
\sqrt{\alpha^2_2 - \frac{\alpha_3^2}{\sin^2{\theta'}}}\ d\theta'
+ \int^r_{a(1-\epsilon^2)}
\sqrt
{-\frac{(m\gamma)^2K^2}{\alpha^2_1}
 - \frac{2m\gamma K}{r'}
 - \frac{\alpha^2_2}{{r'}^2}}\ dr' .
\label{eq:KeplerHJ.13}
\end{equation}
The lower limits and some signs have been chosen arbitrarily
so that the Jacobi ``new momenta'' 
$\beta_1$, $\beta_2$ and $\beta_3$
will have conventional meanings. Refer to reference~\cite{Talman-Mechanics-Chapter-8}
for further details.

The purpose of this appendix has been to reformulate classical relativistic particle mechanics as wave
mechanics, from which single particle orbits, treated as ``rays'', are unambiguous and calculable.
Mathematically, the classical orbits are stationary time ``rays'' normal to H-J ``wave-fronts''.

This formalism was established by Jacobi almost a century before Einstein's special relativity
and longer than that before Heisenberg's and Schr\"odinger's non-relativistic quantum mechanics.
It is more or less equivalent to the Bohr-Sommerfeld model of the hydrogen atom.

During the years from 1900 to 1920, Jacobi's wave theory, (developed a century earlier)
as well as being consistent with special relativity, was surprisingly similar to elementary,
non-relativistic quantum mechanics about to be developed--absent, of course, failure of commutation,
particle creation, and Heisenberg uncertainty.

For the present cosmic ray investigation, the H-J theory seems to be especially well matched.
In the context of this paper, the first invaluable feature of the separation of variables
method of solution of the Hamilton-Jacobi solution is that only one ``separable momentum'', the
particle energy energy changes while a particle is circulating about the sun. During this time interval
the other three ``momenta'' remain constant. The second invaluable feature is that, while out of contact
with the sun, the energy remains constant, while the other three momenta change, in accordance with H-J
theory, plus stochastic dependence on stray magnetic fields.

As one result of the existence of these alternating states of motion, the radial dependence of the energy
is highly predictable, with potential energy depending only on radial position, $r$.  Meanwhile, the
transverse deviations from centripetal or centrifugal motion are caused by magnetic fields that, though weak,
are essentially unknown.  Theoretical description of these motions needs to be described by adiabatic
variation of the other three separation ``constants'', as constrained by the known energy variation.

Since the magnetic fields are, at best, known only in some statistical sense, the transverse deviations
are necessarily knowable only in a stochastic sense.  The basically unpredictable wandering caused by
magnetic fields needs to be modeled stochastically.

While in contact with the sun, the motion is governed by the magneto-hydrodynamic properties of
the solar plasma. This, too, is also stochastic, but the energy gain or loss is well represented by
the total energies of individual particles or coherent bunches of particles.

\section{Kulsrud et al. theory of cosmic ray, stellar plasma interaction \label{sec:Kulsrud}}
Never having believed that cosmic rays were galactic, let alone extra-galactic in origin, I have,
until recently, ignored papers assuming extra-solar sources; which is to say ignoring
the entire plasma theory of the sources of cosmic rays.  I have only belatedly come to realize that the
physics of the source of cosmic rays is more or less the same, irrespective of their source location,
be it solar or extra-galactic.  Understanding only accelerators, and certainly not magneto-hydrodynamic,
this attitude has not been particularly appropriate.

Having overcome this impediment, I have found that reference\cite{Felice-Kulsrud} and 
Kulsrud's book\cite{Kulsrud-Chap.-12}, ``Plasma Physics for
Astrophysics}'', in particular Chapter 12, have provided answers to questions that I have been unable even to
formulate effectively myself. Because of its amazing relevance, and to avoid symbol clashes, this appendix
exhibits and comments on paragraphs copied  along with figures from these sources.
What makes this promising is the surprising way in which plasma physics can ``imitate'' laboratory equipment
that itself took half a century to engineer.

This comment applies especially to particle injection into terrestrial accelerators which, though adequately
understood for most practical purposes, continues to be a serious research area to this day.  For both laboratory and
celestial accelerators, beam extraction is easier to understand than injection, and hence less controversial. 

Initially following Kulsrud, this appendix compares laboratory and celestial accelerators, at
least superficially, and as briefly as possible. 

The second paragraph of Chapter~12 of Kulsrud's book begins as follows:\\
\newline
`` Let the number of cosmic rays in the energy range $d\epsilon$ be $N(\epsilon)\,d\epsilon$.
Their energy spectrum $N(\epsilon)$ is very nearly a perfect power law, with exponent -2.7 from
1 to $10^6$\, GeV $$N(\epsilon) \sim e^{-2.7}.$$ Above $10^6$\, GeV it remains a power law, but
the exponent changes to 3.1 and then back
again to 2.7 above $10^6$\,GeV.  It is believed these very high energy cosmic rays are
extra-galactic and full the universe uniformly. Their origin is believed to be in galactic
sources.''\\
\\
It should go without my saying that I agree with the first half of this statement, which conforms
closely with most of this paper, but categorically disagree with the latter half.

Kulsrud's following several paragraphs explain why inter-galactic cosmic ray scattering can be
neglected without any serious concern.  This means that the cosmic ray sources must lie within galaxies.
\emph{ As an aside, my approach goes further,  by saying the sources are almost entirely within the solar system.} At the same time, The
existence of galactic cosmic rays into the solar system seems to be required for ``start up'' of
the solar cosmic ray factory.  Kulsrud discusses cosmic abundance ratios and many other issues in detail.

\begin{figure}[hbt!]
\centering
\includegraphics[scale=0.35]{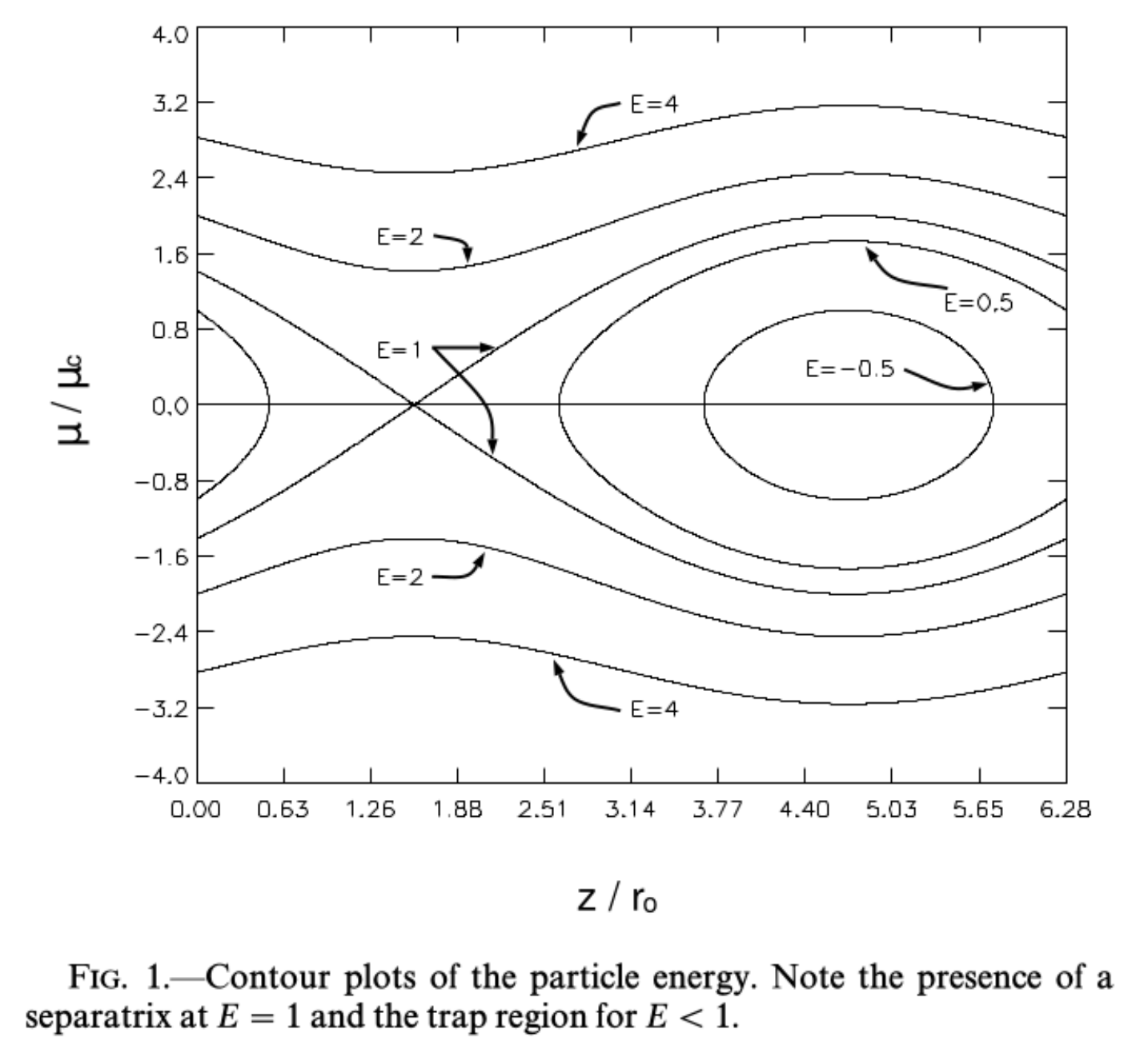}
\includegraphics[scale=0.35]{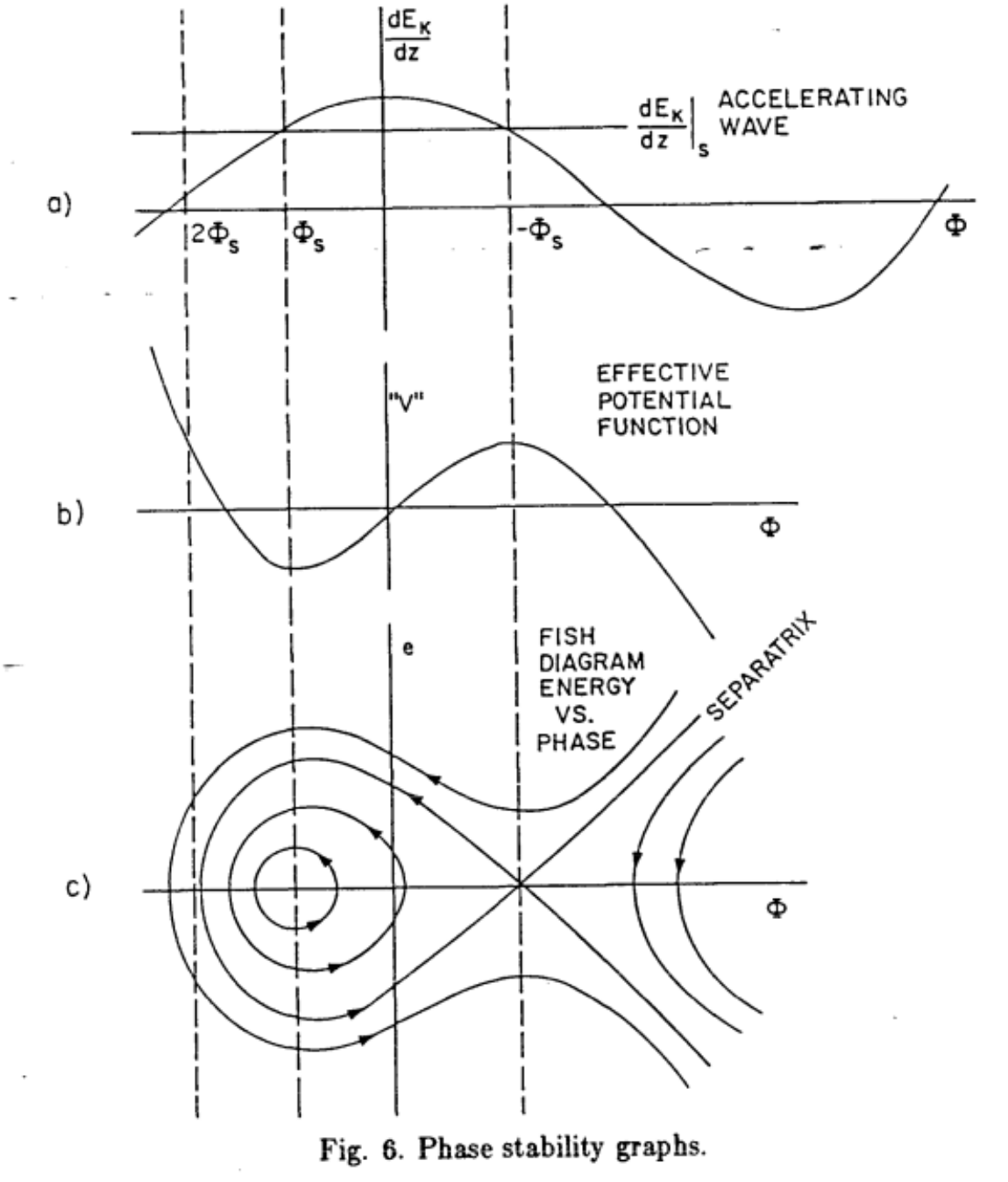}
\caption{\label{fig:Felice-RF-bucket}{\bf Left:\ } Copied from Felice and Kulsrud\,\cite{Felice-Kulsrud}, along with
 its original caption, longitudinal phase space in plasma atmosphere of a star, and
{\bf Right:\ }copied from Loew and Talman\,\cite{Loew-Talman}
longitudinal phase space in a linear proton ``Alvarez'' accelerator.  The similarity of these two figures suggests that
the physics of the acceleration of cosmic rays near a star resembles the physics of acceleration of electrons in a
linear accelerator. In each case there can be charges trapped in moving ``stable buckets''.  A figure on page 347
of reference \cite{Kulsrud-Chap.-12} resembles the upper two figures on the fight, and is followed by appropriate
explanation of the plasma physics.}
\end{figure}

Especially significant is Kulsrud's section 12.2, ``Pitch-angle Scattering of Cosmic Rays by
Alfven Waves''.  This topic is described in greater detail in Felice and Kulsrad\cite{Felice-Kulsrud},
from which Figure~(\ref{fig:Felice-RF-bucket}) is copied.  The abstract to this paper begins with the lines\\
\\
``We study the problem of cosmic-ray di†fusion in the Galactic disk with particular attention paid to
the problem of particle scattering through the  pitch angle in momentum space by wave-particle mirror
interaction (here v is the cosmic-ray velocity parallel to the average Galactic magnetic field).''\\

The left figures is just as applicable to our sun as it would be to any star in any galaxy.
It is matched by Figure~\ref{fig:Felice-RF-bucket}, for which the top figure shows two cells of an
Alvarez proton linac.  Each cell is resonant at the same frequency but is tuned in phase to produce a
slow (i.e. less than the speed of light) traveling electromagnetic wave, moving at exactly the average speed
of ``captured'' bunches of protons. Each captured proton follows an elliptical paths that appears to be
closed, like the two shown, in the frame of reference of the captured protons.
\begin{figure}[hbt!]
  \centering
\includegraphics[scale=0.44]{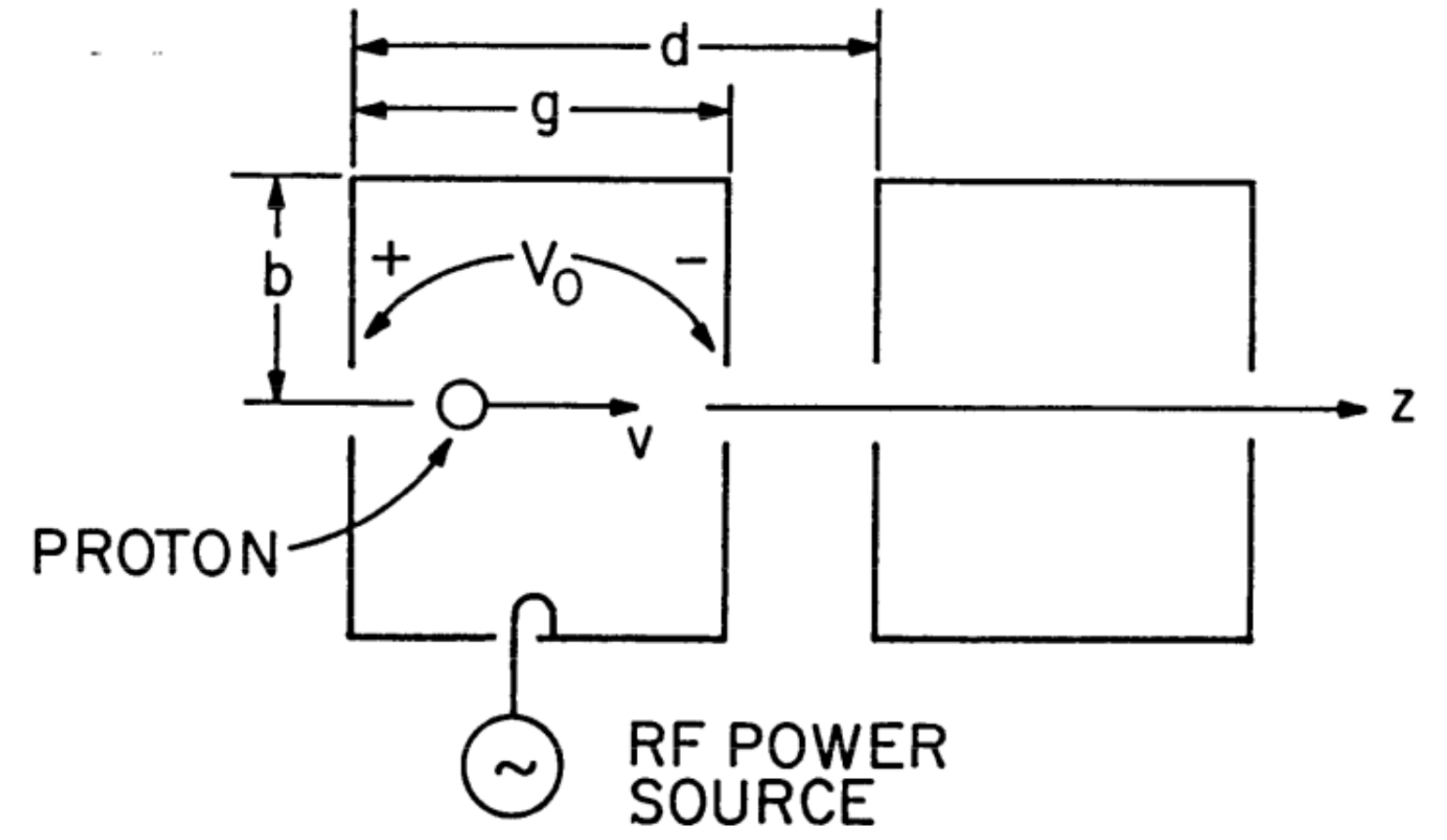}
\includegraphics[scale=0.44]{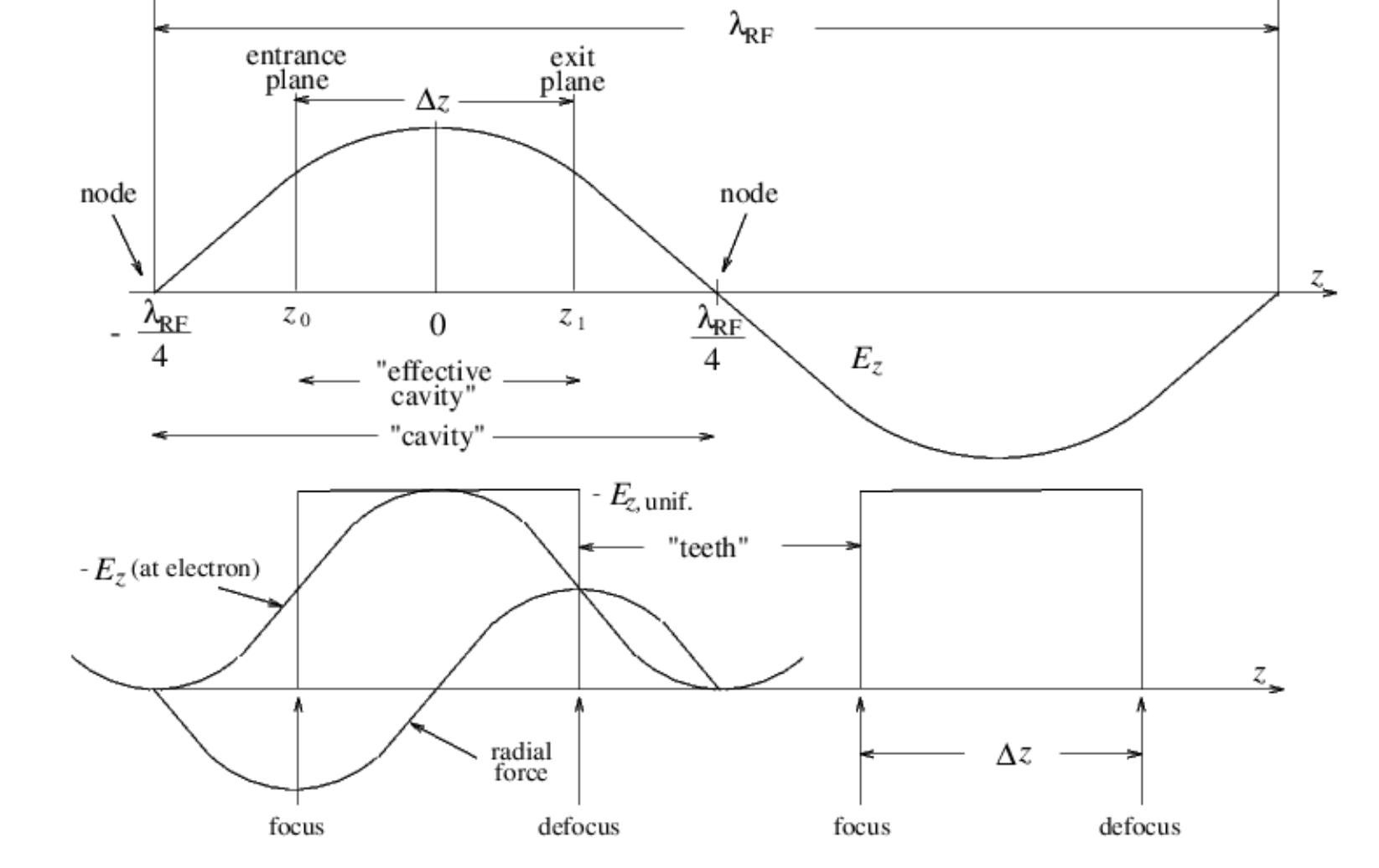}
\caption{\label{fig:UAL-linac-04-01-no-caption}
Copied from Loew and Talman,\cite{Loew-Talman} the {\bf upper} figure 
shows one of the many  RF-cavity accelerating cells of an Alvarez linac
needed to produce a ``slow'' (meaning slower than the speed of light)
electromagnetic traveling wave. Individual cells, though seemingly
independent, are phased to produce the desired wave speed, for example
such that a moving proton is always ``on crest''.
The  {\bf lower} figure resembles Figure~12.1 in the
Kudsrud book\cite{Kulsrud-Chap.-12} which shows plasma fields capable
of accelerating cosmic rays in galactic magneto-hydrodynamic Alfven waves.
This figure, applicable to cells in an Alvarez linac, models a proton's
instantaneous position and replacement of actual accelerating field
by a train of uniform field ``teeth'' of thickness
      $\Delta z \simeq \lambda_{\rm RF}/4$.
Any radial force is modeled by delta functions at entrances and exits.}
\end{figure}
%


\section{Accelerator physics units \label{sec:Units}}
With just a single exception, (with unfortunately confusing consequences) ``accelerator physics units''
and MKSA units are interchangeable,
much in the same way that one meter is 100 centimeters. The exception is that energies are measured
in GeV, rather than eV. What makes this change popular for high energy accelerator and particle
physicists is that, within +/-15\%, the proton mass (expressed as rest energy), or by nuclear A-value,
or by atomic mass units, is roughly equal to 1.  Electric fields are measured in GeV/m, rather than V/m,
which is not inconvenient.  Another easily-remembered practical result is the Bolzmann conversion from
Kelvin temperature to energy; an energy of 1\,MeV corresponds to a temperature of $1.16\times10^{10}$\,K.

However, the units for many other physical quantities become somewhat obscure; especially the magnetic
field, because the Lorentz magnetic force is proportional both to magnetic field and to particle velocity.
Checking that electric and magnetic fields are close to what is expected, it should be possible to
interpret all physical parameters correctly.

Incidentally, these accelerator units are much like ``natural units'' in that
the speed of light is equal to 1. In this paper SI formulas involving the speed of
light retain the factor ``c'', but its numerical value is 1.0.
The fact that the Planck (modified) constant $\hbar$, is not set equal to 1 in
accelerator units complicates this practice. For analyzing weak
interaction processes, natural units are greatly to be favored. 

In the present paper, accelerators with superimposed electric and magnetic fields are 
first described, with bending fractions $\eta_E$ and $\eta_M$ which sum to one. This
is followed immediately by celestial accelerators having predominantly gravitational
bending with similar bending fractions $\eta^*_G$ and $\eta^*_M$, which are proportional
to the bending strengths, but do not sum to 1.

For ${\rm E}\&{\rm M}$ bending, to avoid the direct evaluation of magnetic fields in Tesla units,
electric fields are given in GV/m, while magnetic bending fields are inferred from $\eta_{M1}$
the magnetic bending fraction for beam 1. For celestial ${\rm G}\&{\rm M}$ bending a similar
convention has needed to be devised.

\end{document}